\documentclass[12pt,english]{article}
\usepackage{lmodern}
\usepackage{lmodern}

\usepackage[T1]{fontenc}
\usepackage[latin9]{inputenc}
\usepackage{mathrsfs}
\usepackage{amsmath}
\usepackage{amsthm}
\usepackage{amssymb}
\usepackage{graphicx}
\usepackage{geometry}
\makeatletter
\numberwithin{equation}{section}

\@ifundefined{date}{}{\date{}}
\usepackage{jheppub}
\usepackage{cancel}
\usepackage{leftindex}
\usepackage{babel}
\usepackage{bbold}
\usepackage{tikz}

\usepackage{lineno}

\makeatother

\usepackage{babel}
\begin{document}
\global\long\def\L{\mathcal{L}}%

\global\long\def\bbZ{\mathbb{Z}}%

\global\long\def\bbR{\mathbb{R}}%

\global\long\def\cZ{\mathcal{Z}}%

\global\long\def\S{\mathcal{S}}%

\global\long\def\M{\mathcal{M}}%

\global\long\def\P{\mathcal{P}}%

\global\long\def\C{\mathcal{C}}%

\global\long\def\T{\mathcal{T}}%

\global\long\def\A{\mathcal{A}}%

\global\long\def\U{\mathcal{U}}%

\global\long\def\H{\mathcal{H}}%

\global\long\def\d{\mathrm{d}}%

\global\long\def\vphi{\varphi}%

\global\long\def\id{\mathbb{1}}%

\global\long\def\D{\mathcal{D}}%

\global\long\def\i{\mathrm{i}}%

\global\long\def\e{\mathrm{e}}%

\global\long\def\ket#1{\left|#1\right\rangle }%

\global\long\def\bra#1{\left\langle #1\right|}%

\global\long\def\braket#1#2{\left\langle #1\vert#2\right\rangle }%

\global\long\def\proj#1#2{\vert#1\rangle\langle#2\vert}%

\global\long\def\Ising{\text{Ising}}%

\global\long\def\mcg{{\rm MCG}}%

\global\long\def\voa{\mathscr{V}}%

\global\long\def\Hom{{\rm Hom}}%

\global\long\def\sD{\mathscr{D}}%

\global\long\def\sL{\mathscr{L}}%

\global\long\def\gvec#1{\vec{#1}}%

\global\long\def\Aut{{\rm Aut}}%

\global\long\def\Inv{{\rm Inv}}%

\global\long\def\sS{\mathsf{S}}%

\global\long\def\ii{\alpha}%

\global\long\def\jj{\alpha'}%

\global\long\def\coeff{w}%

\global\long\def\ginf{g_{\infty}}%

\global\long\def\mcgg{{\rm MCG}(\Sigma_{g})}%

\global\long\def\fpdim{{\rm FPdim}}%

\global\long\def\irr{\mathrm{Irr}}%

\newcommand{\enlargeop}[2][1.4]{%
  \mathop{\vcenter{\hbox{\scalebox{#1}{$\displaystyle #2$}}}}\limits%
}

\affiliation{Department of Physics and Astronomy, University of Kentucky,\\
506 Library Drive, Lexington, KY 40506}

\emailAdd{a.barbar@uky.edu}

\abstract{We study the recent proposal of \cite{dymarsky2025_TQFTGravity} which poses a precise holographic duality between a 3d TQFT summed over all topologies and a unitary ensemble of boundary 2d CFTs. In that proposal, the sum over topologies is obtained via genus reduction from topologies with a large genus boundary Riemann surface, while the boundary ensemble is given by all CFTs described by Lagrangian condensations of the bulk TQFT. The main result of this work is to show that each member of this ensemble is weighted by a symmetry factor given by the invertible symmetry group of its categorical symmetry relative to the bulk TQFT as its SymTFT. This is the natural --- uniform up to isomorphism --- measure on the groupoid of Lagrangian algebras that describe the boundary theories. We also write the sum over topologies more explicitly in terms of equivalence classes of Heegaard splittings of 3-manifolds with a given boundary and comment on their weights. The holographic duality in this framework  can then be viewed as a generalization of the Siegel-Weil formula. We discuss the implications of the main result for non-compact TQFTs. In particular, for the Virasoro case, this implies an ensemble of all CFTs at a given central charge in which CFTs are weighted by their full invertible symmetry. Finally, we show how this TQFT gravity framework gives a natural construction of the baby universe Hilbert space.}
\title{Automorphism-weighted ensembles from TQFT gravity}
\author{Ahmed Barbar}

\maketitle

\section{Introduction and summary of results}

\tikzset{every picture/.style={line width=0.75pt}} 

The quest for understanding the holographic duality of 3d gravity
in asymptotically anti-de Sitter space (AdS) faces some interesting
puzzles. First, in the standard AdS/CFT correspondence, we expect
gravity to be dual to a single CFT on the boundary. However, in multiboundary
scenarios, contributions from Euclidean wormholes spoil the factorization
of the partition function. One resolution of this puzzle is that gravity
is dual to an ensemble rather than a single theory. This resolution
seems to be backed up by the results of JT gravity \cite{saad2019_JTGravity}.
The ensemble picture is an illustration of the early ideas considered
by Coleman \cite{coleman1988_babyuniverse} and by Giddings and Strominger
\cite{Giddings:1988cx,giddings1989_BabyUniverses} according to which
the contribution of Euclidean wormholes leads to disorder averaging.
These ideas were recently recast in the context of asymptotically
AdS boundaries by Marolf and Maxfield \cite{marolf2020_TranscendingEnsemble}.
If 3d gravity is dual to an ensemble, the question remains how one
should average over the space of CFTs. One can make progress by considering
a random ensemble of CFT data, for which there are many attempts as
in \cite{Cotler:2020randomCFT,Chandra:2022bqq,diubaldo2023_AdS3RMT2,belin2024_ApproximateCFTs,jafferis2025_3dGravity,boruch2025_RMT},
but this still does not exactly tell us what kind of average over
bona fide CFTs this is.

The other puzzle is that evaluating the path integral of semiclassical
${\rm AdS_{3}}$ gravity with a single boundary by summing over saddles
leads to some pathological results. The resulting density of states
is continuous and negative in some part of the spectrum \cite{MaloneyWitten2007,keller2015_PoincareSeries}.
The continuous spectrum is not really a problem in an ensemble interpretation,
but the negativity problem is a serious blow to any possibility for
a unitary interpretation. The negativity problem has been shown to
be cured by including additional contributions to the classical saddles,
such as off-shell contributions like Seifert manifolds \cite{maxfieldTuriaci2020_seifert}
or conical defects \cite{benjamin2020_PureGravity}. This seems to
tell us that even in the semiclassical limit one should be careful
not to dismiss the sum over all 3-manifolds.

One way to tackle these issues is through the TQFT approach to gravity,
which is now more precisely formulated in terms of Virasoro TQFT (VTQFT)
\cite{Collier:2023Solving3d}, also known as Teichmüller TQFT \cite{ellegaardandersen2014_TQFTQuantum}.
Schematically, we have
\begin{equation}
\boldsymbol{Z}_{\text{grav}}=\sum_{\text{topologies}}\boldsymbol{Z}_{\text{VTQFT}\times\overline{\text{VTQFT}}},
\end{equation}
where we leave implicit the possibility of a nontrivial measure in
this sum. Working with VTQFT has proved to be very fruitful \cite{Collier:2023Solving3d,collier2024_VTQFT_2,post2025_NonrationalVerlinde,yan2025_Puzzles3D,chandra2025_Statistics3d},
but there are obviously certain limitations due to divergences; after
all it is a noncompact TQFT which would need some kind of regularization;
see for example the related Turaev-Viro approaches in \cite{bao2024_QGSymQRG,hartman2025_ConformalTuraevViro,hartman2025_TriangulatingQuantum}.
One can instead try to answer these questions in a more tamed TQFT
hoping eventually to learn something about the Virasoro case. There
are many toy models of ensemble averaging in the literature, either
directly or indirectly involving a bulk TQFT setup \cite{maloney2020_AveragingNarain,afkhami-jeddi2021_FreePartition,dong2021_AveragingModuli,Meruliya:2021RCFTensemble,Ashwinkumar:2021kav,Raeymaekers:2021ypf,Raeymaekers:2023ras}.
However, in most of these examples, the sum over topologies involves
only handlebodies, which in many non-Abelian TQFT examples \cite{castro2012_GravityDual,meruliya2021_PoincareSeries}
suffers from a negativity problem similar in spirit to pure 3d gravity.
Other attempts to sum over all manifolds are mainly in a setting involving
Abelian TQFTs \cite{Nicosanti:2025xwu,Angelinos:2025zek} which are
immune to the negativity problem and are limited as toy models of
VTQFT. 

To understand the holographic duality properly in a general TQFT setting,
we need a way to construct the sum over all topologies and also figure
out what kind of boundary ensemble we should get; i.e. what are the
boundary CFTs? and what are their weights? The picture proposed in
\cite{barbar2025_GlobalSymmetries}, motivated by a connection between
codes and CFTs, is that the sum over all topologies should be dual
to an ensemble of all Lagrangian condensations of the TQFT. A Lagrangian
condensation here means a maximal gauging of a (generally non-invertible)
1-form symmetry of the TQFT reducing it to a trivial TQFT. These are
related to states in the TQFT Hilbert space that describe topological
boundary conditions, which we will denote as $\ket{Z_{\alpha}}$.
So our general expectation now is of the form
\begin{equation}
\sum_{\text{topologies}}\boldsymbol{Z}_{\T}=\sum_{\ii}\coeff_{\ii}Z_{\ii},\qquad\quad w_{\alpha}\geq0,\label{eq: TQFT gravity duality 0}
\end{equation}
where the left-hand side (LHS) describes the ``\emph{TQFT gravity}''
partition function, which is a sum over the partition function ${\bf Z}_{\T}$
of the TQFT $\T$ placed on different topologies, while the RHS describes
an ensemble of CFTs whose partition functions are denoted by $Z_{\alpha}$.
This should be purely understood as a statement in the Hilbert space
of the TQFT where the LHS is a sum over states prepared by putting
the TQFT $\T$ on a particular topology, and the RHS is a sum over
states corresponding to a fixed topology of a handlebody with insertion
of line defects. The statement in (\ref{eq: TQFT gravity duality 0})
is not completely trivial since, in general, the states $\ket{Z_{\alpha}}$
do not form a complete basis in the space of modular invariant states.

This picture was made more precise by Dymarsky and Shapere in \cite{dymarsky2025_TQFTGravity}
with the motivation that the boundary ensemble should be fixed for
all genera of the boundary Riemann surface, while the sum over topologies
at a given genus should be consistent with reduction from its counterpart
at a higher genus.  With this idea one can bootstrap the ensemble
by studying what happens in the large genus limit. In that limit,
the duality simplifies drastically: the sum over topologies becomes
just a sum over handlebodies, and the task of evaluating the weights
of the boundary ensemble becomes much simpler as the normalized versions
of the states $\ket{Z_{\alpha}}$ can be shown to become orthogonal
in that limit. For Abelian TQFTs, as was shown explicitly in \cite{dymarsky2025_TQFTGravity},
all Lagrangian condensations are weighted equally, but in a general
TQFT, the relative weights will have the form
\begin{equation}
\frac{\coeff_{\ii}}{\coeff_{\jj}}=\lim_{g\rightarrow\infty}\frac{\braket{Z_{\jj}}{Z_{\jj}}}{\braket{Z_{\ii}}{Z_{\ii}}},\label{eq: weights_nonabelian_1}
\end{equation}
where the squared norms $\braket{Z_{\ii}}{Z_{\ii}}$ of the states
corresponding to Lagrangian condensations are generally different,
leading to nontrivial weights. These weights can be calculated in
some simple non-Abelian examples (see the End-Matter section of \cite{barbar2025_GlobalSymmetries})
but the general structure of how to compute these weights or what
they mean was still missing.

In this work, we calculate these weights explicitly for general 3d
TQFTs based on semi-simple unitary modular tensor categories and give
them a physical interpretation in terms of symmetries. We also briefly
shed some light on the sum over topologies, writing it explicitly
in terms of equivalence classes of Heegaard splittings, and also make
a connection to the construction of the closed (baby) universe Hilbert
space. Below we summarize the main result of the paper which is related
to the weights of the ensemble.

\subsection{Summary of the main result}

The main result of this paper is that a 3d TQFT summed over all topologies
as proposed in \cite{dymarsky2025_TQFTGravity} is dual to an ensemble
of boundary theories where each member of the ensemble is (inversely)
weighted by a symmetry factor. The bulk TQFT naturally acts as a symmetry
topological field theory (SymTFT) for each member, and so each member
$\alpha$ has a certain categorical symmetry $\C^{(\ii)}$ relative
to the bulk TQFT. What we will show is that the weights are inversely
related to the order of the group of invertible symmetries of this
categorical symmetry, namely (up to an overall normalization)
\begin{equation}
w_{\ii}=\frac{1}{\left|\Inv(\C^{(\alpha)})\right|},\label{eq: weights_intro}
\end{equation}
where $\Inv(\C)$ denotes the group of isomorphism classes of invertible
objects in the category $\C$. In a given boundary theory, these are
the invertible TDLs that commute with the vertex algebra associated
with the bulk TQFT. 

The key point in deriving (\ref{eq: weights_intro}) is to consider
$\braket{Z_{\alpha}}{Z_{\alpha}}$ as a partition function of a 2d
TQFT with a categorical symmetry $\C^{(\alpha)}$, and in the large
genus limit this partition function will effectively count the number
of invertible topological line defects (TDLs) of $\C^{(\alpha)}$. 

  The weights in (\ref{eq: weights_intro}) are not only intrinsic
to the TQFT holographic ensemble duality, but they are also --- as
we will argue --- the natural weights for defining a uniform average
up-to-isomorphism. Each member of the ensemble, being a Lagrangian
condensation, is algebraically described by what is known as a \emph{Lagrangian algebra}
$\A$. We will show that our resulting ensemble average can be written
as
\begin{equation}
<Z>=\left(\sum_{\A}\frac{1}{\left|\Aut(\A)\right|}\right)^{-1}\sum_{\A}\frac{1}{\left|\Aut(\A)\right|}Z_{\A},\label{eq: average Z formula1}
\end{equation}
where $\Aut(\A)$ denotes the automorphism group of $\A$, which will
be defined precisely in the main text. This is the natural way to
average over the groupoid of Lagrangian algebras of a given modular
tensor category. The normalization sum $\sum_{\A}\frac{1}{\left|\Aut(\A)\right|}$
in (\ref{eq: average Z formula1}), which defines the counting measure
in the groupoid sense, is known as the groupoid cardinality \cite{Baez2001}.
It is the analog of the Smith--Minkowski--Siegel ``mass'' formula\footnote{See also \cite{Dymarsky_mass_formula} for related work on the mass
formula and topological boundary conditions.} for lattices \cite{Smith1868_mass_formula,minkowski1885_mass_formula,Siegel1935_mass_formula},
and as we will show later it will be given by the squared norm of
the Hartle-Hawking (HH) state, which is the partition function of
a closed universe
\begin{equation}
\braket{\text{HH}}{\text{HH}}=\sum_{\A}\frac{1}{\left|\Aut(\A)\right|}.
\end{equation}
 The holographic duality we present then can be viewed as  a generalization
of the Siegel-Weil formula \cite{siegel1951_IndefiniteQuadratische,Maass1954_Lecture_notes,weil1964_CertainsGroupes,weil1965_FormuleSiegel}.
The Siegel-Weil formula relates the automorphism-weighted average
over lattice Riemann theta functions (related to the partition functions
of boundary free bosons) to an Eisenstein series (related to the sum
over handlebodies), and explains the holographic duality of the ensemble
average of Narain theories \cite{maloney2020_AveragingNarain,afkhami-jeddi2021_FreePartition}.
In our case, we get an automorphism-weighted average over partition
functions of Lagrangian algebras equivalent to a sum over all 3-manifolds
instead of just handlebodies. 

\subsection{Outline}

The paper is organized as follows. In section \ref{sec:Preliminaries},
we provide some preliminaries for 3d TQFTs, anyon condensation and
SymTFT. This will serve as a brief review of the main ingredients
of our setup as well as establishing the notation and conventions
for later sections. In section \ref{sec:TQFT-gravity}, we first review
the main derivation of TQFT gravity following \cite{dymarsky2025_TQFTGravity}
and elaborate a bit more on the sum over topologies, writing it explicitly
in terms of equivalence classes of Heegaard splittings. We then give
some arguments that the mapping class group invariant subspace of
the infinite genus Hilbert space has a natural interpretation as the
baby universe Hilbert space. Section \ref{sec:Weights-from-symmetries}
provides the main derivation for the weights of the ensemble in terms
of symmetries and the relation to automorphisms of Lagrangian algebras.
In section \ref{sec:Examples}, we present some simple examples to
illustrate the main result of getting the weights from symmetries.
In section \ref{sec:Implications-for-noncompact}, we discuss the
implications of the TQFT gravity framework for noncompact TQFTs.
Finally, in section \ref{sec:Discussion}, we end with some discussion
on the interpretation of this result in light of the principle of
maximum ignorance as well as possible future directions.

\section{Preliminaries\label{sec:Preliminaries}}

\subsection{3d TQFTs}

3d TQFTs are closely connected to 2d rational CFTs (RCFTs) and the
concept of chiral algebras which are more formally known as Vertex
Operator Algebras (VOAs) \cite{frenkel1984_Moonshine,frenkel1989vertex,kac1998vertex}.
In the TQFT/RCFT correspondence \cite{witten1989_QuantumField,moore1989_TamingConformal,fuchs2002_TFTConstruction},
the simple line defects (anyons) of the TQFT are in one to one correspondence
with rational primary operators of the RCFT which are isomorphism
classes of irreducible representations of some VOA. The algebraic
structure of these anyons is captured by a unitary modular tensor
category (MTC). Objects of this category are representations of the
VOA and morphisms between objects are intertwiners. We refer the reader
to appendix \ref{sec:Brief-review-of MTC} for a brief review of the
MTC structure and the related notation used in this section.

In what follows, we will denote the 3d TQFT as well as its category
by $\T$ and its VOA as $\voa_{\T}$, so we have $\T={\rm Rep}(\voa_{\T})$.\footnote{Note that to define the TQFT from an MTC, one should specify the chiral
central charge $c_{-}$ since the category is only sensitive to $c_{-}\mod 8$.} While conventionally $\voa_{\T}$ is taken as the chiral algebra
of an RCFT, we will work in general with $\voa_{\T}$ as the full
non-chiral\footnote{In this context we say the VOA is nonchiral if it contains ${\rm Vir}\times\overline{{\rm Vir}};$
see \cite{Kapustin:2000aa,rosellen2004opealgebrasmodules,Singh:2023_nonchiralVOA}
for definitions of nonchiral VOAs.} algebra of the RCFT. So our analysis is not restricted to the standard
case of $\text{\ensuremath{\voa}}=\voa_{L}\times{\cal \bar{\voa}}_{L}$
, but also covers heterotic cases $\text{\ensuremath{\voa}}=\voa_{L}\times{\cal \bar{\voa}}_{R}$
and non-chiral algebra extensions. However, since our end goal is
to compare to ${\rm Vir}\times\overline{{\rm Vir}}$, one should always
keep in mind the standard case $\text{\ensuremath{\voa}}=\voa_{L}\times{\cal \bar{\voa}}_{L}$
as the main example throughout the text.

\paragraph{Hilbert space.}

The anyons form a natural basis in the Hilbert space of the torus.
We can prepare such a basis by inserting the corresponding anyon around
the nonshrinkable cycle of the solid torus. The characters of the
modules of the VOA $\voa_{\T}$ can be viewed as wavefunctions corresponding
to these states
\begin{equation}
\braket{\tau}a\equiv\chi_{a}(\tau),\label{eq: tau_holomorphic boundary}
\end{equation}
where $\tau$ is the modulus of the torus. The state $\ket{\tau}$
defines a gapless boundary condition similar to that used in the Chern-Simons/Wess-Zumino-Witten
(CS/WZW) correspondence. We will define $\ket{\tau}$ via (\ref{eq: tau_holomorphic boundary})
and call it a ``holomorphic'' boundary condition even though in
the nonchiral case it will depend on both $\tau$ and $\bar{\tau}$.

\begin{figure}
\begin{centering}
\includegraphics[scale=0.8]{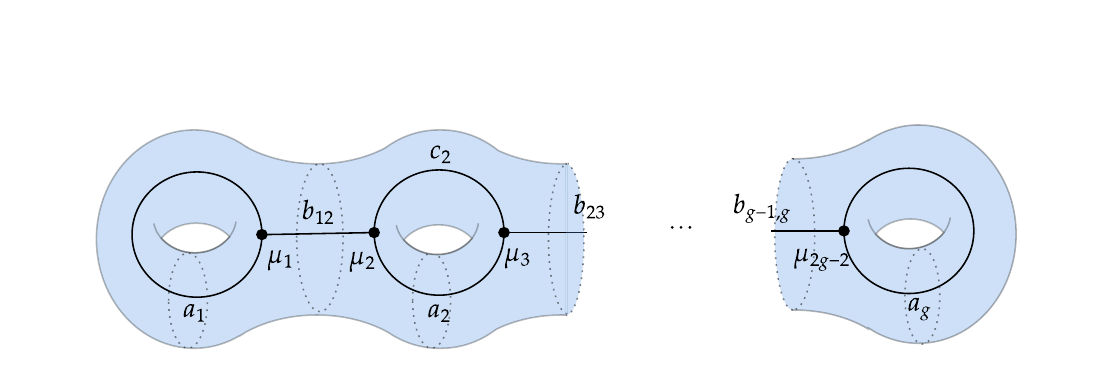}
\par\end{centering}
\caption{A choice of basis states in $\protect\H_{g}$ corresponding to inserting
the shown network of lines inside a handlebody $\protect\sS\Sigma_{g}$.}
\label{fig: anyon_basis_g}
\end{figure}

One can do something similar at higher genus. Let us denote the Hilbert
space of $\T$ on a Riemann surface $\Sigma_{g}$ as $\H_{g}$ where
$g$ denotes the genus of the Riemann surface. Any state in $\H_{g}$
can be prepared by appropriate line insertions inside a handlebody
$\sS\Sigma_{g}$ . A basis for $\H_{\Sigma_{g}}$ can be prepared
by insertion of a spine network of lines labeled by the simple anyons
(subject to fusion rules at the junctions) as shown in figure \ref{fig: anyon_basis_g}.

For convenience we will denote such basis states by the shorthand
\begin{equation}
\ket{\gvec a,\gvec b,\gvec c;\gvec{\mu}}\equiv\ket{a_{1}...a_{g};b_{12}...b_{g-1,g};c_{2}...c_{g-1};\mu_{1}...\mu_{2g-2}}.\label{eq: anyon basis states}
\end{equation}
Using this basis, one can prepare wavefunctions corresponding to genus
$g$ conformal blocks\footnote{These can be defined from sewing torus n-point blocks \cite{Moore:1988qv}.}
of $\voa_{\T}$ by doing the path integral with the corresponding
insertion and using gapless boundary conditions analogous to our genus
1 case. We will denote such wavefunctions as
\begin{equation}
\braket{\Omega}{\gvec a,\gvec b,\gvec c;\gvec{\mu}}\equiv\chi_{\gvec a,\gvec b,\gvec c;\gvec{\mu}}(\Omega),\label{eq: Omega_state}
\end{equation}
where $\Omega$ is the modulus matrix of $\Sigma_{g}$. We note again
that when $\T$ is nonchiral, e.g. $\T=\T_{L}\times\bar{\T}_{L}$,
the blocks (\ref{eq: Omega_state}) will be nonholomorphic in $\Omega$.

The dimension of $\H_{g}$ is given by
\begin{equation}
\dim\mathcal{H}_{g}=\sum_{\gvec a,\gvec b,\gvec c\in\T}N_{a_{1}a_{1}^{\vee}}^{b_{12}}N_{a_{2}c_{2}}^{b_{12}}N_{a_{2}c_{2}}^{b_{23}}\ldots N_{a_{g-1}c_{g-1}}^{b_{g-1,g}}N_{a_{g}a_{g}^{\vee}}^{b_{g-1,g}},
\end{equation}
which can be simplified, using the Verlinde formula to diagonalize
the fusion rules, into
\begin{equation}
\dim\H_{g}=\D_{\T}^{2g-2}\sum_{a}d_{a}^{2-2g},\label{eq: dim of Hg}
\end{equation}
where $d_{a}$ denotes the quantum dimension of the anyon $a$, and
$\D_{\T}=\sqrt{\sum_{a}d_{a}^{2}}$ is the total quantum dimension
of $\T$.

$\H_{g}$ furnishes a representation for the mapping class group (MCG)
of $\Sigma_{g}$. The generators can be explicitly constructed from
$F$,$R$ matrices and the modular data as shown for example in \cite{Turaev+2016,bloomquist2018_TopologicalQuantum}.

\paragraph{A note on basis normalization. }

In $\H_{g}$ we chose an orthogonal basis given by figure \ref{fig: anyon_basis_g}.
For convenience, we will work with a normalized basis, i.e. we will
normalize the vacuum insertion 
\begin{equation}
\braket{0_{g}}{0_{g}}=1,
\end{equation}
such that the vacuum conformal block from the CFT perspective is normalized
to unity (particularly on $S^{2}$). We want to illustrate the meaning
of this choice from the TQFT perspective. The vacuum norm has a geometric
interpretation as the partition function ${\bf Z}$ of the TQFT on
a connected sum of $S^{2}\times S^{1}$
\begin{equation}
\braket{0_{g}}{0_{g}}={\bf Z}_{\T}(\#^{g}S^{2}\times S^{1}).
\end{equation}
However, on a connected sum $M_{1}\#M_{2}$, the TQFT partition function
is given by \cite{witten1989_QuantumField,Turaev+2016}
\begin{equation}
{\bf Z}_{\T}(M_{1}\#M_{2})=\frac{{\bf Z}_{\T}(M_{1}){\bf Z}_{\T}(M_{2})}{{\bf Z}_{\T}(S^{3})},
\end{equation}
which leads to
\begin{equation}
\braket{0_{g}}{0_{g}}=\D_{\T}^{g-1},
\end{equation}
where we used ${\bf Z}_{\T}(S^{3})=\frac{1}{\D_{\T}}$ via modular
S-matrix, and ${\bf Z}_{\T}(S^{2}\times S^{1})={\rm Tr_{\H_{S^{2}}}({\rm I})=1}$. 

The choice of normalization affects the values assigned to connected
sums of closed manifolds versus disjoint manifolds. Choosing $\braket{0_{g}}{0_{g}}=1$
amounts to treating the two cases in the same way, while choosing
$\braket{0_{g}}{0_{g}}=\D_{\T}^{g-1}$ preserves the topological nature
of arbitrarily adding connected sums of $S^{3}$ without changing
the assigned value of the manifold. The latter normalization is more
natural when we do the sum over all manifolds.

\subsection{Algebra objects and anyon condensation}

We want to construct states in $\H_{g}$ that correspond to CFT partition
functions when projected onto $\bra{\Omega}$. In other words, this
would be a linear combination of conformal blocks that are MCG invariant
and satisfy certain physicality constraints (e.g. unique vacuum, positivity
of spectrum and consistency with sewing/factorization). To get a
true CFT, we should be able to construct such a state for arbitrary
genus. Such a construction can be achieved by anyon condensation (or
generalized gauging), specifically Lagrangian anyon condensation which
reduces the TQFT to a trivial TQFT with the boundary theory being
the CFT. The condensation amounts to inserting a specific network
of line defects in the bulk manifold and hence the condensed/gauged
theory can be embedded in the Hilbert space of the parent TQFT. Below
we briefly describe the algebraic construction of anyon condensation.
We refer the reader to \cite{kong2014_anyon_condensation} for further
mathematical details; see also \cite{benini2023_FactorizationGlobal}
for a review of anyon condensation in the context of holography.

\paragraph{Condensable algebra.}

We want to construct an object in $\T$ that acts as the vacuum object
in the condensed phase. This means we want an object that can freely
fuse and braid with itself into itself in an arbitrary way. This can
be reduced to the following diagrammatic conditions: 

Fusion and splitting  \begin{equation}
\label{eq:m_fusion}
    \raisebox{-0.8cm}{\begin{tikzpicture}[scale=1.5,x=0.75pt,y=0.75pt,yscale=-1,xscale=1]

\draw  [color={rgb, 255:red, 0; green, 0; blue, 0 }  ,draw opacity=1 ][fill={rgb, 255:red, 0; green, 0; blue, 0 }  ,fill opacity=1 ] (303.02,82.77) .. controls (303.02,82.07) and (303.6,81.5) .. (304.32,81.5) .. controls (305.04,81.5) and (305.62,82.07) .. (305.62,82.77) .. controls (305.62,83.47) and (305.04,84.03) .. (304.32,84.03) .. controls (303.6,84.03) and (303.02,83.47) .. (303.02,82.77) -- cycle ;
\draw [color={rgb, 255:red, 177; green, 0; blue, 5 }  ,draw opacity=1 ]   (284.01,109.11) -- (304.32,82.77) ;
\draw [color={rgb, 255:red, 177; green, 0; blue, 5 }  ,draw opacity=1 ]   (324.64,109.11) -- (304.32,82.77) ;
\draw [color={rgb, 255:red, 177; green, 0; blue, 5 }  ,draw opacity=1 ]   (304.32,82.77) -- (304.32,56.42) ;
\draw  [color={rgb, 255:red, 247; green, 7; blue, 7 }  ,draw opacity=1 ][fill={rgb, 255:red, 247; green, 7; blue, 7 }  ,fill opacity=1 ] (303.02,82.77) .. controls (303.02,82.07) and (303.6,81.5) .. (304.32,81.5) .. controls (305.04,81.5) and (305.62,82.07) .. (305.62,82.77) .. controls (305.62,83.47) and (305.04,84.03) .. (304.32,84.03) .. controls (303.6,84.03) and (303.02,83.47) .. (303.02,82.77) -- cycle ;
\draw  [color={rgb, 255:red, 0; green, 0; blue, 0 }  ,draw opacity=1 ][fill={rgb, 255:red, 0; green, 0; blue, 0 }  ,fill opacity=1 ] (411.37,82.76) .. controls (411.38,83.46) and (410.79,84.03) .. (410.07,84.03) .. controls (409.35,84.03) and (408.77,83.47) .. (408.77,82.77) .. controls (408.77,82.07) and (409.35,81.5) .. (410.07,81.5) .. controls (410.79,81.5) and (411.37,82.06) .. (411.37,82.76) -- cycle ;
\draw [color={rgb, 255:red, 177; green, 0; blue, 5 }  ,draw opacity=1 ]   (430.34,56.39) -- (410.07,82.77) ;
\draw [color={rgb, 255:red, 177; green, 0; blue, 5 }  ,draw opacity=1 ]   (389.71,56.46) -- (410.07,82.77) ;
\draw [color={rgb, 255:red, 177; green, 0; blue, 5 }  ,draw opacity=1 ]   (410.07,82.77) -- (410.12,109.11) ;
\draw  [color={rgb, 255:red, 247; green, 7; blue, 7 }  ,draw opacity=1 ][fill={rgb, 255:red, 247; green, 7; blue, 7 }  ,fill opacity=1 ] (411.37,82.76) .. controls (411.38,83.46) and (410.79,84.03) .. (410.07,84.03) .. controls (409.35,84.03) and (408.77,83.47) .. (408.77,82.77) .. controls (408.77,82.07) and (409.35,81.5) .. (410.07,81.5) .. controls (410.79,81.5) and (411.37,82.06) .. (411.37,82.76) -- cycle ;
\end{tikzpicture}}
\end{equation}

Bubble removal\begin{equation}
\label{eq:bubble_removal}
    \raisebox{-1.2cm}{\begin{tikzpicture}[scale=1.5,x=0.75pt,y=0.75pt,yscale=-1,xscale=1]

\draw [color={rgb, 255:red, 177; green, 0; blue, 5 }  ,draw opacity=1 ]   (310.62,57.23) -- (310.62,30.89) ;
\draw [color={rgb, 255:red, 177; green, 0; blue, 5 }  ,draw opacity=1 ]   (310.62,103.92) -- (310.62,77.58) ;
\draw  [color={rgb, 255:red, 177; green, 0; blue, 5 }  ,draw opacity=1 ] (300.44,67.41) .. controls (300.44,61.79) and (305,57.23) .. (310.62,57.23) .. controls (316.23,57.23) and (320.79,61.79) .. (320.79,67.41) .. controls (320.79,73.02) and (316.23,77.58) .. (310.62,77.58) .. controls (305,77.58) and (300.44,73.02) .. (300.44,67.41) -- cycle ;
\draw  [color={rgb, 255:red, 247; green, 7; blue, 7 }  ,draw opacity=1 ][fill={rgb, 255:red, 247; green, 7; blue, 7 }  ,fill opacity=1 ] (309.31,77.58) .. controls (309.31,76.88) and (309.9,76.31) .. (310.62,76.31) .. controls (311.34,76.31) and (311.92,76.88) .. (311.92,77.58) .. controls (311.92,78.28) and (311.34,78.84) .. (310.62,78.84) .. controls (309.9,78.84) and (309.31,78.28) .. (309.31,77.58) -- cycle ;
\draw  [color={rgb, 255:red, 247; green, 7; blue, 7 }  ,draw opacity=1 ][fill={rgb, 255:red, 247; green, 7; blue, 7 }  ,fill opacity=1 ] (309.31,57.23) .. controls (309.31,56.53) and (309.9,55.97) .. (310.62,55.97) .. controls (311.34,55.97) and (311.92,56.53) .. (311.92,57.23) .. controls (311.92,57.93) and (311.34,58.5) .. (310.62,58.5) .. controls (309.9,58.5) and (309.31,57.93) .. (309.31,57.23) -- cycle ;
\end{tikzpicture}} \,\, = \,\,\, \, \raisebox{-1.2cm}{\begin{tikzpicture}[scale=1.5,x=0.75pt,y=0.75pt,yscale=-1,xscale=1]

\draw [color={rgb, 255:red, 177; green, 0; blue, 5 }  ,draw opacity=1 ]   (340.46,104.06) -- (340.44,30.67) ;

\end{tikzpicture}}
\end{equation}

Associativity\begin{equation}
\label{eq:m_associativity}
    \raisebox{-1.2cm}{\begin{tikzpicture}[scale=1.5,x=0.75pt,y=0.75pt,yscale=-1,xscale=1]

\draw [color={rgb, 255:red, 177; green, 0; blue, 5 }  ,draw opacity=1 ]   (256.49,89.45) -- (280.33,109.99) ;
\draw [color={rgb, 255:red, 177; green, 0; blue, 5 }  ,draw opacity=1 ]   (230.33,109.99) -- (256.49,89.45) ;
\draw [color={rgb, 255:red, 177; green, 0; blue, 5 }  ,draw opacity=1 ]   (256.49,89.45) -- (276.5,72.31) ;
\draw  [color={rgb, 255:red, 247; green, 7; blue, 7 }  ,draw opacity=1 ][fill={rgb, 255:red, 247; green, 7; blue, 7 }  ,fill opacity=1 ] (256.48,88.14) .. controls (257.18,88.14) and (257.75,88.72) .. (257.76,89.44) .. controls (257.77,90.15) and (257.2,90.74) .. (256.5,90.75) .. controls (255.8,90.75) and (255.23,90.18) .. (255.23,89.46) .. controls (255.22,88.74) and (255.78,88.15) .. (256.48,88.14) -- cycle ;
\draw [color={rgb, 255:red, 177; green, 0; blue, 5 }  ,draw opacity=1 ]   (320.33,109.99) -- (275.54,73.14) ;
\draw [color={rgb, 255:red, 177; green, 0; blue, 5 }  ,draw opacity=1 ]   (301.58,52.43) -- (275.54,73.14) ;
\draw  [color={rgb, 255:red, 247; green, 7; blue, 7 }  ,draw opacity=1 ][fill={rgb, 255:red, 247; green, 7; blue, 7 }  ,fill opacity=1 ] (276.4,74.12) .. controls (275.87,74.58) and (275.06,74.51) .. (274.58,73.97) .. controls (274.11,73.42) and (274.16,72.61) .. (274.69,72.15) .. controls (275.22,71.69) and (276.03,71.76) .. (276.5,72.31) .. controls (276.97,72.85) and (276.92,73.66) .. (276.4,74.12) -- cycle ;

\end{tikzpicture}} \,\, = \,\,\, \, \raisebox{-1.2cm}{\begin{tikzpicture}[scale=1.5,x=0.75pt,y=0.75pt,yscale=-1,xscale=1]

\draw [color={rgb, 255:red, 177; green, 0; blue, 5 }  ,draw opacity=1 ]   (440.33,109.99) -- (416.11,88.94) ;
\draw [color={rgb, 255:red, 177; green, 0; blue, 5 }  ,draw opacity=1 ]   (372.35,52.04) -- (397.28,72.65) ;
\draw [color={rgb, 255:red, 177; green, 0; blue, 5 }  ,draw opacity=1 ]   (392.77,68.85) -- (416.11,88.94) ;
\draw [color={rgb, 255:red, 177; green, 0; blue, 5 }  ,draw opacity=1 ]   (350.33,109.99) -- (397.45,73.73) ;
\draw [color={rgb, 255:red, 177; green, 0; blue, 5 }  ,draw opacity=1 ]   (390.47,110.13) -- (416.11,88.94) ;
\draw  [color={rgb, 255:red, 247; green, 7; blue, 7 }  ,draw opacity=1 ][fill={rgb, 255:red, 247; green, 7; blue, 7 }  ,fill opacity=1 ] (415.24,87.97) .. controls (415.76,87.5) and (416.57,87.56) .. (417.05,88.09) .. controls (417.53,88.62) and (417.5,89.44) .. (416.98,89.91) .. controls (416.46,90.37) and (415.65,90.32) .. (415.17,89.79) .. controls (414.69,89.25) and (414.72,88.44) .. (415.24,87.97) -- cycle ;
\draw  [color={rgb, 255:red, 247; green, 7; blue, 7 }  ,draw opacity=1 ][fill={rgb, 255:red, 247; green, 7; blue, 7 }  ,fill opacity=1 ] (397.8,74.53) .. controls (397.1,74.55) and (396.52,73.99) .. (396.5,73.27) .. controls (396.48,72.55) and (397.03,71.95) .. (397.73,71.93) .. controls (398.43,71.91) and (399.01,72.48) .. (399.03,73.2) .. controls (399.05,73.92) and (398.5,74.52) .. (397.8,74.53) -- cycle ;

\end{tikzpicture}}
\end{equation}

Invariance under $F$-moves for fusion and splitting (crossing symmetry)\begin{equation}
\label{eq:Frobenius}
    \raisebox{-1.4cm}{\begin{tikzpicture}[scale=1.5,x=0.75pt,y=0.75pt,yscale=-1,xscale=1]

\draw [color={rgb, 255:red, 177; green, 0; blue, 5 }  ,draw opacity=1 ]   (270.18,41.02) -- (270.18,91.02) ;
\draw [color={rgb, 255:red, 177; green, 0; blue, 5 }  ,draw opacity=1 ]   (270.11,121.02) -- (270.18,91.02) ;
\draw [color={rgb, 255:red, 177; green, 0; blue, 5 }  ,draw opacity=1 ]   (270.18,91.02) -- (300.18,71.02) ;
\draw  [color={rgb, 255:red, 247; green, 7; blue, 7 }  ,draw opacity=1 ][fill={rgb, 255:red, 247; green, 7; blue, 7 }  ,fill opacity=1 ] (270.17,89.72) .. controls (270.87,89.71) and (271.44,90.29) .. (271.45,91.01) .. controls (271.45,91.73) and (270.89,92.32) .. (270.19,92.32) .. controls (269.49,92.33) and (268.92,91.75) .. (268.91,91.03) .. controls (268.91,90.31) and (269.47,89.73) .. (270.17,89.72) -- cycle ;
\draw [color={rgb, 255:red, 177; green, 0; blue, 5 }  ,draw opacity=1 ]   (300.11,120.66) -- (300.11,80.66) ;
\draw [color={rgb, 255:red, 177; green, 0; blue, 5 }  ,draw opacity=1 ]   (300.18,41.02) -- (300.11,80.66) ;
\draw  [color={rgb, 255:red, 247; green, 7; blue, 7 }  ,draw opacity=1 ][fill={rgb, 255:red, 247; green, 7; blue, 7 }  ,fill opacity=1 ] (301.03,72.01) .. controls (300.51,72.46) and (299.7,72.4) .. (299.22,71.85) .. controls (298.75,71.31) and (298.8,70.5) .. (299.33,70.04) .. controls (299.86,69.58) and (300.67,69.65) .. (301.14,70.19) .. controls (301.61,70.74) and (301.56,71.55) .. (301.03,72.01) -- cycle ;

\end{tikzpicture}} \,\,\, \, \, \, = \,\, \raisebox{-1.4cm}{\begin{tikzpicture}[scale=1.5,x=0.75pt,y=0.75pt,yscale=-1,xscale=1]

\draw  [color={rgb, 255:red, 0; green, 0; blue, 0 }  ,draw opacity=1 ][fill={rgb, 255:red, 0; green, 0; blue, 0 }  ,fill opacity=1 ] (359.37,94.17) .. controls (359.37,93.47) and (359.96,92.9) .. (360.68,92.9) .. controls (361.4,92.9) and (361.98,93.47) .. (361.98,94.17) .. controls (361.98,94.87) and (361.4,95.43) .. (360.68,95.43) .. controls (359.96,95.43) and (359.37,94.87) .. (359.37,94.17) -- cycle ;
\draw [color={rgb, 255:red, 177; green, 0; blue, 5 }  ,draw opacity=1 ]   (340.36,120.51) -- (360.68,94.17) ;
\draw [color={rgb, 255:red, 177; green, 0; blue, 5 }  ,draw opacity=1 ]   (380.99,120.51) -- (360.68,94.17) ;
\draw [color={rgb, 255:red, 177; green, 0; blue, 5 }  ,draw opacity=1 ]   (360.68,94.17) -- (360.68,67.82) ;
\draw  [color={rgb, 255:red, 247; green, 7; blue, 7 }  ,draw opacity=1 ][fill={rgb, 255:red, 247; green, 7; blue, 7 }  ,fill opacity=1 ] (359.37,94.17) .. controls (359.37,93.47) and (359.96,92.9) .. (360.68,92.9) .. controls (361.4,92.9) and (361.98,93.47) .. (361.98,94.17) .. controls (361.98,94.87) and (361.4,95.43) .. (360.68,95.43) .. controls (359.96,95.43) and (359.37,94.87) .. (359.37,94.17) -- cycle ;
\draw  [color={rgb, 255:red, 0; green, 0; blue, 0 }  ,draw opacity=1 ][fill={rgb, 255:red, 0; green, 0; blue, 0 }  ,fill opacity=1 ] (361.96,67.49) .. controls (361.97,68.19) and (361.39,68.76) .. (360.67,68.76) .. controls (359.95,68.77) and (359.36,68.2) .. (359.36,67.5) .. controls (359.35,66.8) and (359.93,66.23) .. (360.65,66.23) .. controls (361.37,66.22) and (361.96,66.79) .. (361.96,67.49) -- cycle ;
\draw [color={rgb, 255:red, 177; green, 0; blue, 5 }  ,draw opacity=1 ]   (380.81,41.02) -- (360.66,67.49) ;
\draw [color={rgb, 255:red, 177; green, 0; blue, 5 }  ,draw opacity=1 ]   (340.18,41.28) -- (360.66,67.49) ;
\draw  [color={rgb, 255:red, 247; green, 7; blue, 7 }  ,draw opacity=1 ][fill={rgb, 255:red, 247; green, 7; blue, 7 }  ,fill opacity=1 ] (361.96,67.49) .. controls (361.97,68.19) and (361.39,68.76) .. (360.67,68.76) .. controls (359.95,68.77) and (359.36,68.2) .. (359.36,67.5) .. controls (359.35,66.8) and (359.93,66.23) .. (360.65,66.23) .. controls (361.37,66.22) and (361.96,66.79) .. (361.96,67.49) -- cycle ;

\end{tikzpicture}}
\end{equation}

Invariance under braiding (commutativity)\begin{equation}
\label{eq:m_commutativity}
    \raisebox{-0.9cm}{\begin{tikzpicture}[scale=1.5,x=0.75pt,y=0.75pt,yscale=-1,xscale=1]

\draw [color={rgb, 255:red, 177; green, 0; blue, 5 }  ,draw opacity=1 ]   (289.53,82.48) -- (289.53,54.82) ;
\draw [color={rgb, 255:red, 177; green, 0; blue, 5 }  ,draw opacity=1 ]   (286.46,96.57) .. controls (284.91,93.84) and (279.03,84.54) .. (289.53,82.48) ;
\draw [color={rgb, 255:red, 177; green, 0; blue, 5 }  ,draw opacity=1 ]   (269.21,107.51) .. controls (289.26,103.56) and (305.73,87.36) .. (289.53,82.48) ;
\draw  [color={rgb, 255:red, 247; green, 7; blue, 7 }  ,draw opacity=1 ][fill={rgb, 255:red, 247; green, 7; blue, 7 }  ,fill opacity=1 ] (288.23,82.48) .. controls (288.23,81.78) and (288.81,81.21) .. (289.53,81.21) .. controls (290.25,81.21) and (290.83,81.78) .. (290.83,82.48) .. controls (290.83,83.18) and (290.25,83.75) .. (289.53,83.75) .. controls (288.81,83.75) and (288.23,83.18) .. (288.23,82.48) -- cycle ;
\draw [color={rgb, 255:red, 177; green, 0; blue, 5 }  ,draw opacity=1 ]   (290.13,101.1) .. controls (291.03,102.78) and (304.72,107.89) .. (309.85,107.51) ;

\end{tikzpicture}} \,\, = \,\,\, \, \raisebox{-0.9cm}{\begin{tikzpicture}[scale=1.5,x=0.75pt,y=0.75pt,yscale=-1,xscale=1]

\draw  [color={rgb, 255:red, 0; green, 0; blue, 0 }  ,draw opacity=1 ][fill={rgb, 255:red, 0; green, 0; blue, 0 }  ,fill opacity=1 ] (363.94,81.83) .. controls (363.94,81.13) and (364.53,80.56) .. (365.25,80.56) .. controls (365.97,80.56) and (366.55,81.13) .. (366.55,81.83) .. controls (366.55,82.53) and (365.97,83.1) .. (365.25,83.1) .. controls (364.53,83.1) and (363.94,82.53) .. (363.94,81.83) -- cycle ;
\draw [color={rgb, 255:red, 177; green, 0; blue, 5 }  ,draw opacity=1 ]   (344.93,108.18) -- (365.25,81.83) ;
\draw [color={rgb, 255:red, 177; green, 0; blue, 5 }  ,draw opacity=1 ]   (385.56,108.18) -- (365.25,81.83) ;
\draw [color={rgb, 255:red, 177; green, 0; blue, 5 }  ,draw opacity=1 ]   (365.25,81.83) -- (365.25,55.49) ;
\draw  [color={rgb, 255:red, 247; green, 7; blue, 7 }  ,draw opacity=1 ][fill={rgb, 255:red, 247; green, 7; blue, 7 }  ,fill opacity=1 ] (363.94,81.83) .. controls (363.94,81.13) and (364.53,80.56) .. (365.25,80.56) .. controls (365.97,80.56) and (366.55,81.13) .. (366.55,81.83) .. controls (366.55,82.53) and (365.97,83.1) .. (365.25,83.1) .. controls (364.53,83.1) and (363.94,82.53) .. (363.94,81.83) -- cycle ;

\end{tikzpicture}}
\end{equation}

In a unitary MTC, there is no simple object that satisfies these relations
under fusion except the trivial/vacuum anyon; however, if we consider
non-simple objects we will have more freedom to consider junctions
that could satisfy these conditions. Hence, we consider a non-simple
object $\A$ with some junction morphisms $m\in\Hom(\A\otimes\A,\A)$
and $m^{\vee}\in\Hom(\A,\A\otimes\A)$ that allow us to achieve such
diagrammatics. We also require that $\A$ can be mapped to the vacuum
of the parent phase when we interface the two phases, so we must have
a unit morphism $\eta\in\Hom(0,\A)$ and a co-unit $\eta^{\vee}\in\Hom(\A,0)$,
shown graphically as \begin{equation}
\label{eq:unit_morphism}
    \raisebox{-0.8cm}{\begin{tikzpicture}[scale=1.5,x=0.75pt,y=0.75pt,yscale=-1,xscale=1]

\draw [color={rgb, 255:red, 177; green, 0; blue, 5 }  ,draw opacity=1 ]   (310.1,99.31) -- (310.1,72.97) ;
\draw  [color={rgb, 255:red, 247; green, 7; blue, 7 }  ,draw opacity=1 ][fill={rgb, 255:red, 255; green, 255; blue, 255 }  ,fill opacity=1 ] (308.8,99.31) .. controls (308.8,98.61) and (309.38,98.04) .. (310.1,98.04) .. controls (310.82,98.04) and (311.41,98.61) .. (311.41,99.31) .. controls (311.41,100.01) and (310.82,100.58) .. (310.1,100.58) .. controls (309.38,100.58) and (308.8,100.01) .. (308.8,99.31) -- cycle ;
\draw  [draw opacity=0][fill={rgb, 255:red, 255; green, 255; blue, 255 }  ,fill opacity=1 ] (308.8,99.31) .. controls (308.8,98.61) and (309.38,98.04) .. (310.1,98.04) .. controls (310.82,98.04) and (311.41,98.61) .. (311.41,99.31) .. controls (311.41,100.01) and (310.82,100.58) .. (310.1,100.58) .. controls (309.38,100.58) and (308.8,100.01) .. (308.8,99.31) -- cycle ;

\draw [color={rgb, 255:red, 177; green, 0; blue, 5 }  ,draw opacity=1 ]   (360.1,74.23) -- (360.29,100.58) ;
\draw  [color={rgb, 255:red, 247; green, 7; blue, 7 }  ,draw opacity=1 ][fill={rgb, 255:red, 255; green, 255; blue, 255 }  ,fill opacity=1 ] (361.41,74.22) .. controls (361.41,74.92) and (360.83,75.5) .. (360.11,75.5) .. controls (359.39,75.51) and (358.81,74.94) .. (358.8,74.24) .. controls (358.8,73.54) and (359.38,72.97) .. (360.09,72.97) .. controls (360.81,72.96) and (361.4,73.52) .. (361.41,74.22) -- cycle ;
\draw  [draw opacity=0][fill={rgb, 255:red, 255; green, 255; blue, 255 }  ,fill opacity=1 ] (361.41,74.22) .. controls (361.41,74.92) and (360.83,75.5) .. (360.11,75.5) .. controls (359.39,75.51) and (358.81,74.94) .. (358.8,74.24) .. controls (358.8,73.54) and (359.38,72.97) .. (360.09,72.97) .. controls (360.81,72.96) and (361.4,73.52) .. (361.41,74.22) -- cycle ;

\end{tikzpicture}}
\end{equation} The unit and co-unit should be compatible with $m$ and $m^{\vee}$,
and hence we should also have the following diagrams \begin{equation}
\label{eq:unit_compatability}
    \raisebox{-0.9cm}{\begin{tikzpicture}[scale=1.5,x=0.75pt,y=0.75pt,yscale=-1,xscale=1]

\draw [color={rgb, 255:red, 177; green, 0; blue, 5 }  ,draw opacity=1 ]   (278.77,98.88) -- (291.26,82.68) ;
\draw  [color={rgb, 255:red, 247; green, 7; blue, 7 }  ,draw opacity=1 ][fill={rgb, 255:red, 255; green, 255; blue, 255 }  ,fill opacity=1 ] (277.61,98.75) .. controls (277.61,98.05) and (278.19,97.48) .. (278.91,97.48) .. controls (279.63,97.48) and (280.21,98.05) .. (280.21,98.75) .. controls (280.21,99.45) and (279.63,100.01) .. (278.91,100.01) .. controls (278.19,100.01) and (277.61,99.45) .. (277.61,98.75) -- cycle ;
\draw  [draw opacity=0][fill={rgb, 255:red, 255; green, 255; blue, 255 }  ,fill opacity=1 ] (277.61,98.75) .. controls (277.61,98.05) and (278.19,97.48) .. (278.91,97.48) .. controls (279.63,97.48) and (280.21,98.05) .. (280.21,98.75) .. controls (280.21,99.45) and (279.63,100.01) .. (278.91,100.01) .. controls (278.19,100.01) and (277.61,99.45) .. (277.61,98.75) -- cycle ;

\draw  [color={rgb, 255:red, 0; green, 0; blue, 0 }  ,draw opacity=1 ][fill={rgb, 255:red, 0; green, 0; blue, 0 }  ,fill opacity=1 ] (289.96,82.68) .. controls (289.96,81.98) and (290.54,81.41) .. (291.26,81.41) .. controls (291.98,81.41) and (292.56,81.98) .. (292.56,82.68) .. controls (292.56,83.38) and (291.98,83.94) .. (291.26,83.94) .. controls (290.54,83.94) and (289.96,83.38) .. (289.96,82.68) -- cycle ;
\draw [color={rgb, 255:red, 177; green, 0; blue, 5 }  ,draw opacity=1 ]   (311.58,109.02) -- (291.26,82.68) ;
\draw [color={rgb, 255:red, 177; green, 0; blue, 5 }  ,draw opacity=1 ]   (291.26,82.68) -- (291.26,56.33) ;
\draw  [color={rgb, 255:red, 247; green, 7; blue, 7 }  ,draw opacity=1 ][fill={rgb, 255:red, 247; green, 7; blue, 7 }  ,fill opacity=1 ] (289.96,82.68) .. controls (289.96,81.98) and (290.54,81.41) .. (291.26,81.41) .. controls (291.98,81.41) and (292.56,81.98) .. (292.56,82.68) .. controls (292.56,83.38) and (291.98,83.94) .. (291.26,83.94) .. controls (290.54,83.94) and (289.96,83.38) .. (289.96,82.68) -- cycle ;

\end{tikzpicture}} \,\, = \,\,\, \,\,\,\, \raisebox{-0.9cm}{\begin{tikzpicture}[scale=1.5,x=0.75pt,y=0.75pt,yscale=-1,xscale=1]

\draw [color={rgb, 255:red, 177; green, 0; blue, 5 }  ,draw opacity=1 ]   (350.58,108.7) -- (350.58,56.63) ;

\end{tikzpicture}} \,\,\,\,\,\,\, = \,\, \raisebox{-0.9cm}{\begin{tikzpicture}[scale=1.5,x=0.75pt,y=0.75pt,yscale=-1,xscale=1]

\draw  [color={rgb, 255:red, 0; green, 0; blue, 0 }  ,draw opacity=1 ][fill={rgb, 255:red, 0; green, 0; blue, 0 }  ,fill opacity=1 ] (408.96,82.35) .. controls (408.96,81.65) and (409.54,81.08) .. (410.26,81.08) .. controls (410.98,81.08) and (411.56,81.65) .. (411.56,82.35) .. controls (411.56,83.05) and (410.98,83.62) .. (410.26,83.62) .. controls (409.54,83.62) and (408.96,83.05) .. (408.96,82.35) -- cycle ;
\draw [color={rgb, 255:red, 177; green, 0; blue, 5 }  ,draw opacity=1 ]   (389.95,108.7) -- (410.26,82.35) ;
\draw [color={rgb, 255:red, 177; green, 0; blue, 5 }  ,draw opacity=1 ]   (423.07,98.96) -- (410.26,82.35) ;
\draw [color={rgb, 255:red, 177; green, 0; blue, 5 }  ,draw opacity=1 ]   (410.26,82.35) -- (410.26,56.01) ;
\draw  [color={rgb, 255:red, 247; green, 7; blue, 7 }  ,draw opacity=1 ][fill={rgb, 255:red, 247; green, 7; blue, 7 }  ,fill opacity=1 ] (408.96,82.35) .. controls (408.96,81.65) and (409.54,81.08) .. (410.26,81.08) .. controls (410.98,81.08) and (411.56,81.65) .. (411.56,82.35) .. controls (411.56,83.05) and (410.98,83.62) .. (410.26,83.62) .. controls (409.54,83.62) and (408.96,83.05) .. (408.96,82.35) -- cycle ;
\draw  [color={rgb, 255:red, 247; green, 7; blue, 7 }  ,draw opacity=1 ][fill={rgb, 255:red, 255; green, 255; blue, 255 }  ,fill opacity=1 ] (422.04,99.4) .. controls (422.04,98.7) and (422.62,98.14) .. (423.34,98.14) .. controls (424.06,98.14) and (424.64,98.7) .. (424.64,99.4) .. controls (424.64,100.1) and (424.06,100.67) .. (423.34,100.67) .. controls (422.62,100.67) and (422.04,100.1) .. (422.04,99.4) -- cycle ;
\draw  [draw opacity=0][fill={rgb, 255:red, 255; green, 255; blue, 255 }  ,fill opacity=1 ] (422.04,99.4) .. controls (422.04,98.7) and (422.62,98.14) .. (423.34,98.14) .. controls (424.06,98.14) and (424.64,98.7) .. (424.64,99.4) .. controls (424.64,100.1) and (424.06,100.67) .. (423.34,100.67) .. controls (422.62,100.67) and (422.04,100.1) .. (422.04,99.4) -- cycle ;

\end{tikzpicture}}
\end{equation} Applying $\eta$ and $\eta^{\vee}$ to the bubble removal (\ref{eq:bubble_removal})
leads to the normalization \begin{equation}
\label{eq:algebra_normalization}
    \raisebox{-0.45cm}{\begin{tikzpicture}[scale=1.5,x=0.75pt,y=0.75pt,yscale=-1,xscale=1]

\draw [color={rgb, 255:red, 177; green, 0; blue, 5 }  ,draw opacity=1 ]   (330.02,97.91) -- (329.88,69.29) ;
\draw  [color={rgb, 255:red, 247; green, 7; blue, 7 }  ,draw opacity=1 ][fill={rgb, 255:red, 255; green, 255; blue, 255 }  ,fill opacity=1 ] (328.72,97.91) .. controls (328.72,97.21) and (329.3,96.64) .. (330.02,96.64) .. controls (330.74,96.64) and (331.33,97.21) .. (331.33,97.91) .. controls (331.33,98.61) and (330.74,99.18) .. (330.02,99.18) .. controls (329.3,99.18) and (328.72,98.61) .. (328.72,97.91) -- cycle ;
\draw  [draw opacity=0][fill={rgb, 255:red, 255; green, 255; blue, 255 }  ,fill opacity=1 ] (328.72,97.91) .. controls (328.72,97.21) and (329.3,96.64) .. (330.02,96.64) .. controls (330.74,96.64) and (331.33,97.21) .. (331.33,97.91) .. controls (331.33,98.61) and (330.74,99.18) .. (330.02,99.18) .. controls (329.3,99.18) and (328.72,98.61) .. (328.72,97.91) -- cycle ;

\draw  [color={rgb, 255:red, 247; green, 7; blue, 7 }  ,draw opacity=1 ][fill={rgb, 255:red, 255; green, 255; blue, 255 }  ,fill opacity=1 ] (331.27,70.45) .. controls (331.27,71.15) and (330.69,71.72) .. (329.97,71.72) .. controls (329.25,71.73) and (328.67,71.17) .. (328.66,70.47) .. controls (328.66,69.77) and (329.23,69.2) .. (329.95,69.19) .. controls (330.67,69.19) and (331.26,69.75) .. (331.27,70.45) -- cycle ;
\draw  [draw opacity=0][fill={rgb, 255:red, 255; green, 255; blue, 255 }  ,fill opacity=1 ] (331.27,70.45) .. controls (331.27,71.15) and (330.69,71.72) .. (329.97,71.72) .. controls (329.25,71.73) and (328.67,71.17) .. (328.66,70.47) .. controls (328.66,69.77) and (329.23,69.2) .. (329.95,69.19) .. controls (330.67,69.19) and (331.26,69.75) .. (331.27,70.45) -- cycle ;

\end{tikzpicture}} \,\, = \,\, \dim \cal{A},
\end{equation}where $\dim\A$ is the quantum dimension of the object $\A$. If we
write the object $\A$ in terms of simple anyons as
\begin{equation}
\A=\bigoplus_{a}n_{a}a,\qquad n_{a}\in\mathbb{Z}_{\geq0},
\end{equation}
then 
\begin{equation}
\dim\A=\sum_{a}n_{a}d_{a}.
\end{equation}
Finally, we require the uniqueness of the vacuum anyon when we condense,
so we require $n_{0}=1$.

The above diagrams define what is known as a commutative connected
special symmetric Frobenius algebra, and hence $\A$ is called an
algebra object.

Given that $\A$ is a direct sum of simple anyons, we have morphisms
from these simple objects denoted as\begin{equation}
\label{eq:morphism_to_simples}
    \raisebox{-0.8cm}{\begin{tikzpicture}[scale=1.5,x=0.75pt,y=0.75pt,yscale=-1,xscale=1]

\draw [color={rgb, 255:red, 177; green, 0; blue, 5 }  ,draw opacity=1 ]   (330.03,110.06) -- (330.03,60.06) ;
\draw    (330.12,90.41) -- (330.03,110.06) ;
\draw  [color={rgb, 255:red, 155; green, 155; blue, 155 }  ,draw opacity=1 ][fill={rgb, 255:red, 155; green, 155; blue, 155 }  ,fill opacity=1 ] (330.11,88.11) .. controls (331.7,88.13) and (332.98,89.16) .. (332.99,90.43) -- (327.17,90.37) .. controls (327.21,89.1) and (328.51,88.09) .. (330.11,88.11) -- cycle ;

\draw (334.13,86.82) node [anchor=north west][inner sep=0.75pt]  [font=\small]  {$\alpha $};
\draw (326.68,110.87) node [anchor=north west][inner sep=0.75pt]  [font=\small]  {$a$};

\end{tikzpicture}} \qquad \alpha \in \Hom (a,\A). 
\end{equation}

In these components, the product morphism is given by\begin{equation}
\label{eq:m_components}
    \raisebox{-1.2cm}{\begin{tikzpicture}[scale=1.5,x=0.75pt,y=0.75pt,yscale=-1,xscale=1]

\draw  [color={rgb, 255:red, 0; green, 0; blue, 0 }  ,draw opacity=1 ][fill={rgb, 255:red, 0; green, 0; blue, 0 }  ,fill opacity=1 ] (279.11,100.52) .. controls (279.11,99.82) and (279.7,99.26) .. (280.42,99.26) .. controls (281.13,99.26) and (281.72,99.82) .. (281.72,100.52) .. controls (281.72,101.22) and (281.13,101.79) .. (280.42,101.79) .. controls (279.7,101.79) and (279.11,101.22) .. (279.11,100.52) -- cycle ;
\draw [color={rgb, 255:red, 177; green, 0; blue, 5 }  ,draw opacity=1 ]   (258.61,125.51) -- (280.42,100.52) ;
\draw [color={rgb, 255:red, 177; green, 0; blue, 5 }  ,draw opacity=1 ]   (301.75,125.66) -- (280.42,100.52) ;
\draw [color={rgb, 255:red, 177; green, 0; blue, 5 }  ,draw opacity=1 ]   (280.42,100.52) -- (280.42,74.18) ;
\draw  [color={rgb, 255:red, 247; green, 7; blue, 7 }  ,draw opacity=1 ][fill={rgb, 255:red, 247; green, 7; blue, 7 }  ,fill opacity=1 ] (279.11,100.52) .. controls (279.11,99.82) and (279.7,99.26) .. (280.42,99.26) .. controls (281.13,99.26) and (281.72,99.82) .. (281.72,100.52) .. controls (281.72,101.22) and (281.13,101.79) .. (280.42,101.79) .. controls (279.7,101.79) and (279.11,101.22) .. (279.11,100.52) -- cycle ;
\draw    (270.11,112.39) -- (258.61,125.51) ;
\draw    (290.66,112.63) -- (301.72,125.62) -- (301.75,125.66) ;
\draw  [color={rgb, 255:red, 155; green, 155; blue, 155 }  ,draw opacity=1 ][fill={rgb, 255:red, 155; green, 155; blue, 155 }  ,fill opacity=1 ] (271.48,110.58) .. controls (272.74,111.55) and (273.13,113.16) .. (272.37,114.17) -- (267.78,110.6) .. controls (268.58,109.61) and (270.22,109.6) .. (271.48,110.58) -- cycle ;
\draw  [color={rgb, 255:red, 155; green, 155; blue, 155 }  ,draw opacity=1 ][fill={rgb, 255:red, 155; green, 155; blue, 155 }  ,fill opacity=1 ] (289.08,110.78) .. controls (290.25,109.7) and (291.89,109.57) .. (292.77,110.49) -- (288.51,114.44) .. controls (287.66,113.49) and (287.91,111.86) .. (289.08,110.78) -- cycle ;
\draw    (280.42,74.18) -- (280.31,89.92) ;
\draw  [color={rgb, 255:red, 155; green, 155; blue, 155 }  ,draw opacity=1 ][fill={rgb, 255:red, 155; green, 155; blue, 155 }  ,fill opacity=1 ] (280.42,87.35) .. controls (282.01,87.38) and (283.29,88.42) .. (283.29,89.7) -- (277.47,89.59) .. controls (277.52,88.32) and (278.83,87.32) .. (280.42,87.35) -- cycle ;

\draw (261.5,103.93) node [anchor=north west][inner sep=0.75pt]  [font=\small]  {$\alpha $};
\draw (293.38,105.06) node [anchor=north west][inner sep=0.75pt]  [font=\small]  {$\beta $};
\draw (284.63,85.56) node [anchor=north west][inner sep=0.75pt]  [font=\small]  {$\gamma $};
\draw (254.25,126.18) node [anchor=north west][inner sep=0.75pt]  [font=\small]  {$a$};
\draw (300.63,126.73) node [anchor=north west][inner sep=0.75pt]  [font=\small]  {$b$};
\draw (278,67.63) node [anchor=north west][inner sep=0.75pt]  [font=\small]  {$c$};

\end{tikzpicture}} \,\, = \,\,\, \, \sum_{\mu} \, m^{c \gamma ; \mu}_{a \alpha,b \beta} \raisebox{-1.2cm}{\begin{tikzpicture}[scale=1.5,x=0.75pt,y=0.75pt,yscale=-1,xscale=1]

\draw  [color={rgb, 255:red, 0; green, 0; blue, 0 }  ,draw opacity=1 ][fill={rgb, 255:red, 0; green, 0; blue, 0 }  ,fill opacity=1 ] (355.86,100.66) .. controls (355.86,99.96) and (356.45,99.4) .. (357.17,99.4) .. controls (357.88,99.4) and (358.47,99.96) .. (358.47,100.66) .. controls (358.47,101.36) and (357.88,101.93) .. (357.17,101.93) .. controls (356.45,101.93) and (355.86,101.36) .. (355.86,100.66) -- cycle ;
\draw [color={rgb, 255:red, 0; green, 0; blue, 0 }  ,draw opacity=1 ]   (335.36,125.65) -- (357.17,100.66) ;
\draw [color={rgb, 255:red, 0; green, 0; blue, 0 }  ,draw opacity=1 ]   (378.5,125.8) -- (357.17,100.66) ;
\draw [color={rgb, 255:red, 0; green, 0; blue, 0 }  ,draw opacity=1 ]   (357.17,100.66) -- (357.17,74.32) ;
\draw  [color={rgb, 255:red, 0; green, 0; blue, 0 }  ,draw opacity=1 ][fill={rgb, 255:red, 0; green, 0; blue, 0 }  ,fill opacity=1 ] (355.86,100.66) .. controls (355.86,99.96) and (356.45,99.4) .. (357.17,99.4) .. controls (357.88,99.4) and (358.47,99.96) .. (358.47,100.66) .. controls (358.47,101.36) and (357.88,101.93) .. (357.17,101.93) .. controls (356.45,101.93) and (355.86,101.36) .. (355.86,100.66) -- cycle ;
\draw    (346.86,112.53) -- (335.36,125.65) ;
\draw    (367.41,112.77) -- (378.47,125.76) -- (378.5,125.8) ;
\draw    (357.17,74.32) -- (357.06,90.06) ;

\draw (331,126.33) node [anchor=north west][inner sep=0.75pt]  [font=\small]  {$a$};
\draw (377.38,126.87) node [anchor=north west][inner sep=0.75pt]  [font=\small]  {$b$};
\draw (354.75,67.77) node [anchor=north west][inner sep=0.75pt]  [font=\small]  {$c$};
\draw (361.13,96.06) node [anchor=north west][inner sep=0.75pt]  [font=\small]  {$\mu $};

\end{tikzpicture}}
\end{equation} and the diagrammatic conditions on $\A$ give us consistency conditions
for the components $m_{a\alpha,b\beta}^{c\gamma;\mu}$ and $\stackrel{\vee}{m}_{a\alpha,b\beta}^{c\gamma;\mu}$.
In a unitary theory, we can choose a gauge where $m^{\vee}=m^{\dagger}$
which is usually called unitary gauge.

Note that the algebra is not uniquely specified by the underlying
object, but rather by the triplet data $(\A,m,\eta)$ up to isomorphism.
In other words, one can have two non-isomorphic algebras that share
the same underlying non-simple object but differ in their product
morphisms. However, for simplicity, we will abuse notation and refer
to the algebra itself by $\A$. 

\paragraph{Lagrangian condensation.}

\begin{figure}
\begin{centering}
\includegraphics[scale=0.8]{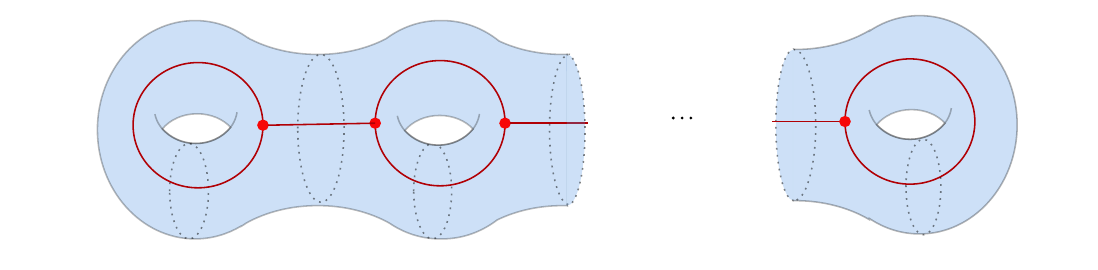}
\par\end{centering}
\caption{Condensation of $\protect\A$ can be reduced to inserting the shown
network of $\protect\A$ lines in a handlebody. This prepares the
state $\protect\ket{Z_{\protect\A}}$.}
\label{fig: A_spine}
\end{figure}

To obtain a trivial TQFT after the condensation, the algebra object
must have a maximal quantum dimension, i.e.~ $\dim\A=\D_{\T}$. Such
an algebra is called a Lagrangian algebra. To implement the condensation
in a manifold $\M$, we insert a fine mesh of $\A$ dual to the triangulation
of $\M$. The diagrammatic properties satisfied by $\A$ ensure that
this construction is independent of the choice of triangulation. On
a handlebody $\sS\Sigma_{g}$, this mesh can be reduced to insertion
of the spine diagram labeled by $\A$ as shown in figure \ref{fig: A_spine}
which prepares for us a state in $\H_{g}$. In our chosen basis of
anyons in $\H_{g}$, the vacuum state of the condensed phase is written
as
\begin{equation}
\ket{Z_{\A}}=\sum_{\gvec{\alpha},\gvec{\beta},\gvec{\gamma}}\sum_{\gvec a,\gvec b,\gvec c\in\A}\sum_{\gvec{\mu}}m_{a_{1}\alpha_{1},a_{1}^{\vee}\beta_{1}}^{b_{12}\gamma_{12};\mu_{1}}\stackrel{\vee}{m}_{a_{2}\alpha_{2},c_{2}\beta_{2}}^{b_{12}\gamma_{12};\mu_{2}}m_{a_{2}c_{2}}^{b_{23};\mu_{3}}\ldots m_{a_{g-1}c_{g-1}}^{b_{g-1,g};\mu_{2g-3}}\stackrel{\vee}{m}_{a_{g}a_{g}^{\vee}}^{b_{g-1,g};\mu_{2g-2}}\ket{\gvec a,\gvec b,\gvec c;\gvec{\mu}}.\label{eq: A state}
\end{equation}
This state is invariant under the action of the mapping class group
of $\Sigma_{g}$. We will normalize this state such that $\braket{0_{g}}{Z_{\A}}=\braket{0_{g}}{0_{g}}=1$,
where $\ket{0_{g}}$ denotes the state corresponding to the all vacuum
insertion. The partition function of the CFT corresponding to this
state is simply given by
\begin{equation}
Z_{\A}(\Omega)=\braket{\Omega}{Z_{\A}}.\label{eq: Z_A}
\end{equation}
On the torus, we simply have
\begin{equation}
\braket{\tau}{Z_{\A}}=\sum_{a}n_{a}\chi_{a}(\tau).
\end{equation}
If $\T=\T_{L}\times\bar{\T_{L}}$, there is always a canonical Lagrangian
condensation which is just the diagonal condensation
\begin{equation}
\A=\bigoplus_{(a,\bar{a})\in\T_{L}\times\bar{\T_{L}}}(a,\bar{a}).
\end{equation}
In this case, one can always find a gauge where the product junctions
can be written as $m_{(a,\bar{a})(b,\bar{b})}^{(c,\bar{c});\mu}=1$
\cite{FFIS:2003Ribbon}.

\paragraph{CFT perspective.}

One can think of the Lagrangian algebra as defining a CFT in the following
sense. The direct summands of $\A$ give us the primary operators
relative to the vertex algebra $\voa_{\T}$ while the product morphism
$m$ tells us their OPE structure. The associativity and the Frobenius
conditions ensure that these OPEs satisfy crossing symmetry. The Lagrangian
condition imposes completeness of the spectrum, which is equivalent
to $S$ modular invariance, and the commutativity condition ensures
mutual locality of the given primaries. In other words, bootstrapping
Lagrangian algebras for RCFTs can be viewed as a baby version of the
full bootstrap program for general CFTs.

From the vertex algebra point of view, anyon condensation implements
extension of the algebra by extra currents.\footnote{Abelian anyons are known as simple currents in this language.}
Lagrangian anyon condensation is then a maximal extension into a vertex
algebra with only one primary operator, with the character of that
primary being the partition function itself. These vertex algebras
are usually called self-dual vertex algebras or holomorphic VOAs in
the chiral case \cite{frenkel1989vertex}.

\subsection{SymTFT picture\label{subsec:SymTFT-picture}}

\begin{figure}
\begin{centering}
\includegraphics[scale=0.7]{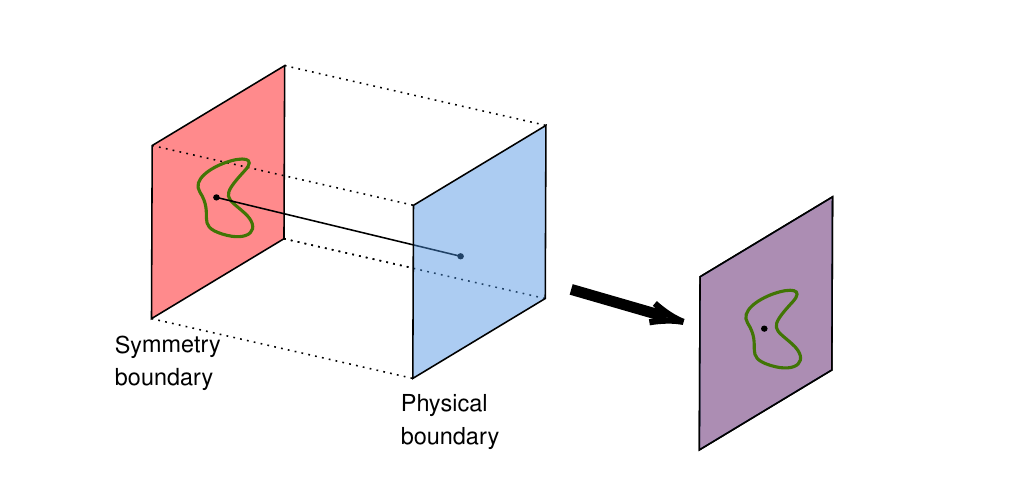}
\par\end{centering}
\caption{A depiction of the sandwich construction where the bulk TQFT is placed
on an interval. The red boundary has a topological boundary condition
with the green line as a defect living on the topological boundary.
The blue boundary represents the physical boundary where the boundary
QFT lives. We can have lines in the bulk extending between the two
boundaries giving rise to local or defect operators in the boundary
QFT. Upon collapsing the interval, we get the boundary QFT with possible
insertions of topological defect lines and local or defect operators.}
\label{fig: sandwich_figure}
\end{figure}

Lagrangian algebras are in one to one correspondence with topological
boundary conditions (gapped boundaries) in the TQFT \cite{kong2014_anyon_condensation}.
Pictorially, we can fatten the network of $\A$ lines into a surface;
then this surface will have a topological boundary condition \cite{kaidi2022_HigherCentral}.
This surface is an interface between the original TQFT and the trivial
TQFT, so we can construct this surface by condensing in half the space.
This gives us a SymTFT construction \cite{Gaiotto:2020iye,ji2020_CategoricalSymmetry,freed2022_TopologicalSymmetry,apruzzi2023_SymmetryTFTs},
also called the sandwich construction. In the sandwich construction,
one starts from the TQFT on $\Sigma\times[0,1]$ and places a topological
boundary condition on one boundary and a boundary condition that
describes the physical theory on the other. The topological boundary,
also called the symmetry boundary, hosts the symmetry line defects
of the theory that lives at the physical boundary as shown in figure
\ref{fig: sandwich_figure}. Compactifying the interval brings us
back to just the physical boundary theory with possible insertions
of topological defects (and/or defect or local operators). In our
case $\ket{Z_{\A}}$ is our topological boundary condition, $\ket{\Omega}$
is the physical boundary condition and the sandwich gives us the partition
function of the boundary CFT as the overlap $\braket{\Omega}{Z_{\A}}$.

For a given Lagrangian algebra $\A$, the symmetry category at the
corresponding topological boundary is the category of right $\A$-modules
which we will denote by $\T_{\A}$. Physically, we can think of this
as the category of confined anyons in the condensed trivial phase
where they live on the gapped domain wall between the two phases.
For diagonal condensations, the topological defect lines will be the
usual Verlinde lines \cite{verlinde1988_FusionRules}. Note that the
confined anyons are not necessarily simple anyons of the bulk TQFT
since the bulk anyons can split or be identified on the gapped boundary.
Given the symmetry category $\T_{\A}$, the bulk TQFT is what is known
as the Drinfeld center of that category \cite{Drinfeld:1986in,JOYAL199320,kassel1995_DoubleConstruction}
denoted as ${\cal Z}(\T_{\A})$ which physically we can think of as
the analog of what a gauge theory is in the case of ordinary group
symmetries.

Given a fixed boundary condition on the physical boundary, which we
denoted previously by $\ket{\Omega}$, the boundary theories obtained
from the different topological boundary conditions are all related
to each other by generalized orbifold/gauging \cite{Gaiotto:2020iye}.

\begin{figure}
\begin{centering}
\includegraphics{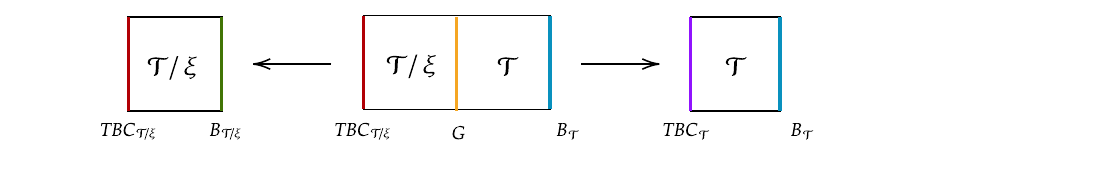}
\par\end{centering}
\caption{An illustration of the club sandwich construction. The middle figure
shows a gapped interface $G$ between theory $\protect\T$ and a condensed
phase $\protect\T/\xi$. Collapsing $G$ to the right gives us a sandwich
construction for $\protect\T$, while collapsing it to the left gives
us a sandwich for $\protect\T/\xi$. The theory at the physical boundary
is the same for all these sandwiches, and the above diagram shows
the relations between boundary conditions in the different scenarios.}
 \label{fig: club_sandwich}
\end{figure}

Note that the SymTFT construction is more general; the physical boundary
can correspond to any quantum field theory with a given symmetry category
$\C$ such that $\T$ is given by the Drinfeld center $\mathcal{Z}(\C)$,
where all dynamical information is encoded in the physical boundary
condition. For our holographic purposes we will be interested in gapless
boundary conditions particularly ones associated with conformal blocks
of the vertex algebra, which we have denoted before by the state $\ket{\Omega}$
since our goal is to mimic the Virasoro case but with a rational vertex
algebra (see \cite{bao2024_QGSymQRG} for a similar discussion about
boundary conditions). Other gapless boundary conditions can be thought
of as starting from a bigger TQFT with the holomorphic boundary condition
we discussed and then interfacing it with a condensed phase in half
the sandwich; this construction is called the ``club sandwich''
\cite{bhardwaj2023_ClubSandwich}. Then one merges the physical boundary
of the original sandwich with the interface giving us a sandwich construction
with the bulk being the condensed TQFT and the boundary condition
is now more general. This is depicted in figure \ref{fig: club_sandwich}.

\section{TQFT gravity\label{sec:TQFT-gravity}}

\subsection{Review of the main duality}

The holographic proposal of \cite{dymarsky2025_TQFTGravity} starts
from a TQFT $\T$ summed over all topologies with a boundary $\Sigma$
and shows that it is dual to an ensemble average of all CFTs constructed
from condensation of Lagrangian algebras. Let us denote the state
corresponding to condensing a given Lagrangian algebra $\A_{\ii}$
by $\ket{Z_{\ii}}$, where $\ii$ for now is just a label for the
different Lagrangian algebras (or different topological boundary conditions).
The duality then reads
\begin{equation}
\sum_{\text{topologies}}\Psi_{0}(\Omega)=\sum_{\ii}\coeff_{\ii}Z_{\ii}(\Omega),\label{eq: duality1}
\end{equation}
where in this notation $\Psi_{0}(\Omega)$ is the wavefunctional
of $\T$ on a given manifold with boundary $\Sigma_{g}$ and modulus
$\Omega$, and $Z_{\ii}(\Omega)\equiv\braket{\Omega}{Z_{\ii}}$ as
defined in (\ref{eq: Z_A}). The sum over topologies on the left-hand
side (LHS) is obtained from starting by summing over handlebodies
in a large genus limit and then performing genus reduction, while
the unnormalized coefficients $\coeff_{\ii}$ are obtained from the
large genus limit as
\begin{equation}
\coeff_{\ii}=\frac{1}{\braket{Z_{\ii}}{Z_{\ii}}|_{g\rightarrow\infty}}.
\end{equation}
In this section, we will review the derivation given in \cite{dymarsky2025_TQFTGravity}
in the context of a general non-Abelian TQFT.

We start by noting that algebra objects define surface operators (condensation
defects) via higher gauging \cite{Roumpedakis:2022_higher_gauging}.
For Lagrangian algebras, these surface operators act as (un-normalized)
projectors onto the respective topological boundary condition states.
For an algebra object $\A_{\ii}$, the corresponding surface operator
is given by
\begin{equation}
\S_{\ii}\equiv\proj{Z_{\ii}}{Z_{\ii}}.
\end{equation}
The fusion algebra of such surface operators can be written as
\begin{equation}
\S_{\ii}\times\S_{\ii}=\braket{Z_{\ii}}{Z_{\ii}}\S_{\ii},
\end{equation}
\begin{equation}
\S_{\ii}\times\S_{\jj}=\braket{Z_{\ii}}{Z_{\jj}}\S_{\ii\jj},
\end{equation}
where
\begin{equation}
\S_{\ii\jj}\equiv\proj{Z_{\ii}}{Z_{\jj}}.
\end{equation}
The fusion coefficients $\braket{Z_{\ii}}{Z_{\jj}}$ are partition
functions of 2d TQFTs as was argued in \cite{Roumpedakis:2022_higher_gauging}.
The above surface operators not only commute with the representation
of the mapping class group operators, denoted below by $\U$, but
they are also invariant under their action
\begin{equation}
\U_{\gamma}\,\S=\S\,\U_{\gamma}=\S,
\end{equation}
where $\gamma\in\mcgg$ and $\mcgg$ denotes the mapping class group
(MCG) of $\Sigma_{g}$.

In what follows, for convenience, we will work with projectors built
out of normalized states, so we define
\begin{equation}
P_{\ii}\equiv\frac{1}{\braket{Z_{\ii}}{Z_{\ii}}}\proj{Z_{\ii}}{Z_{\ii}},\quad P_{\ii\jj}\equiv\frac{1}{\sqrt{\braket{Z_{\ii}}{Z_{\ii}}\braket{Z_{\jj}}{Z_{\jj}}}}\proj{Z_{\ii}}{Z_{\jj}},
\end{equation}
which lead to
\begin{equation}
P_{\ii}\times P_{\ii}=P_{\ii},\qquad P_{\ii}\times P_{\jj}=\frac{\braket{Z_{\ii}}{Z_{\jj}}}{\sqrt{\braket{Z_{\ii}}{Z_{\ii}}\braket{Z_{\jj}}{Z_{\jj}}}}P_{\ii\jj}.
\end{equation}
At an arbitrarily large genus $g$, it is conjectured that all MCG
invariant states are linear combinations of the physical invariants
$\ket{Z_{\alpha}}$ defined by Lagrangian algebras. The motivation
for this is that states corresponding to Lagrangian algebras can always
be defined at any arbitrary genus due to the constraints satisfied
by the algebra, while any linearly independent state at a particular
genus $g$ is not guaranteed to survive the MCG constraints at some
higher genus $g+k$. Given this conjecture, if we show that
\begin{equation}
\lim_{g\rightarrow\infty}\frac{\braket{Z_{\ii}}{Z_{\jj}}}{\sqrt{\braket{Z_{\ii}}{Z_{\ii}}\braket{Z_{\jj}}{Z_{\jj}}}}=\delta_{\ii\jj},\label{eq: orth_condition}
\end{equation}
then the projectors $P_{\ii}$ will form a complete idempotent basis
as $g\rightarrow\infty$. We will prove (\ref{eq: orth_condition})
in section \ref{sec:Weights-from-symmetries}. Using this, we can
write the projector onto the MCG invariant subspace of $\H_{g}$
in terms of $P_{\ii}$ at $g\rightarrow\infty$ as
\begin{equation}
\frac{1}{|\mcgg|}\sum_{\gamma\in\mcgg}\U_{\gamma}=\sum_{\ii}P_{\ii},\label{eq: mcg_average}
\end{equation}
where from here on, for convenience, we will denote $g$ as the regulator
of the $g\rightarrow\infty$ limit and it will be implicit that the
equality in (\ref{eq: mcg_average}) and other similar equations hold
in the limit $g\rightarrow\infty$. Any actual finite genus will be
denoted by $\tilde{g}$. 

Acting by the projector (\ref{eq: mcg_average}) on the vacuum and
capping it off from the left by $\bra{\Omega_{g}}$ we get
\begin{equation}
\frac{1}{|\mcgg/\Gamma^{(g)}|}\sum_{\gamma\in\mcgg/\Gamma^{(g)}}\bra{\Omega_{g}}\U_{\gamma}\ket{0_{g}}=\sum_{\ii}\coeff_{\ii}Z_{\ii}(\Omega_{g}),\label{eq: duality2}
\end{equation}
where $\coeff_{\ii}=\frac{1}{\braket{Z_{\ii}}{Z_{\ii}}|_{g}}$  and
$\Gamma^{(g)}\subset\mcgg$ is the stabilizer group of the handlebody
$\sS\Sigma_{g}$ associated with $\ket{0_{g}}$ which is known as
the \emph{handlebody group} \cite{Suzuki_1977,hensel2020primer};
this is the subgroup of $\mcg(\Sigma_{g})$ that extends inside the
handlebody. Note that this is different than the full stabilizer group
of the state $\ket{0_{g}}$, where the latter is larger due to the
nontrivial kernel of the TQFT representation of the $\mcg(\Sigma_{g})$.
 With this definition, the LHS of (\ref{eq: duality2}) so far is
a formal average over all handlebodies at genus $g$. Now we want
to do genus reduction from genus $g$ to genus $\tilde{g}$ by taking
the pinching limit for $\Omega_{g}$ for $g-\tilde{g}$ cycles, effectively
making them into $g-\tilde{g}$ independent tori and then sending
the imaginary part of their modulus to infinity. In this limit, the
partition functions in the RHS of (\ref{eq: duality2}) will degenerate
to their counterpart at genus $\tilde{g}$. We can see this explicitly
at the level of wavefunctions $\braket{\Omega_{g}}Z$. Using the definition
of $\ket{\Omega_{g}}$ in (\ref{eq: Omega_state}) and the fact that
vacuum dominates at low temperature, we get
\begin{equation}
\ket{\Omega_{g}}\rightarrow\ket{0_{g-\tilde{g}},\Omega_{\tilde{g}}},
\end{equation}
where the state $\ket{0_{g-\tilde{g}},\Omega_{\tilde{g}}}$ is an
element of the product subspace $\H_{g-\tilde{g}}\otimes\H_{\tilde{g}}\subset\H_{g}$.

\begin{figure}

\begin{centering}
\includegraphics[scale=0.5]{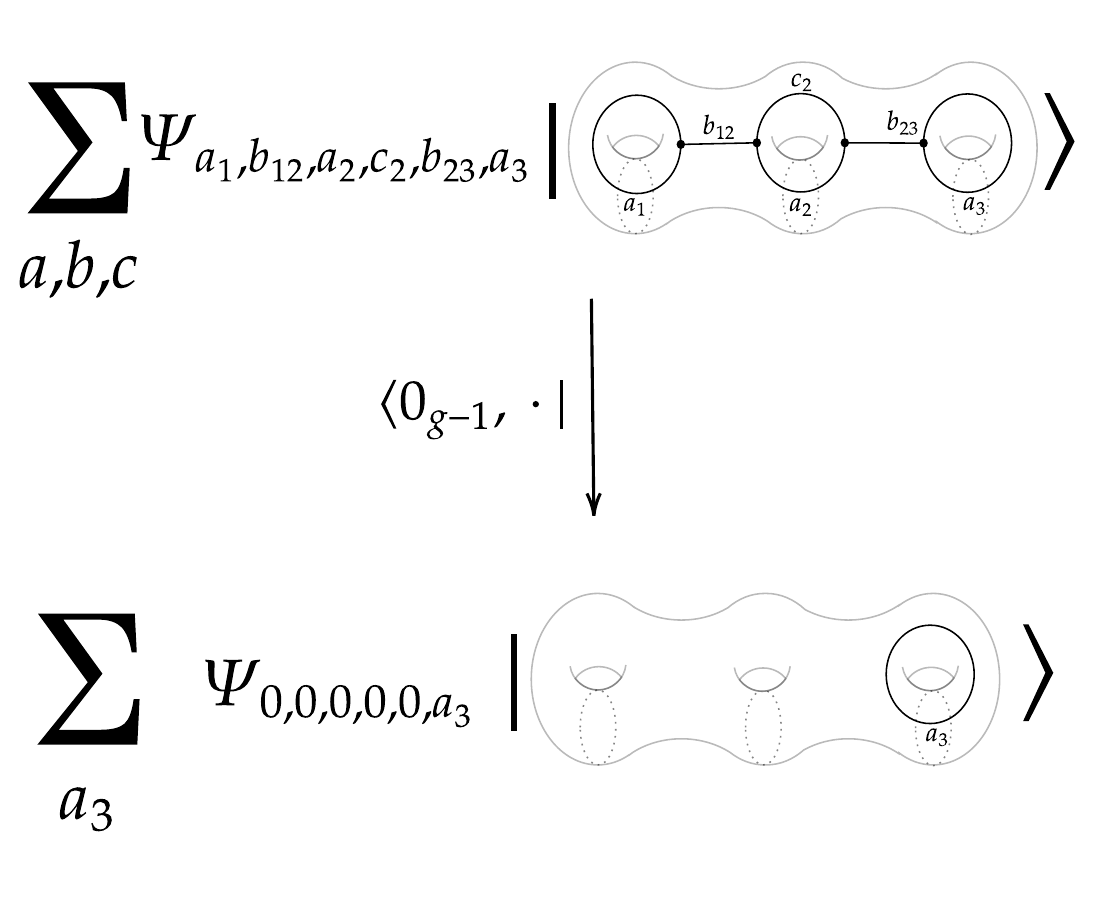}\caption{Illustration of projection via $\protect\bra{0_{g-\tilde{g}},\cdot}$
in a simple case for $g=3$ and $\tilde{g}=1$. The resulting state
can be viewed as a state in $\protect\H_{\tilde{g}=1}$.}
\label{fig: projection map_genus_reduction}
\par\end{centering}
\end{figure}

Overlapping any ket state with $\bra{0_{g-\tilde{g}},\Omega_{\tilde{g}}}$
then implements a projection map 
\begin{equation}
\bra{0_{g-\tilde{g}},\cdot}\equiv\Phi_{\tilde{g}}:\H_{g}\rightarrow\H_{\tilde{g}},
\end{equation}
 where $\bra{0_{g-\tilde{g}},\cdot}$ denotes overlapping with the
first $g-\tilde{g}$ cycles with the vacuum insertion on those cycles
leaving the final $\tilde{g}$ cycles intact.\footnote{The intermediate link between the $g-\tilde{g}$ cycles and $\tilde{g}$
cycles must be vacuum; otherwise, the overlap will vanish. }  So starting from a state $\ket{\Psi}\in\H_{g}$, we effectively
get a state $\ket{\tilde{\Psi}}\in{\cal H}_{\tilde{g}}$ given by
$\braket{0_{g-\tilde{g}},\cdot}{\Psi}$. This is illustrated for a
simple case in figure \ref{fig: projection map_genus_reduction}.
Hence, after genus reduction we get
\begin{equation}
\frac{1}{|\mcgg/\Gamma^{(g)}|}\sum_{\gamma\in\mcgg/\Gamma^{(g)}}\braket{\Omega_{\tilde{g}}}{\gamma}_{\tilde{g}}=\sum_{\ii}\coeff_{\ii}Z_{\ii}(\Omega_{\tilde{g}}),\label{eq: duality_gtilde}
\end{equation}
where we denote 
\begin{equation}
\ket{\gamma}_{\tilde{g}}\equiv\bra{0_{g-\tilde{g}},\cdot}{\cal U}_{\gamma}\ket{0_{g}}.
\end{equation}

What remains now is to show that the left hand side of (\ref{eq: duality_gtilde})
would give us wavefunctions corresponding to evaluating the TQFT path
integral on all topologies ending on $\Sigma_{\tilde{g}}$ with boundary
condition given by $\ket{\Omega_{\tilde{g}}}$. The argument in \cite{dymarsky2025_TQFTGravity}
was that $\bra{0_{g}}{\cal U}_{\gamma}\ket{0_{g}}$ for all $\gamma\in\mcgg$
will give us all possible Heegaard splittings of closed 3-manifolds,
and so genus reduction should similarly produce all 3-manifolds with
boundary. We will elaborate a bit more on this argument in the next
subsection, showing explicitly that this will give us all 3-manifolds
with $\Sigma_{\tilde{g}}$ boundary.

The MCG average discussed in the derivation above should be viewed
in a formal sense. As noted earlier, the LHS of (\ref{eq: duality2})
is a formal average over all genus $g$ handlebodies. In a given TQFT,
this average will reduce to an average over certain equivalence classes
of handlebodies since the TQFT representation of $\mcgg$ is not faithful
in general and hence the kernel of the representation factors out
of the LHS average in (\ref{eq: duality2}).\footnote{The coset can be further reduced using the full stabilizer of |$0_{g}\rangle$
within the MCG representation image, which may be larger than the
image of the geometric handlebody group.} For a generic non-Abelian TQFT, the representation of $\mcgg$ for
$g>1$ can still have an infinite image \cite{jian2020_EstablishingStronglycoupled}
and thus even the reduced sum may need to be regularized. It is not
clear how to do such regularization for the full MCG average (the
MCG itself is known to be non-amenable \cite{jian2020_EstablishingStronglycoupled,romaidis2024_CFTCorrelators}).
Nonetheless, one could possibly regularize the coset sum in a similar
fashion to \cite{MaloneyWitten2007,Cynthia_cusps2025}. It would be
interesting to test this in simple non-Abelian TQFTs with infinite
MCG image like the doubled Fibonacci TQFT and see if one could regularize
the coset sum.

An important aspect of (\ref{eq: duality2}) is that topological boundary
conditions that are related by an anyon permutation symmetry have
equal weights. This is a consequence of the fact that anyon permutations
commute with the action of the mapping class group.\footnote{Note that this statement is more general than the case at hand where
the weights are related to the norms (which are clearly the same for
two states related by anyon permutations).}  

\subsection{The sum over topologies\label{subsec:The-sum-over topologies}}

\begin{figure}

\begin{centering}
\includegraphics[scale=0.8]{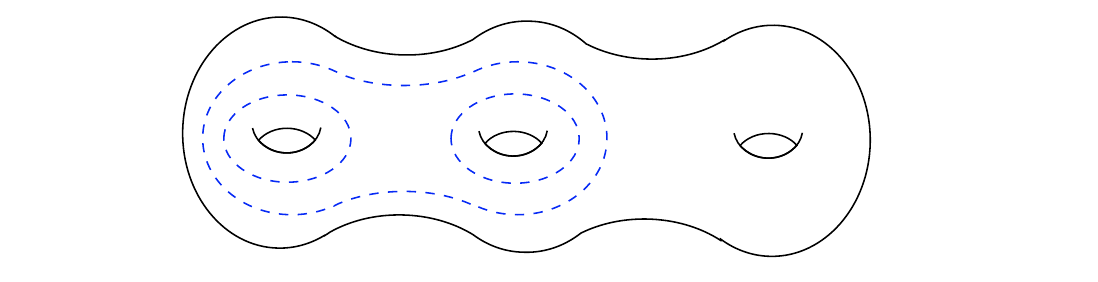}\caption{A visualization of a compression body with an outer boundary $\partial_{+}C=\Sigma_{3}$
and an inner boundary $\partial_{-}C=\Sigma_{2}$.}
\label{fig: compression_body}
\par\end{centering}
\end{figure}

We now want to show how the states $\ket{\gamma}_{\tilde{g}}$ give
rise to all manifolds with boundary $\Sigma_{\tilde{g}}$. The main
idea is to show that any such state can be obtained by the following
(generalized) Heegaard splitting  procedure:
\begin{enumerate}
\item Start from a handlebody $\sS\Sigma_{g}$ corresponding to state $\ket{0_{g}}$
and carve out a handlebody $\sS\Sigma_{\tilde{g}}$ from inside $\sS\Sigma_{g}$.
This gives us what is known as a \emph{compression body} $C$ with
inner boundary $\partial_{-}C=\Sigma_{\tilde{g}}$ and outer boundary
$\partial_{+}C=\Sigma_{g}$. This is visualized in figure \ref{fig: compression_body}.
\item Get another handlebody related to the original $\sS\Sigma_{g}$ by
a boundary mapping class group transformation $\gamma$. Let us denote
it as $\sS_{\gamma}\Sigma_{g}$.
\item Glue $\sS_{\gamma}\Sigma_{g}$ with $C$ across $\partial_{+}C$,
i.e.~ $C\cup_{\Sigma_{g}}\sS_{\gamma}\Sigma_{g}$. This gluing can
be also denoted simply as $C\cup_{\gamma}\sS\Sigma_{g}$.
\end{enumerate}
 The result of this procedure is some 3-manifold with a boundary $\Sigma_{\tilde{g}}$.
For arbitrary $\gamma\in\mcgg$ and arbitrarily large $g$, this procedure
should be enough to generate all 3-manifolds with boundary $\Sigma_{\tilde{g}}$
(possibly many many times) based on the fact that all triangulated
manifolds admit a (generalized) Heegaard splitting \cite{Moise1952_triangulation,scharlemann2016lecture}.
We will elaborate on this point later in this section.

We can see that the procedure above is equivalent to the genus reduction
by starting from the definition of $\ket{\gamma}_{\tilde{g}}$ and
inserting a resolution of the identity operator of $\H_{g}$
\begin{align}
\ket{\gamma} & =\bra{0_{g-\tilde{g}},\cdot}{\rm I}_{g}\,{\cal U}_{\gamma}\ket{0_{g}}\\
 & =\sum_{\gvec a,\gvec b,\gvec c,\gvec{\mu}}\braket{0_{g-\tilde{g}},\cdot}{\gvec a,\gvec b,\gvec c;\gvec{\mu}}\bra{\gvec a,\gvec b,\gvec c;\gvec{\mu}}U_{\gamma}\ket{0_{g}}\\
 & \equiv CU_{\gamma}\ket{0_{g}},
\end{align}
where 
\begin{equation}
C\equiv\sum_{\gvec a,\gvec b,\gvec c,\gvec{\mu}}\braket{0_{g-\tilde{g}},\cdot}{\gvec a,\gvec b,\gvec c;\gvec{\mu}}\bra{\gvec a,\gvec b,\gvec c;\gvec{\mu}}
\end{equation}
is our desired compression body, and we used the basis notation of
eq. (\ref{eq: anyon basis states}) corresponding to figure \ref{fig: anyon_basis_g}.
The compression body can be viewed as a map $C:\H_{g}\rightarrow\H_{\tilde{g}}$
given by the composition $\Phi_{\tilde{g}}\circ{\rm I}_{g}$. Physically,
this compression body is a wormhole between $\Sigma_{g}$ and $\Sigma_{\tilde{g}}$
which can be obtained from the Euclidean wormhole $\Sigma_{g}\times I$
by degenerating $\Sigma_{g}$ to $\Sigma_{\tilde{g}}$ on one end
but not the other. 

Let us illustrate this with a simple example of $g=2$ and $\tilde{g}=1$.
We will drop the fusion junction label $\mu$ for notational convenience
and label our basis states on $\Sigma_{2}$ by $\ket{abc}$ where,
in the notation of figure \ref{fig: anyon_basis_g}, $a\equiv a_{1}$,
$b\equiv b_{12}$ and $c\equiv a_{2}$. First, let us construct the
state $\ket{\gamma}_{\tilde{g}=1}$ explicitly from genus reduction.
We get 
\begin{equation}
U_{\gamma}\ket{000}=\sum_{abc}\left({\cal U}_{\gamma}\right)_{(abc)(000)}\ket{abc},
\end{equation}
\begin{equation}
\ket{\gamma}_{\tilde{g}=1}=\sum_{c}\left({\cal U}_{\gamma}\right)_{(00c)(000)}\ket c,
\end{equation}
and we denote $\left({\cal U}_{\gamma}\right)_{(abc)(a'b'c')}\equiv\bra{abc}{\cal U}_{\gamma}\ket{a'b'c'}$.
The Euclidean wormhole of $\Sigma_{g=2}$ is just the identity operator
\begin{equation}
\sum_{abc}\ket{abc}\bra{abc}.
\end{equation}
Degenerating to $\Sigma_{\tilde{g}=1}$ on one end gives us
\begin{equation}
C=\sum_{c}\ket c\bra{00c},
\end{equation}
where $\ket c$ denotes a state in $\H_{\tilde{g}=1}$. We can see
that $C$ acts as an identity operator when restricted to $\H_{\tilde{g}=1}$.
Applying the gluing of $\Sigma_{g}$ to $\partial_{+}C$ gives us
our desired result as
\begin{align}
C\cup_{\gamma}\sS\Sigma_{2} & =C{\cal U}_{\gamma}\ket{000}\\
 & =\sum_{c}\ket c\bra{00c}\sum_{a'b'c'}\left({\cal U}_{\gamma}\right)_{(a'b'c')(000)}\ket{a'b'c'}\\
 & =\sum_{c}\left({\cal U}_{\gamma}\right)_{(00c)(000)}\ket c\\
 & =\ket{\gamma}_{\tilde{g}=1}.
\end{align}

As we see, the genus reduction method is powerful enough to generate
our desired sum over topologies at any arbitrary genus, which from
the ensemble perspective calculates the unnormalized averaged partition
function which we denote as $<Z(\Omega_{\tilde{g}})>'$. One can get
multi-point moments of the partition function by taking the splitting
limit of $\tilde{g}$ into disconnected surfaces of even lower genera.
All of these can be obtained from (\ref{eq: mcg_average}) by applying
the appropriate bra and ket states to the projector. For example for
$n$-point torus moments this would amount to 
\begin{equation}
\frac{1}{|\mcgg|}\sum_{\gamma\in\mcgg}\bra{0_{g-n},\tau_{1},\tau_{2},\dots,\tau_{n}}\U_{\gamma}\ket{0_{g}}=\sum_{\ii}\frac{Z_{\ii}(\tau_{1})Z_{\ii}(\tau_{2})\dots Z_{\ii}(\tau_{n})}{\braket{Z_{\ii}}{Z_{\ii}}|_{g}},
\end{equation}
 where similar to our previous notation, $\bra{0_{g-n},\tau_{1},\tau_{2},\dots,\tau_{n}}$
denotes a factorizable bra-state in $\H_{g-n}\otimes\H_{1}^{\otimes n}\subset\H_{g}$.
More generally, we can include orientation-reversed versions of the
tori

\begin{align}
\frac{1}{|\mcgg|}\sum_{\gamma\in\mcgg}\bra{0_{g-n},\tau_{1},\tau_{2},\dots,\tau_{n}}\U_{\gamma}\ket{0_{g-m},\bar{\tau}_{1}',\bar{\tau}_{2}',\dots,\bar{\tau}_{m}'}\nonumber \\
=\sum_{\ii}\frac{Z_{\ii}(\tau_{1})Z_{\ii}(\tau_{2})\dots Z_{\ii}(\tau_{n})Z_{\ii}(\bar{\tau}_{1}')Z_{\ii}(\bar{\tau}_{2}')\dots Z_{\ii}(\bar{\tau}_{m}')}{\braket{Z_{\ii}}{Z_{\ii}}|_{g}}.
\end{align}

An important amplitude to consider is the vacuum to vacuum amplitude
\begin{equation}
\frac{1}{|\mcgg|}\sum_{\gamma\in\mcgg}\bra{0_{g}}\U_{\gamma}\ket{0_{g}}=\sum_{\ii}\frac{1}{\braket{Z_{\ii}}{Z_{\ii}}|_{g}}.\label{eq: sum closed manifolds bare}
\end{equation}
 Let us understand this sum a bit better and try to write it explicitly
as a sum over distinct homeomorphism classes of closed 3-manifolds.
First, note that the distinct elements of the LHS are described by
elements of the double-coset space $\Gamma^{(g)}\backslash\mcgg/\Gamma^{(g)}$.
Writing the sum only over the distinct elements, we get
\begin{equation}
\left(\sum_{\gamma\in\Gamma^{(g)}\backslash\mcgg/\Gamma^{(g)}}\frac{1}{|\Aut(\gamma)|}\right)^{-1}\sum_{\gamma\in\Gamma^{(g)}\backslash\mcgg/\Gamma^{(g)}}\frac{1}{|\Aut(\gamma)|}{\bf Z}_{\T}(M_{\gamma})=\sum_{\ii}\frac{1}{\braket{Z_{\ii}}{Z_{\ii}}|_{g}},\label{eq: sum over closed manifolds Aut}
\end{equation}
 where $M_{\gamma}$ is the closed 3-manifold obtained by the Heegaard
gluing $\sS\Sigma_{g}\cup_{\gamma}\sS\Sigma_{g}$, ${\bf Z}_{\T}$
is the TQFT partition function in our chosen normalization, and $\Aut(\gamma)\equiv\Gamma^{(g)}\cap\gamma\Gamma^{(g)}\gamma^{-1}$.
Note that in going from (\ref{eq: sum closed manifolds bare}) to
(\ref{eq: sum over closed manifolds Aut}), we have used the fact
that the formal size of each double-coset $|\Gamma^{(g)}\gamma\Gamma^{(g)}|$
is given by 
\begin{equation}
|\Gamma^{(g)}\gamma\Gamma^{(g)}|=\frac{|\Gamma^{(g)}|^{2}}{|\Gamma^{(g)}\cap\gamma\Gamma^{(g)}\gamma^{-1}|},
\end{equation}
and
\begin{equation}
|\mcgg|=\sum_{\gamma\in\Gamma^{(g)}\backslash\mcgg/\Gamma^{(g)}}|\Gamma^{(g)}\gamma\Gamma^{(g)}|.
\end{equation}
Geometrically, $\Aut(\gamma)$ is the group that preserves the Heegaard
splitting $\sS\Sigma_{g}\cup\sS_{\gamma}\Sigma_{g}$ which is known
as the \emph{Goeritz group} ${\cal G}_{g}$ of the splitting \cite{goeritz1933abbildungen,johnson2011_MappingClass,cho2014genusgoeritzgroups2,sekino2022goeritzgroupheegaardsplitting}.
It is the subgroup of the MCG of the splitting surface that extends
to both handlebodies $\sS\Sigma_{g}$ and $\sS_{\gamma}\Sigma_{g}$,
and hence can be expressed as the intersection of their handlebody
groups, which are $\Gamma^{(g)}$ and $\gamma\Gamma^{(g)}\gamma^{-1}$
respectively in our case. To illustrate the Goeritz group in a simple
example, consider the $g=1$ case. The MCG of the torus is ${\rm SL(2,\mathbb{Z})}$,
and the handlebody group $\Gamma^{(1)}$ is the group generated by
$\{-I,T\}$. \footnote{Note that in this example we are working with the MCG itself rather
than its action on the torus modulus $\tau$ or its TQFT representation.
On the modulus $\tau$, the MCG action is by ${\rm PSL}(2,\mathbb{Z})$
since $-I$ acts trivially, and the stabilizer of the $c>1$ Virasoro
vacuum $|\chi_{0}^{\text{Vir.}}(\tau)|^{2}$ is just the group generated
by $T$, usually denoted as $\Gamma_{\infty}$. In a given compact
TQFT, the representation of $-I$ can act nontrivially as the charge
conjugation matrix $C$ and so the image of the handlebody group in
that case is generated by $\{C,T\}$.} The $g=1$ splitting of $S^{3}$ involves the $S$ transformation,
so 
\begin{equation}
{\cal G}_{1}(S^{3})=\Gamma^{(1)}\cap S\Gamma^{(1)}S^{-1}=\{I,-I\}\simeq\mathbb{Z}_{2},
\end{equation}
while for $S^{2}\times S^{1}$ it is just the identity map, hence
\begin{equation}
{\cal G}_{1}(S^{2}\times S^{1})=\Gamma^{(1)}\cap\Gamma^{(1)}=\Gamma^{(1)}\simeq\mathbb{Z}\times\mathbb{Z}_{2}.
\end{equation}

The elements of the double-coset sum in (\ref{eq: sum over closed manifolds Aut})
are unique Heegaard splittings of genus $g$ but are they unique as
3-manifolds? At any finite genus, there could be two different Heegaard
splittings that produce the same 3-manifold. However, in the $g\rightarrow\infty$
limit, this degeneracy will go away. This is a consequence of the
Reidemeister-Singer theorem \cite{reidemeister1933_ZurDreidimensionalen,Singer1933ThreedimensionalMA,lei2000stability}
which states that any two genus $g$ Heegaard splittings of the same
manifold become equivalent under a finite number of \emph{stabilizations},
where a stabilization \cite{Hass_2009} means adding an extra handle,
i.e. going from genus $g$ to genus $g+1$. Therefore, in the $g\rightarrow\infty$
limit, the set of double-cosets $\Gamma^{(g)}\backslash\mcgg/\Gamma^{(g)}$
bijects onto the set of homeomorphism classes of closed orientable
3-manifolds \cite{hensel2020primer,scharlemann2000heegaardsplittingscompact3manifolds}.

In a similar fashion to the closed-manifolds' case, where now instead
the Heegaard splitting involves a compression body and a handlebody,
the sum over homeomorphism classes of compact orientable 3-manifolds
with a boundary of genus $\tilde{g}$ should\footnote{The equivalence between Heegaard splittings under stabilization should
hold for manifolds with boundaries as well; see for example \cite{scharlemann2000heegaardsplittingscompact3manifolds}.}  be described by a similar double-coset sum in the $g\rightarrow\infty$
limit, namely
\begin{equation}
\left(\sum_{\gamma\in\Gamma^{(g,\tilde{g})}\backslash\mcgg/\Gamma^{(g)}}\frac{1}{|\Aut(\gamma)|}\right)^{-1}\sum_{\gamma\in\Gamma^{(g,\tilde{g})}\backslash\mcgg/\Gamma^{(g)}}\frac{1}{|\Aut(\gamma)|}\ket{\gamma}_{\tilde{g}}=\sum_{\ii}\frac{\ket{Z_{\alpha}}_{\tilde{g}}}{\braket{Z_{\ii}}{Z_{\ii}}|_{g}},
\end{equation}
where $\Gamma^{(g,\tilde{g})}$ (\emph{the relative compression body group}
\cite{oertel2007mappingclassgroupscompression}) is the subgroup of
$\mcgg$ that extends to the compression body $C$ with $\partial_{+}C=\Sigma_{g}$,
$\partial_{-}C=\Sigma_{\tilde{g}}$ and leaves $\partial_{-}C$ invariant
pointwise, while $\Aut(\gamma)\equiv\Gamma^{(g,\tilde{g})}\cap\gamma\Gamma^{(g)}\gamma^{-1}$
is the Goeritz group of the generalized Heegaard splittings $C\cup_{\gamma}\sS\Sigma_{g}$.\footnote{More generally, in principle, one should be able to describe manifolds
with disjoint boundaries using gluings of more general compression
bodies and using the corresponding double-coset.}  So each connected manifold $M_{\gamma}$ with boundary $\partial M_{\gamma}=\Sigma_{\tilde{g}}$,
labeled by $\gamma\in\Gamma^{(g,\tilde{g})}\backslash\mcgg/\Gamma^{(g)}$,
is weighted by
\begin{equation}
\mu_{\tilde{g}}(M_{\gamma})=\left(\sum_{\gamma'\in\Gamma^{(g,\tilde{g})}\backslash\mcgg/\Gamma^{(g)}}\frac{1}{|\Gamma^{(g,\tilde{g})}\cap\gamma'\Gamma^{(g)}\gamma'{}^{-1}|}\right)^{-1}\frac{1}{|\Gamma^{(g,\tilde{g})}\cap\gamma\Gamma^{(g)}\gamma^{-1}|},\qquad g\rightarrow\infty.\label{eq: Goeritz measure}
\end{equation}
Note that in the case where $\tilde{g}\rightarrow g$ we get back
the sum over only handlebodies in (\ref{eq: duality2}) as $\Gamma^{(g,g)}$
from our definition is just the identity element of $\mcgg$.

What remains now is to understand the implications of this measure
for the sum over 3-manifolds, which is related to the Goeritz group
in the large genus limit. It is not a surprise that the sum over 3-manifolds
we got is weighted by the Goeritz group; after all, we were averaging
uniformly over all labeled Heegaard splittings and so restricting
to the equivalence classes leads to an automorphism weighted average,
with the Goeritz group being the natural automorphism group in this
context.\footnote{This is similar to the $1/|\Aut|$ sum over bordisms in TQFT considered
in \cite{Banerjee:2022pmw}, but here the automorphisms are given
by the Goeritz group of the large genus Heegaard splittings since
we are using those to represent 3-bordisms.} This is of course an ill-defined measure since the Goeritz group
is infinite in this case for all manifolds and it is not obvious how
to regularize it; however one would like to understand the meaning
of using the large genus Goeritz group in the light of the relation
between 3d gravity and VTQFT. We leave this task for future work.

Finally, we want to note that even though we used the full MCG in
our formal derivations, a given TQFT will not be sensitive to all
manifolds since the representation of MCG is not faithful in general.
In cases where the image of the MCG is finite, as for example in Abelian
TQFTs or the Ising TQFT, the TQFT will only distinguish a finite number
of equivalence classes of manifolds at any finite genus. In these
cases, the sum over topologies (regularized at a finite Heegaard splitting
genus) is tractable, and the effective weights over the equivalence
classes of manifolds will be finite. A very simple example is the
$\mathbb{Z}_{2}$ gauge theory or toric code (TC) TQFT. At $g=1$,
the image of MCG is ${\rm SL}(2,\mathbb{Z}_{2})$ and hence, the set
of all lens spaces will collapse to just two classes: $S^{2}\times S^{1}$
and $S^{3}$. The image $\rho$ of their Goeritz group at the $g=1$
splitting is $\rho({\cal G}_{1}(S^{2}\times S^{1}))=\mathbb{Z}_{2}$
and $\rho({\cal G}_{1}(S^{3}))=\{I\}$. Hence, the renormalized sum
over all $g=1$ splittings of closed 3-manifolds in the toric code
case can be simply reduced to
\begin{equation}
\frac{1}{|\mathbb{Z}_{2}|}{\bf Z}_{{\rm TC}}(S^{2}\times S^{1})+{\bf Z}_{{\rm TC}}(S^{3})=\frac{1}{2}(1)+\frac{1}{2}.
\end{equation}
In general, the toric code TQFT, being a $\mathbb{Z}_{2}$ gauge theory,
is only sensitive to ${\rm H}_{1}(M,\mathbb{Z}_{2})$, so the sum
over closed manifolds can be made equivalent to a sum over connected
sums of $S^{2}\times S^{1}$ and $S^{3}$. The image of the large
genus Goeritz group can be calculated explicitly in these cases; we
will leave this for future work.

\subsection{The baby universe Hilbert space}

In the previous subsection, we have seen that all possible moments
of the ensemble are secretly encoded in the infinite genus $g$ Hilbert
space, or more precisely in the MCG invariant subspace which we will
denote as $\H_{g}^{{\rm MCG}}$. This suggests that there should be
a natural interpretation of $\H_{g}^{{\rm MCG}}$, which we interpret
here as the baby universe Hilbert space. 

To illustrate this, let us briefly revisit the construction of Marolf
and Maxfield (MM) \cite{marolf2020_TranscendingEnsemble} of the baby
universe Hilbert space. They start by considering the gravitational
path integral on a manifold with multiple asymptotic boundaries given
some arbitrary boundary conditions. They cut this path integral in
a way such that all the connected components of the asymptotic boundaries
are either entirely in the future or entirely in the past relative
to the cut, hence defining a state in the baby (closed) universe Hilbert
space $\H_{\text{BU}}$. In their notation, if we have $n$ connected
components of asymptotic boundaries with a set of boundary conditions
labeled by $\{J_{1},J_{2},\dots,J_{n}\}$, then if we make a cut such
that all these boundaries are in the past, we get a state that we
can denote by 
\begin{equation}
\ket{Z[J_{1}]Z[J_{2}]\dots Z[J_{n}]}\in\H_{\text{BU}}.\label{eq: ZJn MM state BU}
\end{equation}
Such a state if overlapped with the no-boundary (Hartle-Hawking) state
$\ket{\text{HH}}$ defines the unnormalized correlator of the boundary
ensemble as
\begin{equation}
<Z[J_{1}]Z[J_{2}]\dots Z[J_{n}]>'\equiv\braket{{\rm HH}}{Z[J_{1}]Z[J_{2}]\dots Z[J_{n}]},
\end{equation}
where the prime notation denotes the unnormalized version of the correlator.
Now MM construct $\H_{\text{BU}}$ as the completion of all states
of the form (\ref{eq: ZJn MM state BU}) while quotienting by null
states. They then proceed to define an operator $\widehat{Z[J]}$
such that
\begin{equation}
\ket{Z[J_{1}]Z[J_{2}]\dots Z[J_{n}]}=\widehat{Z[J_{1}]}\widehat{Z[J_{2}]}\dots\widehat{Z[J_{n}]}\ket{\text{HH}},
\end{equation}
which implies that all such $\widehat{Z[J]}$ operators commute. Hence,
one can then define common eigenstates for all $\widehat{Z[J]}$,
which are the so-called $\alpha$-states where factorization is restored
\begin{equation}
\widehat{Z[J]}\ket{\alpha}=Z_{\alpha}[J]\ket{\alpha},
\end{equation}
and $\ket{\alpha}$ form an orthonormal basis in $\H_{\text{BU}}$.
The $\alpha$ label is therefore a label for members of the ensemble
as concluded in \cite{marolf2020_TranscendingEnsemble}. Using these
states, one can write the normalized amplitudes in terms of $Z_{\alpha}[J]$
\begin{align}
<Z[J_{1}]Z[J_{2}]\dots Z[J_{n}]> & \equiv\frac{\bra{\text{HH}}\widehat{Z[J_{1}]}\widehat{Z[J_{2}]}\dots\widehat{Z[J_{n}]}\ket{\text{HH}}}{\braket{\text{HH}}{\text{HH}}}\\
 & =\sum_{\alpha}p_{\alpha}Z_{\alpha}[J_{1}]Z_{\alpha}[J_{2}]\dots Z_{\alpha}[J_{n}],
\end{align}
where
\begin{equation}
p_{\alpha}\equiv\frac{\left|\braket{\alpha}{\text{HH}}\right|^{2}}{\braket{\text{HH}}{\text{HH}}}
\end{equation}
are the probabilities of the ensemble properly normalized to add up
to unity. 

Now let us get back to our TQFT gravity setup. Our ``gravitational''
path integral is not the path integral of the TQFT because we want
to glue the cuts with the $\mcg$ projector rather than the identity
operator. However, the $\mcg$ projector is the identity operator
on $\H_{g}^{\text{\ensuremath{\mcg}}}$, so this is our ``gravitational''
Hilbert space for a closed universe. So we are prompted to identify
$\H_{{\rm BU}}\simeq\H_{g}^{\mcg}$ (in the large $g$ limit).

Let us illustrate this further. Our no-boundary state in $\H_{g}$
is given by $\ket{0_{g}}$. Projecting this onto $\H_{g}^{\mcg}$
defines for us the Hartle-Hawking (HH) state as 
\begin{equation}
\ket{\text{HH}}\equiv\lim_{g\rightarrow\infty}\frac{1}{|\mcgg|}\sum_{\gamma\in\mcgg}\U_{\gamma}\ket{0_{g}}.
\end{equation}
Since we are using a projector, the vacuum to vacuum amplitude in
(\ref{eq: sum closed manifolds bare}) is simply equal to $\braket{\text{HH}}{\text{HH}}$
as expected. Similarly, all the amplitudes we calculated in the previous
subsection can be defined this way, where now the states of the form
$\ket{0_{g-n},\tau_{1},\dots,\tau_{n}}$ are analogs of the states
corresponding to asymptotic boundaries but in $\H_{g}$. Projecting
these onto $\H_{g}^{\mcg}$ defines for us the MM states (\ref{eq: ZJn MM state BU}),
for example
\begin{equation}
\ket{Z[\tau_{1}]Z[\tau_{2}]\dots Z[\tau_{n}]}\equiv\lim_{g\rightarrow\infty}\frac{1}{|\mcgg|}\sum_{\gamma\in\mcgg}\U_{\gamma}\ket{0_{g-n},\tau_{1},\dots,\tau_{n}}.
\end{equation}

It should be now clear that the $\alpha$-eigenstates that span $\H_{\text{BU}}$
should be matched with the normalized topological boundary condition
states in the infinite genus limit, i.e.
\begin{equation}
\ket{\alpha}\equiv\lim_{g\rightarrow\infty}\frac{\ket{Z_{\alpha}}}{\sqrt{\braket{Z_{\alpha}}{Z_{\alpha}}}},
\end{equation}
which now truly justifies our choice of label $\alpha$. We remind
the reader that so far we have only claimed in (\ref{eq: orth_condition})
that $\ket{Z_{\alpha}}$ are orthogonal, but we shall prove this in
the next section. The probabilities in our case are then given by
\begin{equation}
p_{\alpha}=\lim_{g\rightarrow\infty}\left(\sum_{\alpha}\frac{1}{\braket{Z_{\alpha}}{Z_{\alpha}}}\right)^{-1}\frac{1}{\braket{Z_{\alpha}}{Z_{\alpha}}}.
\end{equation}
One can also define an operator similar to $\widehat{Z[J]}$ but there
is no explicit need to do so in our case since we already know the
$\alpha$-states a priori and can define (\ref{eq: ZJn MM state BU})
directly in our TQFT Hilbert space.

To summarize, we see that the TQFT gravity approach gives a direct
construction for $\H_{\text{BU}}$ as the MCG invariant subspace $\H_{g}^{\mcg}$
in the large genus limit. Why does this construction work?  One can
argue that the need to go to the limit of large genus is to go to
a Hilbert space that knows (through degeneration) about all possible
topology changes, while the need to project to $\H_{g}^{\mcg}$ is
a manifestation of the mismatch between the partition function of
gravity, which is MCG invariant, versus the partition function of
the TQFT as noted in \cite{Collier:2023Solving3d}. 

We see that the ensemble arises because we start in the no-boundary
HH state in a baby universe Hilbert space with $\dim\H_{\text{BU}}>1$.
As expected in this baby-universe picture, starting from an $\alpha$-state
would lead to a factorizable answer. Note that the identification
of topological boundary conditions (TBCs) with $\alpha$-states from
this factorization perspective is not new and has been mentioned in
\cite{benini2023_FactorizationGlobal,bao2024_QGSymQRG,Torres:2025jcb}.\footnote{See \cite{Torres:2025jcb} for a different approach to the baby-universe
than the approach discussed here.} The nontrivial part that the TQFT gravity framework provides in this
context of TBCs is answering the question of what the baby-universe
Hilbert space is given the fact that states corresponding to TBCs
are not orthogonal in general, nor do they span the Hilbert space
of the TQFT. These states only become orthogonal in the large $g$
limit, and they span the MCG invariant subspace (subject to the completeness
conjecture discussed earlier). Hence, we conclude that 
\begin{equation}
\H_{{\rm BU}}\simeq\lim_{g\rightarrow\infty}\H_{g}^{\mcg}.
\end{equation}

\section{Ensemble weights from symmetries\label{sec:Weights-from-symmetries}}

\subsection{Main derivation}

We want to evaluate the overlap between any two TBC states $\braket{Z_{\jj}}{Z_{\ii}}$
at an arbitrary genus $g$, which corresponds to the overlap of states
of the form given in (\ref{eq: A state}).  For states corresponding
to diagonal condensation, one can choose $m_{ab}^{c;\mu}=1$ for all
permissible junctions \cite{FFIS:2003Ribbon}, and in that case we
find that the squared norm of such a state is the dimension of the
chiral Hilbert space which can be evaluated using (\ref{eq: dim of Hg}).
The squared norm of course in this case is the trace of the identity
operator in the chiral Hilbert space. For general condensations, the
direct evaluation of such overlaps, or traces of products of surface
operators in the chiral part, is generally difficult.  Instead of
this direct way, a more convenient and insightful approach is to use
the SymTFT picture and the construction of $\C$-symmetric 2d TQFTs
\cite{bhardwaj2018_FiniteSymmetries,Huang:2021Construction_2dTQFT}.
We can understand $\braket{Z_{\jj}}{Z_{\ii}}$ from the SymTFT picture
as follows. Here we can put a gapped boundary at the physical boundary,
which defines for us a 2d TQFT that has a symmetry category $\C^{(\alpha)}$.
So we have a 2d TQFT with a partition function given by
\begin{equation}
\mathcal{Z}_{\text{2d TQFT}}=\braket{Z_{\jj}}{Z_{\ii}}.
\end{equation}
Obviously here there is an ambiguity of which boundary to call the
physical boundary versus the symmetry boundary, but one can choose
either of them to play either role and get the same answer. In a general
2d TQFT, one needs to know local operators and their OPEs.\footnote{These OPE coefficients can be obtained from the product and coproduct
junctions of the algebra objects associated with each gapped boundary
as was shown in \cite{Cordova:2024TopCosets} for the case of using
the same boundary condition on both sides.} One can always find an idempotent basis that diagonalizes the OPEs
such that \cite{Huang:2021Construction_2dTQFT}
\begin{equation}
\mathcal{O}_{i}\mathcal{O}_{j}=\frac{1}{\tilde{d}_{i}}\delta_{ij}\mathcal{O}_{i},
\end{equation}
where $\tilde{d}_{i}\equiv<\mathcal{O}_{i}>|_{S^{2}}$ is called the
quantum dimension of $\mathcal{O}_{i}$. Such a basis corresponds
to simple objects in the category of boundary conditions of the 2d
TQFT where $\tilde{d}$ are their quantum dimensions. Using this basis,
one can write the partition function on $\Sigma_{g}$ as 
\begin{equation}
\mathcal{Z}_{\text{2d TQFT}}(\Sigma_{g})=\mathcal{\tilde{D}}^{2g-2}\sum_{i}\tilde{d}_{i}^{2-2g},\label{eq: 2dTQFT part function generic}
\end{equation}
where $\tilde{D}=\sqrt{\sum_{i}\tilde{d}_{i}^{2}}$ and $i$ runs
over the simple local operators. Note that the partition function
of a 2d TQFT is always defined up to an Euler counterterm $\lambda^{2g-2}$
which is unphysical. Here we conveniently choose the Euler counterterm
such that $\mathcal{Z}_{\text{2d TQFT}}(S^{2})=1$ to be consistent
with our chosen 3d TQFT normalization and to easily determine the
relative Euler counterterm between different $\C$-symmetric TQFTs
which is a physical quantity unlike the absolute version. One can
always absorb the global Euler counterterm into the definition of
$\tilde{d}$, but for our purposes we want to choose a normalization
such that the quantum dimensions $\tilde{d}\geq1$, and hence we are
keeping the global $\mathcal{\tilde{D}}^{2g-2}$ factor explicit.

Let us first focus on the case where both boundaries of the SymTFT
sandwich have the same boundary condition. This will give us what
is called the \emph{regular $\cal{C}$-symmetric} 2d TQFT. In a regular
$\C$-symmetric TQFT, the quantum dimensions of the idempotent basis
operators are simply given by the quantum dimensions of the topological
lines of $\C$ themselves \cite{Huang:2021Construction_2dTQFT}. So
by knowing $\C$, we can calculate the partition function directly
from the quantum dimension data of $\C$. So we simply get
\begin{equation}
\mathcal{Z}_{\C}=\mathcal{D_{\C}}^{2g-2}\sum_{a\in\C}d_{a}^{2-2g}.\label{eq: 2dTQFT_part2}
\end{equation}
Thus, for a topological boundary condition $\ket Z$ that gives rise
to a symmetry category $\C$, the squared norm of the state $\braket ZZ|_{\Sigma_{g}}$
is given by (\ref{eq: 2dTQFT_part2}), where now our Euler counterterm
normalization reflects the fact that in our 3d TQFT we have $\braket ZZ|_{\Sigma_{0}\simeq S^{2}}=1$.

Now as $g\rightarrow\infty$, the sum will be dominated by the invertible
simple objects ($d=1$) and hence we have up to leading order
\begin{equation}
\braket{Z_{\ii}}{Z_{\ii}}|_{\Sigma_{g\ggg1}}\approx\D_{\T}^{g-1}|{\rm Inv}(\C^{(\alpha)})|,
\end{equation}
where we used the fact that $D_{\C}^{2}=D_{\T}$ since $\T$ is the
Drinfeld center of $\C$, and ${\rm Inv}(\C^{(\alpha)})$ denotes
the group of invertible simple objects in $\C^{(\alpha)}$. \footnote{This is sometimes also known as the Picard group of $\C^{(\alpha)}$
, see for example \cite{Fuchs:2004_simple_currents}.} From this we get our main result
\begin{equation}
\D_{\T}^{g-1}\coeff_{\ii}=\frac{1}{|{\rm Inv}(\C^{(\alpha)})|}.\label{eq: weights main equation}
\end{equation}

Now we want to return back to the proof of (\ref{eq: orth_condition})
which leads to the orthogonality of the projectors $P_{\alpha}$.
This involves evaluating $\braket{Z_{\jj}}{Z_{\ii}}$ which from the
above discussion can be considered a partition function of a $\C^{(\alpha)}$-symmetric
2d TQFT albeit not the \emph{regular} one. It was shown in \cite{bhardwaj2018_FiniteSymmetries,Huang:2021Construction_2dTQFT}
that any $\C$-symmetric 2d TQFT can be obtained from the regular
one by gauging an algebra object in $\C$. If we denote the algebra
object by $A'$, then the category of boundary conditions in the
gauged theory is given by the category of right $A'$ modules, $\C_{A'}$.
 We can show that in our chosen normalization 
\begin{equation}
\D_{\C}^{2}=\dim A'_{\text{min}}\,\D_{\C_{A'}}^{2},\label{eq: total quantum dim for gauged 2d}
\end{equation}
where $A'_{\text{min}}$ is a Morita representative of the Morita
class of $A'$ with the minimum quantum dimension (see appendix \ref{sec: FPdim of module categories}
for more details). With this, we have the leading order in the large
$g$ limit as
\begin{equation}
\frac{\braket{Z_{\ii}}{Z_{\jj}}}{\sqrt{\braket{Z_{\ii}}{Z_{\ii}}\braket{Z_{\jj}}{Z_{\jj}}}}\Bigg\vert_{g\ggg1}\approx\frac{1}{(\dim A'_{\text{min}})^{g-1}}\frac{\xi_{\C_{A'}}}{\sqrt{|{\rm Inv}(\C^{(\alpha)})||{\rm Inv}(\C^{(\alpha')})|}},\label{eq: general overlaps inf g}
\end{equation}
where $\xi_{\C_{A'}}$ denotes the number of simple objects in $\C_{A'}$
with unit quantum dimension in our normalization.\footnote{The category $\C_{A'}$ in general is not necessarily a tensor category
and so we cannot always speak of ``invertible'' objects. Instead,
what we want is the objects of dimension 1 in our chosen normalization
for the dimensions. When $\C$ is braided and $A'$ is a commutative
special symmetric Frobenius algebra object then $\C_{A'}$ is a tensor
category. See appendix \ref{sec: FPdim of module categories} for
more details.}  Since $\dim A'_{\text{min}}>1$ for any nontrivial algebra object,
(\ref{eq: orth_condition}) will be satisfied. This also tells us
that at any finite $g$, the lowest bound on the suppression in $g$
is $O(2^{1-g})$ since the lowest nontrivial algebra would have $\dim A'_{\text{min}}=2$.

Finally, we note that so far we have been working with the normalized
basis $\braket{0_{g}}{0_{g}}=1$, which leads to (\ref{eq: weights main equation}).
Writing the holographic duality (\ref{eq: duality1}) with the schematic
sum over topologies now takes the form 
\begin{equation}
\D_{\T}^{g-1}\sum_{{\rm topologies}}\Psi_{0}(\Omega)=\sum_{\alpha}\frac{1}{|{\rm Inv}(\C^{(\alpha)})|}Z_{\alpha}(\Omega).
\end{equation}
Absorbing the factor of $\D_{\T}^{g-1}$ into the TQFT wavefunction
restores the natural normalization of the TQFT where $\braket{0_{g}}{0_{g}}=\D_{\T}^{g-1}$
instead of $\braket{0_{g}}{0_{g}}=1$. Note that this is an overall
normalization regardless of the genus reduction to any genus $\tilde{g}$
due to the fact that all our calculations are embedded in $\H_{g}$
for large fixed $g$. Hence, the natural normalization leads to
\begin{equation}
\sum_{M,\partial M=\Sigma_{\tilde{g}}}\mu(M){\bf Z}_{\T}(M;\Omega_{\tilde{g}})=\sum_{\alpha}\frac{1}{|{\rm Inv}(\C^{(\alpha)})|}Z_{\alpha}(\Omega_{\tilde{g}}),\label{eq: duality sum over top full measure}
\end{equation}
where the LHS is a sum over all connected orientable 3-manifolds $M$
with boundary $\Sigma_{\tilde{g}}$ labeled by an element of the double-coset
space we discussed in section \ref{subsec:The-sum-over topologies},
${\bf Z}_{\T}(M;\Omega_{\tilde{g}})$ is the naturally normalized
TQFT partition function on $M$, and the measure $\mu(M)$ is the
Goeritz group measure (\ref{eq: Goeritz measure}). 

\subsection{Algebra automorphisms\label{subsec:Algebra-automorphisms}}

As explained in section \ref{subsec:SymTFT-picture}, for a given
Lagrangian algebra $\A$ in $\T$, the category of right $\A$-modules
$\T_{\A}$ is the category of topological defect lines (TDLs) that
preserve the vertex algebra of $\T$. So our ensemble average when
written as a sum over algebra objects can be written as
\begin{equation}
\sum_{\A}\frac{1}{|\Inv(\T_{\A})|}Z_{\A}(\Omega).
\end{equation}
The group $\Inv(\T_{\A})$ is known to be isomorphic to the automorphism
group of the Lagrangian algebra $\A$ \cite{davydov2011_InvertibleDefects},
and so the ensemble average can be written as
\begin{equation}
\sum_{\A}\frac{1}{|\Aut(\A)|}Z_{\A}(\Omega).\label{eq: AutA_sum}
\end{equation}
Before we discuss this formula further, let us elaborate on what $\Aut(\A)$
means here. Any given simple object $a$ in the category $\T$, being
an irreducible representation of the vertex algebra, has endomorphisms
${\rm End}(a)\simeq\mathbb{C}{\bf id}_{a}$. Since the object $\A$
in $\T$ is a direct sum of simple objects, $\A=\bigoplus_{a}n_{a}a$,
its endomorphisms are ${\rm End}(\A)\simeq\bigoplus_{a}{\rm Mat_{n_{a}}(\mathbb{C}){\bf id}_{a}}$.
The group $\Aut(\A)$ is then defined as the invertible endomorphisms
of object $\A$ which preserve the algebra structure, i.e.~they preserve
the product morphism $m$ and the unit morphism $\eta$. In other
words, $\varphi\in\Aut(\A)$ if $m\circ(\varphi\otimes\varphi)=\varphi\circ m$
and $\varphi\circ\eta=\eta$ \cite{Fuchs:2005qv}. In the CFT picture,
the simple anyons are primaries of the vertex algebra associated to
$\T$, and so $\varphi$ would correspond to a linear transformation
on $\H_{\text{CFT}}$, acting on operators by conjugation, which preserves
the vacuum primary and the OPE coefficients of these primary operators.
This directly means that this is an invertible symmetry transformation
that commutes with the vertex algebra.

For multiplicity free Lagrangian algebras, i.e.~$n_{a}\in\{0,1\}$,
one needs to only know the allowed nonzero $m_{ab}^{c}$ without explicitly
knowing their value since in this case the actual value drops out
of the conditions on the candidate automorphism map $\vphi$. In other
words, for a map that acts by 
\begin{equation}
\varphi=\bigoplus_{a}\varphi_{a}{\bf id}_{a},\qquad\varphi_{a}\in\mathbb{C}^{\times},\qquad\varphi_{0}=1,
\end{equation}
the condition $m\circ(\varphi\otimes\varphi)=\varphi\circ m$ simply
reads
\begin{equation}
\varphi_{a}\varphi_{b}=\varphi_{c},\qquad m_{ab}^{c}\neq0.\label{eq: vphi aut condition}
\end{equation}
The possible solutions to these equations are basically elements
of the group characters of the Abelian subgroup of the fusion ring
induced from the nonzero $m_{ab}^{c}$. For example, in diagonal Lagrangian
algebras $\A=\bigoplus_{a}(a,\bar{a})$, the product junctions can
be written as $m_{(a,\bar{a})(b,\bar{b})}^{(c,\bar{c});(\mu,\bar{\mu})}=1$
for $N_{ab}^{c}\neq0$, where $N_{ab}^{c}$ are the fusion coefficients
of the chiral part \cite{FFIS:2003Ribbon}. So elements of $\Aut(\A)$
in this case are given by the invertible lines of the chiral part.
Let us label them by $\vphi^{(i)}\in\Aut(\A)$, then from the Verlinde
formula the explicit solutions of (\ref{eq: vphi aut condition})
are given by
\begin{equation}
\vphi_{a}^{(i)}=\frac{S_{ia}}{S_{0a}},
\end{equation}
where $i$ denotes an invertible anyon in the chiral part. We see
that this is the action of the invertible Verlinde lines as expected
in a diagonal RCFT.

Now let us return to eq. (\ref{eq: AutA_sum}). This is a sum over
isomorphism classes of Lagrangian algebras, so this is the natural
measure on the groupoid of Lagrangian algebras of $\T$, and hence,
this is a uniform measure up to isomorphisms. The holographic duality
(\ref{eq: duality sum over top full measure}) then gives us a generalization
of the Siegel-Weil formula where in this case $\A$ is the analog
of the lattice and $Z_{\A}$ is the analog of the lattice $\Theta$
function. We will elaborate more on the Siegel-Weil formula in section
\ref{subsec:Virasoro-TQFT}. The mass formula, or in the groupoid
language the cardinality of the groupoid, then gives us the vacuum
to vacuum amplitude (\ref{eq: sum over closed manifolds Aut}) as
(if we restore the natural TQFT normalization $\braket{0_{g}}{0_{g}}=\D^{g-1}$)
\begin{equation}
\braket{\text{HH}}{\text{HH}}=\sum_{\A}\frac{1}{|\Aut(\A)|}.
\end{equation}

In some cases we can have more than one topological boundary condition
giving rise to the same boundary CFT because of a hidden duality in
the physical boundary condition $\ket{\Omega}$. In other words, the
state $\ket{\Omega}$ for any genus can be invariant under a subgroup
$G$ of the anyon permutation symmetry and hence we can have $\braket{\Omega}{Z_{\A_{I}}}=\braket{\Omega}{Z_{\A_{J}}}$
for $\A_{J}=h\cdot\A_{I}$ for any element $h\in G$. For a given
Lagrangian algebra $\A$, if we now denote its categorical automorphism
group discussed earlier by $\Aut_{\T}(\A)$, and define its stabilizer
group with respect to $G$ as $\Aut_{G}(\A)$, then the automorphism
group that preserves $Z_{\A}(\Omega)$ satisfies the following short
exact sequence
\begin{equation}
1\rightarrow\Aut_{\T}(\A)\rightarrow\Aut(Z_{\A}(\Omega))\rightarrow\Aut_{G}(\A)\rightarrow1.
\end{equation}
If we want to restrict the sum in (\ref{eq: AutA_sum}) to equivalence
classes of $Z_{\A}(\Omega)$ we get
\begin{equation}
|G|\sum_{[A]_{\sim G}}\frac{1}{|\Aut(Z_{\A}(\Omega))|}Z_{\A}(\Omega).
\end{equation}
 From the CFT perspective, $\Aut_{\T}(\A)$ is the group of invertible
topological defect lines that commute with the vertex algebra, while
$\Aut_{G}(\A)$ corresponds to the group of topological line defects
that act on the vertex algebra by an outer-automorphism but leave
the presentation of the theory in terms of the vertex algebra invariant
(i.e.~ they do not change the choice of the Lagrangian algebra object
we choose to describe the CFT). For example if we take a rational
compact boson with a partition function $\sum_{\lambda}|\chi_{\lambda}|^{2}$,
where $\chi_{\lambda}$ are characters of ${\rm U}(1)_{k}$ Chern-Simons
theory, the charge conjugation symmetry (reflection symmetry) on the
right-moving copy changes this presentation into the T-dual presentation
of the theory $\sum_{\lambda}\chi_{\lambda}\bar{\chi}_{-\lambda}$
so it is not an element of the $\Aut_{G}(\A_{\text{diag.}})$ group
for the diagonal Lagrangian algebra $\A_{\text{diag.}}$ in this case.
On the other hand, the simultaneous conjugation of the left- and right-moving
copies is an example of an element of $\Aut_{G}(\A_{\text{diag.}})$.

Note that this extra equivalence between algebra objects --- or topological
boundary conditions --- induced from the choice of gapless boundary
condition is not a priori and so we will not deem it as fundamental
unless there are no gapless boundary conditions that can distinguish
such topological boundary conditions. In that case, from the baby
universe Hilbert space, the equivalence becomes like a gauge redundancy
where the operator $\widehat{Z[J]}$ will have a smaller subset of
distinguishable eigenstates than the set of topological boundary conditions,
i.e. $\H_{{\rm BU}}$ is a quotient of $\H_{g\rightarrow\infty}^{{\rm MCG}}$
in that case. In the Virasoro case this will not be an issue to worry
about since Virasoro characters are distinct as modular functions
unlike the case of some rational chiral algebras where we can have
for example identical characters $\chi_{a}(\tau)=\chi_{f(a)}(\tau)$
for some chiral primaries related by an outer-automorphism $f$.

\section{Examples\label{sec:Examples}}

\subsection{Warm up: Abelian TQFTs\label{subsec:Warm-up:-Abelian}}

In \cite{dymarsky2025_TQFTGravity}, the general Abelian case was
considered in detail and it was shown that the resulting ensemble
has equal weights. In this section, we briefly review this result
in light of the relation between weights and symmetries.

For Abelian TQFTs, all the anyons are invertible and hence form an
Abelian group under fusion. Any Abelian TQFT can be described (although
non-uniquely) by an Abelian Chern-Simons theory with an action given
by
\begin{equation}
S_{CS}=\frac{1}{4\pi}\int_{\M}K_{IJ}A^{I}\wedge\d A^{J},
\end{equation}
with $D$ gauge fields $A^{I}$ with a torus gauge group $\mathbb{R}^{D}/\Lambda$,
and $K$ is an even, non-degenerate integer-valued bilinear form.
The anyons are given by elements of the discriminant group $\sD\equiv\Lambda^{\perp}/\Lambda$
where $\Lambda^{\perp}$ is the dual lattice with respect to the bilinear
form $K$. The twist and braiding (monodromy) of anyons are all determined
by $K$ as follows
\begin{equation}
\theta_{a}=\e^{\pi\i K^{IJ}a_{I}a_{J}},\quad B_{ab}=\e^{2\pi\i K^{IJ}a_{I}b_{J}}.
\end{equation}

The bilinear form $K$ induces a quadratic form ${\bf q}:\sD\rightarrow\mathbb{Q}/\mathbb{Z}$,
where ${\bf q}$ is given explicitly by $\frac{1}{2}K^{-1}\mod 1$.
Lagrangian anyon condensation corresponds to gauging a Lagrangian
subgroup $\mathscr{L}\subset\sD$ (with respect to the induced quadratic
form on $\sD$). This is related to the construction of code CFTs
where $\sL$ can be regarded as an even self-dual code over $\sD$
as was shown in \cite{barbar2025_GlobalSymmetries}. The corresponding
Lagrangian algebra is just a direct sum of anyons of the Lagrangian
subgroup
\begin{equation}
\A_{\sL}=\bigoplus_{\ell\in\sL}\ell.
\end{equation}
The symmetry lines of the boundary theory are the confined anyons
in the condensed phase; these are the anyons that braid nontrivially
with $\sL$ modulo identifications by $\sL$. So these are given by
the group $\sD/\sL\simeq\hat{\sL}$, where $\hat{\sL}\equiv\Hom(\sL,U(1))$
is the Pontryagin dual of $\sL$. This is indeed the same as $\Aut(\A_{\sL})$
from our discussion in section \ref{subsec:Algebra-automorphisms}.
This leads to an ensemble of equal weights.

As alluded to before, in some cases the different topological boundary
conditions can lead to the same CFT on the boundary. In this case
we would have equivalences between Lagrangian subgroups, which are
code equivalences in the code language. If we choose a representative
of each equivalence class $[\sL]$, then we can write the unnormalized
average as 
\begin{equation}
<Z>'=\sum_{[\sL]}\frac{1}{|\Aut(\sL)|}Z_{\sL},\label{eq: code equivalence ensemble}
\end{equation}
where $\Aut(\sL)$ denotes the stabilizer under the group that implements
code equivalences. Also, from now on we will use the primed $<Z>'$
notation to denote the unnormalized version of the averaged partition
function.

Let us illustrate this with a simple example: ${\rm U}(1)_{8}\times\overline{{\rm U}(1)}_{8}$
Chern-Simons theory. Let us label the anyons by the pair $(a,b)\in\mathbb{Z}_{8}\times\mathbb{Z}_{8}$.
The quadratic form is given by ${\bf q}((a,b))=\frac{1}{16}(a^{2}-b^{2})\mod 1$.
There are three Lagrangian subgroups given by
\begin{equation}
\sL_{1}=\{(0,0),(1,1),(2,2),(3,3),(4,4),(5,5),(6,6),(7,7)\}\simeq\mathbb{Z}_{8},
\end{equation}
\begin{equation}
\sL_{1}'=\{(0,0),(1,7),(2,6),(3,5),(4,4),(5,3),(6,2),(7,1)\}\simeq\mathbb{Z}_{8},
\end{equation}
\begin{equation}
\sL_{2}=\{(0,0),(0,4),(4,0),(4,4),(2,2),(2,6),(6,2),(6,6)\}\simeq\mathbb{Z}_{2}\times\mathbb{Z}_{4}.
\end{equation}

Now consider the bulk anyon permutation symmetry\footnote{For our purposes it suffices to restrict to anyon permutations that
preserve chirality.} $\mathbb{Z}_{2}^{L}\times\mathbb{Z}_{2}^{R}$, where each $\mathbb{Z}_{2}$
is just the charge conjugation on the left and right copies of ${\rm U}(1)_{8}$
respectively. Consider the boundary condition (in the convention of
(\ref{eq: tau_holomorphic boundary}))
\begin{equation}
\bra{\tau,\xi}=\sum_{(a,b)}\chi_{a}(\tau,\xi)\bar{\chi}_{b}(\bar{\tau},\bar{\xi})\bra{a,b},
\end{equation}
where 
\begin{equation}
\chi_{a}(\tau,\xi)=\frac{1}{\eta(\tau)}\sum_{n\in\mathbb{Z}}q^{\frac{1}{2k}(kn+a)^{2}}\e^{2\pi\i\xi(kn+a)},\qquad q\equiv\e^{2\pi\i\tau}\label{eq: u(1)_k character}
\end{equation}
 denotes the character of the chiral algebra of ${\rm U}(1)_{k}$
with flavor $\xi$ denoting the value of the ${\rm U}(1)$ gauge field
on the boundary. In our case $\sL_{1}$ and $\sL_{1}'$ correspond
to compact bosons of radii $R=2$ and $R=\frac{1}{2}$ represented
in terms of $\chi_{a}(\tau,\xi)$, while $\sL_{2}$ corresponds to
the self-dual boson at $R=1$. So our ensemble is given by
\begin{equation}
<Z(\tau,\xi)>'\sim Z_{R=2}(\tau;\xi,\bar{\xi})+Z_{R=1/2}(\tau;\xi,\bar{\xi})+Z_{R=1}(\tau;\xi,\bar{\xi}).
\end{equation}
Note that $R=2$ and $R=\frac{1}{2}$ are T-dual radii (we are using
the convention $R\leftrightarrow\frac{1}{R}$); however with nonzero
flavor $\xi$, T-duality acts by
\begin{equation}
Z_{R}(\tau;\xi,\bar{\xi})=Z_{\frac{1}{R}}(\tau;\xi,-\bar{\xi}),
\end{equation}
or at the level of characters as
\begin{equation}
\bar{\chi}_{a}(\bar{\tau},\bar{\xi})=\bar{\chi}_{-a}(\bar{\tau},-\bar{\xi}).
\end{equation}
Hence, our ensemble with the boundary condition $\ket{\tau,\xi}$
distinguishes between partition functions for $\sL_{1}$ and $\sL_{1}'$.
If instead we use a boundary condition with $\xi=0$, $\sL_{1}$ and
$\sL_{1}'$ will give the same partition function and so if we want
to restrict to equivalence classes we would have two orbits, with
representatives $\sL_{1}$ and $\sL_{2}$. Their automorphisms under
$\mathbb{Z}_{2}^{L}\times\mathbb{Z}_{2}^{R}$ are $\mathbb{\Aut}(\sL_{1})=\mathbb{Z}_{2}$
and $\mathbb{\Aut}(\sL_{2})=\mathbb{Z}_{2}\times\mathbb{Z}_{2}$.
Writing the average over equivalence classes we evidently get, up
to an overall factor, 
\begin{equation}
<Z>'\sim Z_{R=2}+\frac{1}{2}Z_{R=1}.
\end{equation}

From the CFT perspective, both theories have the same size for the
symmetry group that commutes with the ${\rm U}(1)_{8}$ algebra. However,
the $R=2$ boson presentation in terms of the primaries of ${\rm U}(1)_{8}$
changes under the right-moving $\mathbb{Z}_{2}$ charge conjugation
symmetry into its T-dual version, while the $R=1$ boson is invariant.
This is why T-duality contributes towards the weight of the self-dual
boson here giving rise to a relative weight of $1/2$.

\subsection{Illustrative example: ${\rm SU}(2)_{4}\times\overline{{\rm SU}(2)_{4}}$}

${\rm SU}(2)_{4}$ has 5 anyons $L_{\lambda}$ labeled by the Dynkin
labels $\lambda\in\{0,1,2,3,4\},$ where $L_{0}$ is the vacuum. The
nontrivial fusion rules are given by
\begin{align}
 & L_{4}\otimes L_{\lambda}=L_{4-\lambda}\\
 & L_{1}\otimes L_{1}=L_{0}\oplus L_{2}, & L_{1}\otimes L_{2}=L_{1}\oplus L_{3}\qquad & L_{1}\otimes L_{3}=L_{2}\oplus L_{4},\\
 & L_{2}\otimes L_{2}=L_{0}\oplus L_{2}\oplus L_{4}, & L_{2}\otimes L_{3}=L_{1}\oplus L_{3}\qquad & L_{3}\otimes L_{3}=L_{0}\oplus L_{2}.
\end{align}
The quantum dimensions are $d_{0}=d_{4}=1$, $d_{1}=d_{3}=\sqrt{3}$
and $d_{2}=2$, giving rise to a total quantum dimension $\mathcal{D}_{{\rm SU}(2)_{4}}=\sqrt{12}$.
The spins of the anyons are given by $h_{\lambda}=\frac{\lambda(\lambda+2)}{24}.$

The doubled ${\rm SU}(2)_{4}$ Chern-Simons theory has two gapped
boundaries corresponding to the A-invariant (${\rm SU}(2)_{4}$ WZW
CFT) $Z_{A}$ and the D-invariant (${\rm SU}(3)_{1}$ WZW CFT) $Z_{D}$
which corresponds to the conformal embedding $su(2)_{4}\subset su(3)_{1}$.
The VOA of the bulk TQFT is basically just left and right copies of
$su(2)_{4}$. So according to our result, the weights should be inversely
proportional to the order of the invertible symmetry group that commutes
with this VOA. For $Z_{A}$, these are just the invertible lines from
the $su(2)_{4}$ Verlinde lines, which form a $\mathbb{Z}_{2}$ group.
For $Z_{D}$, we have the $\mathbb{Z}_{3}$ from the Verlinde lines
of $su(3)_{1}$ which also commutes with $su(2)_{4}\subset su(3)_{1}$.
We also have charge conjugation which does not commute with $su(3)_{1}$
because it exchanges the ${\bf 3}$ and $\bar{{\bf 3}}$ representations
of $su(3)_{1}$ but commutes with $su(2)_{4}\subset su(3)_{1}$ since
${\bf 3}$ and $\bar{{\bf 3}}$ project onto the same representation
$L_{2}$ in $su(2)_{4}$. So our full group for $Z_{D}$ is $S_{3}\simeq\mathbb{Z}_{3}\rtimes\mathbb{Z}_{2}$.

We now want to show how these appear explicitly from the symmetry
category of each topological boundary condition. Let us label the
topological boundary condition states in the TQFT Hilbert space as
$\ket{Z_{A}}$ and $\ket{Z_{D}}$ respectively. The corresponding
Lagrangian algebra objects are given by
\begin{align}
\mathcal{A}_{A} & =L_{0\bar{0}}\oplus L_{1\bar{1}}\oplus L_{2\bar{2}}\oplus L_{3\bar{3}}\oplus L_{4\bar{4}},\\
\mathcal{A}_{D} & =L_{0\bar{0}}\oplus L_{0\bar{4}}\oplus L_{4\bar{0}}\oplus L_{4\bar{4}}\oplus2L_{2\bar{2}}.
\end{align}
Condensing the diagonal invariant algebra gives us the symmetry category
as ${\rm SU}(2)_{4}$ as expected from diagonal condensation. This
gives us 
\begin{align}
\braket{Z_{A}}{Z_{A}}|_{\Sigma_{g}} & =\mathcal{D}_{{\rm SU}(2)_{4}}^{2g-2}\sum_{a\in{\rm SU}(2)_{4}}d_{a}^{2-2g}\nonumber \\
 & =(12)^{g-1}(2+2(3)^{1-g}+2^{2-2g}).
\end{align}
For the D invariant, its symmetry category $\C_{D}$ is a $\mathbb{Z}_{2}$
extension of the Tambara-Yamagami category ${\rm TY}(\mathbb{Z}_{3})$
\cite{TAMBARA1998692}. This $\C_{D}$ category has 6 invertible anyons
forming $S_{3}$ fusion rules, and two anyons $\sigma_{1}$ and $\sigma_{2}$
of quantum dimension $\sqrt{3}$. Let us denote the $S_{3}$ anyons
by the following parametrization 
\begin{equation}
S_{3}=<r,s|r^{3}=s^{2}=(sr)^{2}=1>,
\end{equation}
then the other key fusion rules are
\begin{align}
r\otimes\sigma_{i} & =\sigma_{i}, & \sigma_{1}\otimes\sigma_{2} & =s\oplus sr\oplus sr^{2},\\
s\otimes\sigma_{1} & =\sigma_{2}, & \sigma_{i}\otimes\sigma_{i} & =1\oplus r\oplus r^{2}.
\end{align}

To briefly see how we get $\C_{D}$, we can do the full condensation
of $\A_{D}$ in multiple Abelian steps while keeping track of the
fusion category that includes the confined anyons. We will call this
category the domain wall category for simplicity. We start by condensing
${\rm SU}(2)_{4}$ to ${\rm SU}(3)_{1}$ by condensing the $L_{4}$
anyon, which by the condensation rules of Bais and Slingerland \cite{Bais:2008_anyon_cond}
leads to identifications under $L_{4}$ fusion as $L_{0}\sim L_{4}$,
$L_{1}\sim L_{3}$, and splits the fixed point $L_{2}$ into two anyons
$L_{2}^{+}$ and $L_{2}^{-}$. These four sectors form the domain
wall category ${\rm TY}(\mathbb{Z}_{3})$ which lives at the interface
between ${\rm SU}(2)_{4}$ and ${\rm SU}(3)_{1}$. Hence, for ${\rm SU}(2)_{4}\times\overline{{\rm SU}(2)}_{4}$
going to ${\rm SU}(3)_{1}\times\overline{{\rm SU}(3)}_{1}$, we get
${\rm TY}(\mathbb{Z}_{3})\times\overline{{\rm TY}(\mathbb{Z}_{3})}$
as our domain wall category. Now we want to condense ${\rm SU}(3)_{1}\times\overline{{\rm SU}(3)}_{1}$
(which is an Abelian phase) to the trivial phase by condensing the
diagonal $\mathbb{Z}_{3}$. On the domain wall side, this leads to
the quotient $\frac{{\rm TY}(\mathbb{Z}_{3})\times\overline{{\rm TY}(\mathbb{Z}_{3})}}{\mathbb{Z}_{3}^{{\rm diag.}}},$
which again by the rules of \cite{Bais:2008_anyon_cond} can be shown
to be some $\mathbb{Z}_{2}$ extension of ${\rm TY}(\mathbb{Z}_{3})$.
This must be a nontrivial $\mathbb{Z}_{2}$ extension in order for
the Drinfeld center of this category to match with ${\rm SU}(2)_{4}\times\overline{{\rm SU}(2)}_{4}$.
One can verify that $\mathcal{Z}_{\text{Drinfeld}}(\mathcal{C}_{D})={\rm SU}(2)_{4}\times\overline{{\rm SU}(2)}_{4}$
\cite{Vanhove:2021_C_S3,Ren:2024_C_S3} as required.

With this $\mathcal{C}_{D}$ symmetry category we get
\begin{align}
\braket{Z_{D}}{Z_{D}}|_{\Sigma_{g}} & =\mathcal{D}_{\C_{D}}^{2g-2}\sum_{a\in\C_{D}}d_{a}^{2-2g}\nonumber \\
 & =(12)^{g-1}(6+2(3)^{1-g})\nonumber \\
 & =2^{2g-1}(3^{g}+1).
\end{align}
One could also calculate this squared norm through the intermediate
condensation to the Abelian phase ${\rm SU}(3)_{1}\times\overline{{\rm SU}(3)}_{1}$
similar to the method utilized in \cite{barbar2025_GlobalSymmetries}.
This phase has $\mathbb{Z}_{3}\times\mathbb{Z}_{3}$ fusion rules.
There are two invariants related by charge conjugation on one chiral
copy. They correspond to the even self-dual codes $C_{1}=\{00,11,22\}$
and $C_{2}=\{00,12,21\}$. Both of these codes correspond to the D
invariant in the parent phase (they are code equivalent). The operator
$\mathcal{I}$ defining the induced inner product on ${\rm SU}(3)_{1}$
after condensing from ${\rm SU}(2)_{4}$ is an unnormalized projector
on the charge conjugation invariant space. This can be written explicitly
as
\begin{equation}
\mathcal{I}=2^{g-1}(I+\mathcal{S}_{\mathbb{Z}_{2}^{\text{c.c.}}}),
\end{equation}
where $\mathcal{S}_{\mathbb{Z}_{2}^{\text{c.c.}}}$ is the surface
operator that implements the charge conjugation 0-form symmetry. Applying
this to our ${\rm SU}(3)_{1}\times\overline{{\rm SU}(3)}_{1}$ phase,
we get
\begin{align}
\braket{Z_{D}}{Z_{D}}|_{\Sigma_{g}} & =2^{2g-2}\bra{Z_{C_{1}}}I\otimes I+I\otimes\mathcal{S}_{\mathbb{Z}_{2}^{\text{c.c.}}}+\mathcal{S}_{\mathbb{Z}_{2}^{\text{c.c.}}}\otimes I+\mathcal{S}_{\mathbb{Z}_{2}^{\text{c.c.}}}\otimes\mathcal{S}_{\mathbb{Z}_{2}^{\text{c.c.}}}\ket{Z_{C_{1}}}\nonumber \\
 & =2^{2g-2}(3^{g}+1+1+3^{g})\nonumber \\
 & =2^{2g-1}(3^{g}+1),
\end{align}
which matches what we got from the 2d TQFT calculation.

With these results for $\braket{Z_{A}}{Z_{A}}$ and $\braket{Z_{D}}{Z_{D}}$,
we see that as $g\rightarrow\infty$ we indeed get $\coeff_{A}\sim\frac{1}{|\mathbb{Z}_{2}|}$
and $\coeff_{D}\sim\frac{1}{|S_{3}|}$. Normalizing the weights to
sum to one, we get
\begin{equation}
<Z>=\frac{3}{4}Z_{A}+\frac{1}{4}Z_{D}.
\end{equation}

For completeness, we give a demonstration of eq. (\ref{eq: general overlaps inf g})
in this simple example. To get the overlap $\braket{Z_{D}}{Z_{A}}$,
we can start from the regular ${\rm SU}(2)_{4}$-symmetric 2d TQFT
and gauge the algebra object $0\oplus4$. This is exactly like the
domain wall story in the condensation of $L_{4}$ in ${\rm SU}(2)_{4}$,
so the gauged 2d TQFT will have $\mathrm{TY(\mathbb{Z}_{3})}$ as
its category of boundary conditions, which has $\D_{\mathrm{TY}(\mathbb{Z}_{3})}^{2}=6=\frac{1}{2}\D_{{\rm SU}(2)_{4}}^{2}$.
The full overlap is given by
\begin{align}
\braket{Z_{D}}{Z_{A}}|_{\Sigma_{g}} & =\mathcal{D}_{\mathrm{TY}(\mathbb{Z}_{3})}^{2g-2}\sum_{a\in\mathrm{TY}(\mathbb{Z}_{3})}d_{a}^{2-2g}\nonumber \\
 & =6^{g-1}\left(3+3^{1-g}\right),
\end{align}
which agrees with (\ref{eq: general overlaps inf g}). Similarly,
we could have started from the regular $\C_{D}$-symmetric 2d TQFT
and gauged the algebra corresponding to the $\mathbb{Z}_{2}$ reflection
of the $S_{3}$ invertible part to get $\mathrm{TY(\mathbb{Z}_{3})}$.

\subsection{$\mathbb{Z}_{2}$ orbifold of the compact boson\label{subsec: Z2_orbifold}}

The chiral algebra of the $\mathbb{Z}_{2}$ orbifold of the compact
boson of rational radius corresponds to ${\rm O}(2)_{2N}$ Chern-Simons
theory which is obtained by gauging the charge conjugation 0-form
symmetry in ${\rm U}(1)_{2N}$. Let us start from the data of ${\rm U}(1)_{2N}$.
We have $2N$ Abelian lines forming the $\mathbb{Z}_{2N}$ group under
fusion, so we can label them as $\phi_{\lambda}$ where $\lambda\in\mathbb{Z}_{2N}$.
The spins are given by $h_{\lambda}=\frac{\lambda^{2}}{4N}$. This
quadratic form is preserved by $\lambda\rightarrow\omega\lambda\mod{2N}$
where
\begin{equation}
\omega^{2}=1\mod{4N,\qquad\omega\in\mathbb{Z}_{2N}.}
\end{equation}
These transformations form  the group of anyon permutation symmetries
of ${\rm U}(1)_{2N}$. The modular invariants in the doubled theory
are classified by factors of $N$. So for each factor $\delta$, we
get a compact boson with a squared radius $R^{2}=\frac{\delta^{2}}{N}$.
These correspond to surface operators in the chiral ${\rm U}(1)_{2N}$
\cite{fuchs2002_TFTConstruction,kapustin2010_SurfaceOperators,Roumpedakis:2022_higher_gauging}.
Surface operators corresponding to $\delta$ such that $\gcd(\delta,\frac{N}{\delta})\neq1$
are non-invertible. Note that ${\rm U}(1)_{2N}$ distinguishes between
a compact boson and its T-dual since the charge conjugation permutes
anyons of ${\rm U}(1)_{2N}$ acting as $\lambda\leftrightarrow2N-\lambda$.
Let us also focus on the case when $N$ is square free such that all
modular invariants are in the same orbit with respect to the anyon
permutation group.  This does not matter that much for the $\mathrm{U}(1)_{2N}$
case but will be important later when we discuss $\mathrm{O}(2)_{2N}$
and its orbits in that case. The ensemble average for ${\rm U}(1)_{2N}$
can be written as
\begin{equation}
<Z>'=\sum_{\substack{pq=N\\
{\rm gcd}(p,q)=1
}
}\frac{1}{2N}Z_{R^{2}=\frac{p}{q}}^{\text{circ.}},
\end{equation}
where all radii have the same order $2N$ for their automorphism
groups\footnote{The actual structure of the automorphism group for the $R^{2}=\frac{p}{q}$
boson here is inherited from its Lagrangian subgroup, which is generated
by the $(q,\bar{q})$ and $(p,-\bar{p})$ anyons. This gives us $\mathbb{Z}_{2p}\times\mathbb{Z}_{q}$
for $q$ odd or $\mathbb{Z}_{p}\times\mathbb{Z}_{2q}$ for odd $p$.} as expected in the Abelian case. Note that at this point we are keeping
$R$ and $\frac{1}{R}$ as distinct members since ${\rm U}(1)_{2N}$
can tell the difference, i.e. they are different topological boundary
conditions.

Now let us go to ${\rm O}(2)_{2N}$ by gauging the charge conjugation
0-form symmetry in ${\rm U}(1)_{2N}$. ${\rm O}(2)_{2N}$ has $N+7$
anyons, so let us explain how they are related to the $2N$ anyons
of ${\rm U}(1)_{2N}$. The ${\rm U}(1)_{2N}$ pairs of anyons that
get exchanged under charge conjugation will group together to form
a single anyon of the ${\rm O}(2)$ theory, so this gives us $N-1$
anyons denoted by $\phi_{k}$ where $k=1,\dots,N-1$. On the other
hand, the invariant anyons, $\lambda=0$ and $\lambda=N$, will split
into even and odd sectors
\begin{align}
\phi_{0} & \rightarrow1\oplus j,\\
\phi_{N} & \rightarrow\phi_{N}^{(1)}\oplus\phi_{N}^{(2)}.
\end{align}
 Finally, we have four twisted sectors $\sigma_{i},\tau_{i}$ where
$i=1,2$. The anyons, their spins, torus characters and quantum dimensions
can be summarized as follows (using the notation of \cite{dijkgraaf1989_OperatorAlgebra})
\begin{align}
1: & \quad\chi=\frac{1}{2}\chi_{0}+\frac{1}{2}\vartheta\left[\begin{array}{c}
0\\
1/2
\end{array}\right], & h & =0, & d=1\\
j: & \quad\chi=\frac{1}{2}\chi_{0}-\frac{1}{2}\vartheta\left[\begin{array}{c}
0\\
1/2
\end{array}\right], & h & =1, & d=1\\
\phi_{N}^{(i)}: & \quad\chi=\frac{1}{2}\chi_{N},\quad(i=1,2), & h & =\frac{1}{4}N, & d=1\\
\phi_{k}: & \quad\chi=\chi_{k},\quad(k=1,\dots,N-1), & h & =\frac{k^{2}}{4N}, & d=2\\
\sigma_{i}: & \quad\chi=\frac{1}{2}\vartheta\left[\begin{array}{c}
1/4\\
0
\end{array}\right]+\frac{1}{2}\vartheta\left[\begin{array}{c}
1/4\\
1/2
\end{array}\right],\quad(i=1,2), & h & =\frac{1}{16}, & d=\sqrt{N}\\
\tau_{i}: & \quad\chi=\frac{1}{2}\vartheta\left[\begin{array}{c}
1/4\\
0
\end{array}\right]-\frac{1}{2}\vartheta\left[\begin{array}{c}
1/4\\
1/2
\end{array}\right],\quad(i=1,2), & h & =\frac{9}{16}, & d=\sqrt{N},
\end{align}
where
\begin{equation}
\vartheta\left[\begin{array}{c}
\alpha\\
\beta
\end{array}\right]=\frac{1}{\eta(\tau)}\sum_{n\in\mathbb{Z}}q^{(n+\alpha)^{2}}\e^{2\pi\i n\beta},
\end{equation}
and $\chi_{a}$ in the above equations denotes the unflavored version
of the ${\rm U}(1)_{k}$ characters given in (\ref{eq: u(1)_k character})
for $k=2N$. Note that for $N=1$, ${\rm O}(2)_{2}$ is the same as
${\rm U}(1)_{8}$ which we have dealt with in the Abelian section.

The fusion rules are slightly different for the case when $N$ is
even versus when it is odd. They are summarized in \cite{dijkgraaf1989_OperatorAlgebra}.

The modular invariants of the double of ${\rm O}(2)_{2N}$ are just
all the compact boson theories we can construct from ${\rm U}(1)_{2N}$
and their $\mathbb{Z}_{2}$ orbifolds. ${\rm O}(2)_{2N}$ now does
not distinguish between T-dual compact bosons, so they are represented
by one topological boundary condition, while the $\mathbb{Z}_{2}$
orbifold theories correspond to two topological boundary conditions
which are related by exchanging $\sigma_{1}\leftrightarrow\sigma_{2},\,\tau_{1}\leftrightarrow\tau_{2}$
and $\phi_{N}^{(1)}\leftrightarrow\phi_{N}^{(2)}$ in the chiral or
anti-chiral copy. This latter permutation is a 0-form symmetry for
the ${\rm O}(2)_{2N}$ Chern-Simons theory which on the CFT side corresponds
to a $\mathbb{Z}_{2}$ shift symmetry for the orbifold boson \cite{thorngren2021_FusionCategoryII}.

Let us first focus on the case when $N$ is prime. In this case there
are three topological boundary conditions, which we can label as $\ket{Z^{\text{orb.}}}$,$\ket{\tilde{Z}^{\text{orb.}}}$
and $\ket{Z^{\text{circ.}}}$. The first two correspond to the orbifold
boson at $R=\sqrt{N}$, while $\ket{Z^{\text{circ.}}}$ corresponds
to the compact (circle) boson. The corresponding algebra objects are
given by
\begin{equation}
\A^{\text{orb.}}=1\bar{1}\oplus j\bar{j}\oplus\bigoplus_{k=1}^{N-1}\left(\phi_{k}\bar{\phi}_{k}\right)\oplus\phi_{N}^{(1)}\bar{\phi}_{N}^{(1)}\oplus\phi_{N}^{(2)}\bar{\phi}_{N}^{(2)}\oplus\sigma_{1}\bar{\sigma}_{1}\oplus\sigma_{2}\bar{\sigma}_{2}\oplus\tau_{1}\bar{\tau_{1}}\oplus\tau_{2}\bar{\tau_{2}},
\end{equation}
\begin{equation}
\tilde{\A}^{\text{orb.}}=1\bar{1}\oplus j\bar{j}\oplus\bigoplus_{k=1}^{N-1}\left(\phi_{k}\bar{\phi}_{k}\right)\oplus\phi_{N}^{(1)}\bar{\phi}_{N}^{(2)}\oplus\phi_{N}^{(2)}\bar{\phi}_{N}^{(1)}\oplus\sigma_{1}\bar{\sigma}_{2}\oplus\sigma_{2}\bar{\sigma}_{1}\oplus\tau_{1}\bar{\tau_{2}}\oplus\tau_{2}\bar{\tau_{1}},
\end{equation}
\begin{equation}
\A^{\text{circ.}}=1\bar{1}\oplus1\bar{j}\oplus j\bar{1}\oplus j\bar{j}\oplus2\bigoplus_{k=1}^{N-1}\left(\phi_{k}\bar{\phi}_{k}\right)\oplus\phi_{N}^{(1)}\bar{\phi}_{N}^{(1)}\oplus\phi_{N}^{(2)}\bar{\phi}_{N}^{(2)}\oplus\phi_{N}^{(1)}\bar{\phi}_{N}^{(2)}\oplus\phi_{N}^{(2)}\bar{\phi}_{N}^{(1)}.
\end{equation}
For $Z^{\text{orb.}}$, being the diagonal invariant, the symmetry
category is just given by the Verlinde lines which are just the anyons
of the chiral part. The invertible ones correspond to $1,\phi_{N}^{(1)},\phi_{N}^{(2)}$
and $j$, forming either $\mathbb{Z}_{2}\times\mathbb{Z}_{2}$ for
$N$ even or $\mathbb{Z}_{4}$ when $N$ is odd. The permutation invariant
$\tilde{Z}^{\text{orb.}}$ has the same structure but with permuted
chiral labels $\sigma_{1}\leftrightarrow\sigma_{2}$, $\tau_{1}\leftrightarrow\tau_{2}$
and $\phi_{N}^{(1)}\leftrightarrow\phi_{N}^{(2)}$. 

For $Z^{\text{circ.}}$, we can find the category of line defects
of the topological boundary in a similar way to the ${\rm SU}(2)_{4}$
examples. First, we can look at the domain wall defects that live
on the interface between a chiral ${\rm O}(2)_{2N}$ and ${\rm U}(1)_{2N}$
by condensing $j$ in ${\rm O}(2)_{2N}$. The category of domain wall
defects $\C^{\text{U}(1)_{2N}}$ will have $2N$ unconfined lines
$\phi_{\lambda}$, forming the $\mathbb{Z}_{2N}$ lines of $\mathrm{U}(1)_{2N}$,
and two confined non-invertible twist defects $\sigma_{1}$ and $\sigma_{2}$.
The nontrivial fusion rules for the case of even $N$ are given by
\begin{equation}
\sigma_{1}\otimes\sigma_{2}=\bigoplus_{\lambda\,\text{odd}}\phi_{\lambda},\qquad\sigma_{i}\otimes\sigma_{i}=\bigoplus_{\lambda\,\text{even}}\phi_{\lambda},
\end{equation}
\begin{equation}
\phi_{\lambda\,\text{even}}\otimes\sigma_{i}=\sigma_{i},\qquad\phi_{\lambda\,\text{odd}}\otimes\sigma_{1}=\sigma_{2}.
\end{equation}
Now we combine the chiral and antichiral parts of the condensed ${\rm U}(1)$
phases and gauge the diagonal $\mathbb{Z}_{2N}$. Similar to the ${\rm SU}(2)_{4}$
case, we get a symmetry category as a $\mathbb{Z}_{2}$ extension
of $\C^{\text{U}(1)_{2N}}$ where the $\mathbb{Z}_{2}$ acts on the
$\mathbb{Z}_{2N}$ by charge conjugation, giving rise to the dihedral
group $D_{4N}$ as the invertible group while just trivially doubling
the twist defects $\sigma_{i}$.\footnote{The formal reason we get this $\mathbb{Z}_{2}$ extension in both
of our examples is a special case of the general result discussed
in \cite{Frohlich:2006duality}. Condensing an algebra $B$ in the
chiral part gives the bimodule category $_{B}\C_{B}$ as the defect
category in the corresponding full CFT. When $B$ corresponds to Abelian
condensation, then the category $_{B}\C_{B}$ (the category of line
defects) can be written as an extension of the module category $\C_{B}$
(the domain wall category) by the automorphism group of $B$. } Hence, we get
\begin{equation}
<Z>'=\frac{1}{4}Z_{R=\sqrt{N}}^{\text{orb.}}+\frac{1}{4}\tilde{Z}_{R=\sqrt{N}}^{\text{orb.}}+\frac{1}{4N}Z_{R=\sqrt{N}}^{\text{circ.}}.
\end{equation}
For $N=1$, this reduces to our ${\rm U}(1)_{8}$ example where $Z_{R=\sqrt{N}}^{\text{orb.}}$
corresponds to the compact boson at $R=2$ and $\tilde{Z}_{R=\sqrt{N}}^{\text{orb.}}$
is its T-dual.\footnote{For $N=1$, the relative weights agree with the $\mathrm{U}(1)_{8}$
case, but the overall automorphism factors will be enhanced since
the twist fields $\sigma_{i}$ (which have quantum dimension $\sqrt{N}$)
become invertible at $N=1$.} For $N=2$, we have ${\rm O(2)_{4}}\simeq\Ising\times\Ising$ which
was the example studied in \cite{barbar2025_GlobalSymmetries}.

Finally, the case where $N$ is square free is very similar, but now
we have one orbit for the circle branch and another for the orbifold
branch, where each theory in the same orbit shares the same weight.
Thus in this case, restricting to the equivalence classes of CFTs,
\begin{equation}
<Z>'=\sum_{\substack{pq=N,\,p>q\\
{\rm gcd}(p,q)=1
}
}\left(\frac{1}{2}Z_{R^{2}=\frac{p}{q}}^{\text{orb.}}+\frac{1}{4N}Z_{R^{2}=\frac{p}{q}}^{\text{circ.}}\right).
\end{equation}
As $N\rightarrow\infty$, we see that the orbifold branch dominates.
This can be viewed as a regularized version of averaging over the
full orbifold branch and the circle branch with their relative symmetry
factor. 

\subsection{ADE classification: ${\rm SU}(2)_{k}$ and minimal models}

\paragraph{${\rm SU}(2)_{k}$ Chern-Simons theory.}

The physical modular invariants of doubled ${\rm {\rm SU}}(2)_{k}$
Chern-Simons theory follow the famous ADE classification \cite{Cappelli:1986hf,Cappelli:1987xt}.
However, the number of linearly independent modular invariant combinations
of left and right torus characters is given by the floor of  half
the number of factors of $k+2$, which is related to the classification
problem for ${\rm U}(1)_{2N}$ with $N=k+2$ \cite{Cappelli:1987xt}.
Hence, the general Poincaré series does not always lead to a combination
of only physical invariants as was shown in \cite{meruliya2021_PoincareSeries}.
 These unphysical modular invariants cannot be extended consistently
to all genera since they do not correspond to algebra objects in ${\rm SU}(2)_{k}$,
and there is evidence that they disappear completely in some cases
at relatively small genera \cite{korinman2019_DecompositionWittenReshetikhinTuraev}.
The TQFT gravity proposal cures this problem by construction and leads
to a linear combination of the physical invariants with positive weights
as shown before.

Let us briefly describe the ADE classification for ${\rm SU}(2)_{k}$.
The $A$ series corresponds to the diagonal invariant, which is just
the ${\rm SU}(2)_{k}$ Wess--Zumino--Witten (WZW) CFT. The $D$
series corresponds to the $\mathbb{Z}_{2}$ orbifold of the ${\rm SU}(2)_{k}$
WZW, which is possible only for $k$ even since the $\mathbb{Z}_{2}$
symmetry is anomalous otherwise. The $D_{2\ell}$ series corresponds
to a simple-current extended algebra invariant while the $D_{2\ell+1}$
is a permutation invariant. Finally, there are three exceptional cases:
$E_{6}$ and $E_{8}$ are exceptional conformal embeddings corresponding
to $su(2)_{10}\subset so(5)_{1}$ and $su(2)_{28}\subset(G_{2})_{1}$
respectively, while the $E_{7}$ is an exceptional permutation of
the $D_{10}$ invariant of $su(2)_{16}$. They are related to their
respective ${\rm SU}(2)_{k}$ WZW CFTs by generalized gauging of non-invertible
symmetries (see for example \cite{diatlyk2024_GaugingNoninvertible}).

We now want to find the weights of the ADE invariants from their symmetries.
The TDLs of the ADE theories arising from the $su(2)_{k}$ algebra
are related to Ocneanu graph algebras \cite{petkova2001_ManyFaces,petkova2001_GeneralizedTwisted,coquereaux2002_TwistedPartition,chui2003_IntegrableConformal},
which are related to the ADE Dynkin graphs. Remarkably, the group
of invertible TDLs is isomorphic to automorphisms of the associated
Dynkin graphs \cite{Lienart:2000jw}.\footnote{This seems to be special to the ADE graphs of $su(2)_{k}$. For other
affine Lie algebras there are similar graphs analogous to the ADE,
and in general the map between invertible symmetries of the CFT and
graph symmetries is not one to one. I thank Noah Snyder for a discussion
about this.} We thus get the general formula for any given level
\begin{equation}
<Z>'=\frac{1}{|\Gamma_{A}|}Z_{A}+\frac{1}{|\Gamma_{D}|}Z_{D}+\frac{1}{|\Gamma_{E}|}Z_{E},\label{eq: ADE weights su2}
\end{equation}
where the sum is over the allowed invariants at that particular level,
and $\Gamma$ denotes the automorphism group of the corresponding
Dynkin graph. For example, in our ${\rm SU}(2)_{4}$ example, the
$S_{3}$ symmetry we got for the $D$ invariant was precisely the
graph symmetry of the $D_{4}$ diagram. For any even $k$ beyond $k=4$,
we get equal weights for the $A$ and $D$ invariants since their
graph symmetries are just $\mathbb{Z}_{2}$. 

\paragraph{Virasoro minimal models ($c<1$ pure gravity).}

The unitary minimal models $\M(p,p+1)$ have an ADE classification
closely related to that of $su(2)_{k}$ \cite{Cappelli:1986hf,Cappelli:1987xt}.
This is related to the fact that they can be formulated in terms of
a coset construction
\begin{equation}
\M(k+2,k+3)\simeq\frac{su(2)_{k}\otimes su(2)_{1}}{su(2)_{k+1}},
\end{equation}
which, in terms of TQFT language, amounts to starting from ${\rm SU(2)}_{k}\times{\rm SU(2)}_{1}\times{\rm SU(2)}_{-k-1}$
and gauging (condensing) the diagonal $\mathbb{Z}_{2}$ 1-form symmetry.
So we have an ADE classification related to the ${\rm SU(2)}$ copies
at levels $k$ and $k+1$: $AA$, $AD$ (or $DA$), $AE$ (or $EA$).
The symmetries follow a similar pattern to the ${\rm SU(2)}$ case
(\ref{eq: ADE weights su2}) \cite{Ruelle:1998zu}, but now these
are the full set of invertible symmetries of the CFTs since the chiral
algebra here is $c<1$ Virasoro. This gives a viable proposal for
the boundary ensemble dual to pure $c<1$ gravity generalizing the
results of \cite{castro2012_GravityDual,jian2020_EstablishingStronglycoupled}
beyond the cases of Ising and Tricritical Ising, and solving the negativity
issues that arise in the Poincaré series for higher levels \cite{meruliya2021_PoincareSeries}.

\section{Implications for noncompact TQFTs\label{sec:Implications-for-noncompact}}

So far our main results are strictly speaking valid for compact TQFTs
based on semi-simple MTCs. Going beyond that regime, we have to be
a bit schematic and use a more heuristic approach based on the lessons
we have learned from the compact/rational case. What we have learned
so far can be summarized as follows:
\begin{itemize}
\item We start from a TQFT based on the representation theory of the vertex
algebra $\voa_{L}\times\bar{\voa}_{R}$.
\item The dual ensemble consists of CFTs constructed from topological boundary
conditions of the TQFT; these are all CFTs with $\voa_{L}\times\bar{\voa}_{R}$
as their common maximally extended vertex algebra.
\item The CFTs are inversely weighted by the size of their automorphism
group relative to $\voa_{L}\times\bar{\voa}_{R}$, which is the group
of invertible topological defects that commute with $\voa_{L}\times\bar{\voa}_{R}$.
\end{itemize}
We can now try to apply these lessons to cases where $\voa_{L}\times\bar{\voa}_{R}$
is not rational. However, we will assume that the vacuum must appear
in the physical spectrum of the CFTs akin to the rational case. 

\subsection{$\mathbb{R}$ Chern-Simons theory}

The holographic duality of the Narain ensemble average \cite{maloney2020_AveragingNarain,afkhami-jeddi2021_FreePartition}
can be, at least formally, understood as a case of this TQFT gravity
framework. Consider the Narain ensemble of $D$ bosons. The bulk TQFT
should be understood as a representation of left and right copies
of the Heisenberg VOA $u(1)^{D}\times\bar{u}(1)^{D}$, which can be
schematically taken to be $\mathbb{R}^{D,D}$ Chern-Simons theory.
This is an Abelian theory, so we should look for Lagrangian subgroups
and use the analog of (\ref{eq: code equivalence ensemble}). As noted
in \cite{benini2023_FactorizationGlobal,barbar2025_GlobalSymmetries},
the Lagrangian subgroups are given by even self-dual Lorentzian lattices
$\Lambda\subset\mathbb{R}^{D,D}$, which gives us the symmetry $\mathbb{R}^{D,D}/\Lambda\simeq{\rm U}(1)^{D}\times{\rm U}(1)^{D}$.\footnote{This is in line with recent work that relates $\mathbb{R}$ Chern-Simons
theory to the SymTFT of ${\rm U}(1)$ symmetries \cite{Antinucci:2024zjp}. } These are indeed the symmetries that commute with the $u(1)^{D}\times\bar{u}(1)^{D}$
current algebra. The moduli space of topological boundary conditions
will be given by $\mathrm{O}(D,D,\mathbb{Z})\backslash\mathrm{O}(D,D)$
as this parameterizes the embeddings of a fixed self-dual lattice
in $\mathbb{R}^{D,D}$ modulo the stabilizer subgroup of that lattice.
Since all these topological boundary conditions have the same symmetry
as in any Abelian theory, namely ${\rm U}(1)^{D}\times{\rm U}(1)^{D}$
in this case, we should use the uniform measure on $\mathrm{O}(D,D,\mathbb{Z})\backslash\mathrm{O}(D,D)$
which is the Haar measure. Pushing this sum to the distinct Narain
CFTs under the duality $\mathrm{O}(D)\times\mathrm{O}(D)$ of the
unflavored physical boundary condition\footnote{If we use a nonzero flavor in the physical boundary condition, similar
to our Abelian example in section \ref{subsec:Warm-up:-Abelian},
it will not be invariant under the separate rotations $\mathrm{O}(D)$
for left- and right-momenta, and so in that case we need to use the
full moduli of Narain lattices $\mathrm{O}(D,D,\mathbb{Z})\backslash\mathrm{O}(D,D)$
as was done in \cite{Datta:2021ftn}.} will give us the usual Narain average over $\mathrm{O}(D,D,\mathbb{Z})\backslash\mathrm{O}(D,D)/\left(\mathrm{O}(D)\times\mathrm{O}(D)\right)$
with the Haar measure, which is the same as the Zamolodchikov measure.
 This leads to the ${\rm U}(1)$ gravity result of \cite{maloney2020_AveragingNarain,afkhami-jeddi2021_FreePartition}.
Even though the TQFT gravity sums over all topologies, we can argue
that the bulk theory, which is schematically $\mathbb{R}$ Chern-Simons
theory, is only sensitive to handlebodies and that is why the bulk
sum is a Poincaré series. The genus $g$ Poincaré series converges
only for $D>g+1$ \cite{maloney2020_AveragingNarain}, so the large
genus handlebody sum at fixed $D$ needs to be regularized. A regularized
version of ${\rm U}(1)$ gravity was studied in \cite{aharony2024_HolographicDescription,Dymarsky:2025agh}
where the bulk TQFT was taken to be $D$ copies of $\mathbb{Z}_{k}$
gauge theory for prime $k$ and it was argued there that the full
Narain moduli averaged with the Haar measure is reproduced in the
limit $k\rightarrow\infty$. In that construction, the bulk theory
was only sensitive to handlebodies. It was shown recently in \cite{Barbar:2025krh}
that this behavior persists for the non-square-free case of $k=p^{2}$
for prime $p$ as $p\rightarrow\infty$.

\subsection{Virasoro TQFT\label{subsec:Virasoro-TQFT}}

 For the Virasoro case, we want to envision a TQFT based on all
representations of ${\rm Vir}_{c}\times{\rm \overline{Vir}}_{c}$.
This is different from Virasoro/Teichmüller TQFT at $c>1$ which includes
only the Liouville spectrum (i.e. Virasoro representations with chiral
conformal dimensions $h\geq\frac{c-1}{24}$) as delta-normalizable
states. Nonetheless, if we heuristically consider our results for
a formal TQFT based on $\mathrm{Rep}({\rm Vir}_{c})\times\mathrm{Rep}({\rm \overline{Vir}}_{c})$
then the ensemble should include all CFTs of central charge $c$,
where each CFT is inversely weighted by the size of its full symmetry
group.\footnote{The idea of averaging over CFTs with $1/|\Aut|$ factor was speculated
by Alex Maloney on multiple occasions given the results of ${\rm U}(1)$
gravity (see for example \cite{Maloney:StringsFields2021}).} This tells us right away that highly symmetric theories will be heavily
suppressed; for example, theories with continuous symmetries will
effectively drop out of the ensemble. This seems to be compatible
with a pure 3d gravity interpretation as $c\rightarrow\infty$. In
the large $c$ limit, the space of CFTs is very vast and has an enormous
number of highly symmetric theories as was argued in \cite{belin2025_MeasureSpace}.
For example, we will have a proliferation of products of small-central-charge
theories which will become highly symmetric in the large $c$ limit.
Luckily for us, these highly symmetric theories will crumble under
their own weight. Hence, one might expect that a typical theory of
such an ensemble would have a large gap of order $c$ and a sparse
spectrum of light states. This is motivated by the average solution
of the bootstrap at large $c$ as shown in \cite{Chandra:2022bqq}
as well as wisdom from the result of Narain average where a typical
Narain theory at large $c$ has a primary gap of $O(c)$ relative
to the $u(1)$ vacuum even though we are unable to construct a single
such theory. 

To make more sense of this ensemble of all CFTs, we need to consider
how to weigh families of CFTs that form conformal manifolds. A conformal
manifold is generated by exactly marginal deformations, so all points
share the same symmetry except for some special loci points with enhanced
symmetry, which are points of zero measure on the manifold. Since
we interpret our ensemble average result as uniform up to isomorphisms,
the natural uniform measure to use on conformal manifolds is the Zamolodchikov
measure. Each conformal manifold is then weighted by its respective
symmetry factor, while the special loci points are weighted separately
similar to isolated points in the space of CFTs. Motivated by the
proposal of \cite{belin2025_MeasureSpace}, we can try to motivate
the following measure on the space of CFTs at a given central charge\footnote{In our schematic setup here, we will work with a particular value
of the central charge rather than a small window as done in \cite{belin2025_MeasureSpace}
since our construction fixes a ``bulk TQFT'' based on the representation
of the Virasoro algebra with an a priori given central charge. }
\begin{equation}
\sum_{\M}\frac{1}{|\Aut C_{\M}|}\int\d\mu_{\text{Zam.}}+\sum_{C}\frac{1}{|\Aut(C)|},\label{eq: measureCFT}
\end{equation}
where $\M$ denotes a conformal manifold and $C_{\M}$ denotes a typical
representative CFT from that manifold. This is in contrast to the
proposed measure of \cite{belin2025_MeasureSpace} which weighs all
CFTs with equal weights and does not agree with a pure gravity interpretation.
We interpret the measure in (\ref{eq: measureCFT}) as the correct
maximum ignorance measure where the equal weights are assigned to
``labeled'' CFTs which gives us a symmetry factor weight once we
project onto isomorphism classes. This means that we should view the
CFTs via some algebraic definition and then the space of CFTs should
be a groupoid of some sort. We will comment on this aspect in the
context of chiral CFTs later in this section.  

To make sense of (\ref{eq: measureCFT}) as a normalizable measure,
we need to introduce a cutoff on conformal manifolds with divergent
volumes. This is motivated by the distance conjecture \cite{Perlmutter:2020buo,Baume:2023msm,Ooguri:2024ofs}
where we expect CFTs at infinite distance on a conformal manifold
to have a tower of light states, and hence they will not share the
same symmetry common to all other members of the conformal manifold.
So if we want to implement our 1/|Aut| measure properly they should
be included separately and to account for that we could provide a
cutoff on the integral over the manifold in (\ref{eq: measureCFT}).
The inclusion of a cutoff can be implemented universally by defining
the measure with a minimum gap as was done in \cite{belin2025_MeasureSpace}.

We can now try to apply this for cases where we have at least a rough
classification of CFTs beyond $c<1$.

\paragraph*{$c=1$ pure gravity?}

For $c=1$, even though we do not have a proven classification for
all CFTs, we sort of have a handle on what the space of (unitary)
CFTs looks like. There are two main conformal manifolds: a compact
boson branch, an orbifold branch, and then there are three isolated
orbifold points of the compact boson at the self-dual radius corresponding
to orbifolding by the three exceptional discrete subgroups of ${\rm SU}(2)$:
Tetrahedral (${\rm T}$), Octahedral (${\rm O}$) and Icosahedral
$({\rm I})$ \cite{Ginsparg:1987eb}. As we have argued earlier, the
CFTs at infinite distance of these branches (the decompactified theories)
should be treated separately, and in a suitable regularization their
contribution to the ensemble will be negligible.\footnote{For example, if one considers the decompactification limit of the
rational boson of radius $R^{2}=N$ viewed as the diagonal invariant
of $\mathrm{U}(1)_{2N}$ Chern-Simons theory, we find that the relative
symmetry is $\mathbb{Z}_{2N}$, and since the partition function diverges
as $R$, the symmetry factor will still win.} There is one more known noncompact theory to worry about: the Runkel-Watts
(RW) theory \cite{RunkelWatts2001} which is a $c\rightarrow1$ limit
theory of the AA series of minimal models and is the analog of Liouville
theory at $c=1$. This is a theory with continuous spectrum and no
invertible symmetries. It is not obvious if we should include this
theory in our ensemble because of pathologies like non-normalizability
of the vacuum state and the fact that (similar to Liouville) the vacuum
does not appear in physical OPEs. We will be more conservative and
exclude such theories from the ensemble under the assumption that
they will not appear in a regularized version of the TQFT gravity
framework for the Virasoro case. 

Just to demonstrate (\ref{eq: measureCFT}) in this case, the unnormalized
average in the formal sense would look like
\begin{multline}
<Z>'=\frac{1}{|D_{8}|}\int_{1^{+}}^{R_{\text{max}}}\frac{\d R}{R}\,Z_{R}^{{\rm orb.}}+\frac{1}{|\left({\rm U}(1)\times{\rm U}(1)\right)\rtimes\mathbb{Z}_{2}|}\int_{1^{+}}^{R_{\text{max}}}\frac{\d R}{R}\,Z_{R}^{{\rm circ.}}\\
+\frac{Z_{R=2}^{{\rm orb.}}}{|S_{4}|}+\frac{Z_{R=1}^{\text{circ.}}}{|\left({\rm SU}(2)\times{\rm SU}(2)\right)/\mathbb{\mathbb{Z}}_{2}|}+\frac{Z_{{\rm T}}}{|S_{3}|}+\frac{Z_{{\rm O}}}{|\mathbb{Z}_{2}|}+Z_{{\rm I}}.\label{eq: c=00003D1_average}
\end{multline}
Here the entire circle boson branch has symmetry $\left({\rm U}(1)\times{\rm U}(1)\right)\rtimes\mathbb{Z}_{2}$
except the self-dual radius at $R=1$ which has an enhanced symmetry
$\left({\rm SU}(2)\times{\rm SU}(2)\right)/\mathbb{\mathbb{Z}}_{2}$.
The entire orbifold branch has a $D_{8}$ symmetry except for $Z_{R=2}^{{\rm orb.}}$
(four-state Potts model) which has an enhanced symmetry $S_{4}$ \cite{thorngren2021_FusionCategoryII}
and $Z_{R=1}^{{\rm orb.}}$ which is the same as $Z_{R=2}^{{\rm circ.}}$
and hence already covered by the circle branch. The symmetries of
${\rm T}$, ${\rm O}$ and ${\rm I}$ can be deduced from their chiral
algebras' modular S-matrices which can be found in \cite{thorngren2021_FusionCategoryII}.\footnote{We assume that they do not possess any other invertible symmetries
beyond the automorphisms (outer and inner) of their chiral algebra.} As we can see, the compact boson branch effectively drops out.

\paragraph*{Chiral gravity.}

We can consider the case of having only a chiral copy of Virasoro
for which we get an ensemble of chiral CFTs at a given central charge,
where we should consider $c=24k$ for integer $k$ in order to cancel
the global gravitational anomaly. On the gravity side, this corresponds
to topologically massive gravity where we add a gravitational Chern-Simons
term with a particularly tuned coupling that kills the right-moving
Virasoro asymptotic symmetry \cite{Li:2008ChiralGravity}, and is
thus known as chiral gravity. It was shown in \cite{Maloney:2009ChiralGravity}
that the Poincaré series of chiral gravity leads to the (chiral) extremal
partition function conjectured by Witten \cite{Witten:2007kt}, and
hence evades the negativity problem that plagues the Poincaré series
of pure gravity. However, beyond $k=1$, there is some evidence that
there are no extremal CFTs \cite{Lin:2021bcp,Bae:2016yna}, so chiral
gravity with just a handlebody sum over topologies seems to be unphysical
beyond $k=1$. The TQFT gravity framework would seem to solve this
conundrum for $k>1$ by positing that the true sum over all manifolds
(not just handlebodies) is dual to an ensemble of actual CFTs which
will not be equal to the potentially unphysical extremal partition
function. Note that the ensemble is expected to be close to extremal,
for example theories like $k$ copies of the Monster will be highly
suppressed compared to other less symmetric theories.

For $k=1$, there is the well-known classification proposed by Schellekens
\cite{Schellekens:1992db} which contains 71 theories, classified
by their space of spin 1 currents denoted as $V_{1}$. These theories
were studied later and all theories, except for the Monster CFT \cite{frenkel1984_Moonshine},
have been shown to be unique to their $V_{1}$ spaces \cite{van_Ekeren_2017,van_Ekeren_2018,van_Ekeren_2021}.
It is highly conjectured that the Monster is the only theory with
no continuous symmetries, but yet remains an open problem.

Our ensemble average for $k=1$ is then just the ensemble of theories
with $\dim V_{1}=0$ which is highly conjectured to be just the Monster.
This seems to agree, at least at the level of the partition function
with the Poincaré series for $k=1$. This poses an interesting puzzle
with two possible resolutions. The first is that the extra topologies
do not matter for some reason in this scenario, and it would be important
to understand why if that is the case. The other is that the Poincaré
series could be giving an a priori different answer when viewed as
a linear combination (with some negative coefficients) of $c=24$
CFTs' partition functions than that of the sum over all topologies,
but both agree at the level of the modular functions.\footnote{Note that this is possible regardless of the uniqueness of the Monster,
since all $c=24$ CFTs have the same partition function $J(\tau)$
as the Monster up to an overall additive constant related to their
$\dim V_{1}$, and the Poincaré series does not a priori guarantee
a linear combination of the resulting partition functions with nonnegative
coefficients.} To know for sure, one would in principle need to go to higher genus
and compare the Poincaré series to the sum over all topologies. 

\paragraph*{Siegel-Weil formula for self-dual VOAs.}

Averaging over CFTs with symmetry factors tells us that we should
think about CFTs from some algebraic definition rather than just averaging
over OPE data. One way to do this is within the framework of VOAs.
Chiral CFTs have been extensively studied in such a framework where
they correspond to self-dual VOAs (also known as holomorphic VOAs)
which have only one primary. There are various approaches to generalize
the framework of VOAs to the nonchiral case (see for example \cite{KIRILLOV2002183,rosellen2004opealgebrasmodules,Singh:2023_nonchiralVOA});
however, for concreteness, we will focus on the chiral case when discussing
VOAs in this section.

As alluded to in section \ref{subsec:Algebra-automorphisms}, the
holographic duality of TQFT gravity gives a generalization of the
Siegel-Weil formula. We will now discuss this more concretely in the
context of VOAs. The usual Siegel-Weil formula can be cast in a VOA
context by mapping the lattice to its corresponding lattice VOA via
the known construction of \cite{frenkel1989vertex}. In this case
$|\Aut(V_{\Lambda})|$ is infinite for all $V_{\Lambda}$ so the direct
$\sum_{\Lambda}\frac{1}{|\Aut(V_{\Lambda})|}\Theta_{\Lambda}$ is
ill defined. Instead, the meaningful statement is to extract the common
continuous symmetry of ${\rm U(1)^{c}}$ within the subgroup that
preserves the Heisenberg algebra, which leaves us with only lattice
automorphisms. This gives us the usual Siegel-Weil formula, which
can be rewritten holographically as 
\begin{equation}
\left(\sum_{\Lambda}\frac{1}{|\Aut(\Lambda)|}\right)^{-1}\sum_{\Lambda}\frac{1}{|\Aut(\Lambda)|}Z_{\Lambda}(\Omega_{\tilde{g}})=\sum_{\gamma\in P\backslash{\rm Sp}(2\tilde{g},\mathbb{Z})}\chi_{0}^{u(1)^{c}}(\gamma\cdot\Omega_{\tilde{g}}).
\end{equation}
Here $Z_{\Lambda}$ is the partition function of the free-boson theory
based on lattice $\Lambda$ and $\chi_{0}^{u(1)^{c}}$ is the vacuum
character of the $u(1)^{c}$ algebra. The coset space on the RHS is
the quotient of the symplectic group by the Siegel parabolic subgroup
$P$. This coset sum runs over equivalence classes of handlebodies
with a boundary $\Sigma_{\tilde{g}}$ classified according to choices
of primitive Lagrangian sublattices of $H_{1}(\Sigma_{\tilde{g}},\mathbb{Z})$.\footnote{For $\tilde{g}\ge2$, handlebody fillings of $\Sigma_{\tilde{g}}$
are not fully determined by their associated Lagrangian sublattices
in $H_{1}(\Sigma_{\tilde{g}},\mathbb{Z})$, but rather by the isotopy
classes of essential simple closed boundary curves that bound disks
in the bulk. Elements of the Torelli group, the kernel of the projection
$\mathrm{MCG}(\Sigma_{\tilde{g}})\to\mathrm{Sp}(2\tilde{g},\mathbb{Z})$,
do not necessarily preserve these curves, and hence can change the
handlebody filling while preserving its associated Lagrangian sublattice
\cite{hensel2020primer,Collier:2023Solving3d}.} This is the same result that leads to ${\rm U}(1)$ gravity (but
chiral in this context) so there are no surprises there. Next, we
want to consider a generalization of this beyond lattice VOAs.

The concept of a \emph{genus of VOAs} was first proposed by H\"ohn
\cite{hoehn2003_GeneraVertex} where he defined it as the set of all
VOAs that have the same modular tensor category (MTC) as their representation
category and the same central charge. If we consider holomorphic VOAs,
where the associated MTC and TQFT are trivial, this definition is
exactly the ensemble of all chiral CFTs at a given central charge.
H\"ohn proposed a mass formula for a genus of VOAs ${\rm gen}(V)$
as well as a corresponding average over the VOA partition functions
(see also \cite{Moller:2024bulkgenus} for recent relevant work) as
\begin{equation}
\sum_{W\in{\rm gen}(V)}\frac{1}{|\Aut(W)|},\qquad\sum_{W\in{\rm gen}(V)}\frac{Z_{W}}{|\Aut(W)|},
\end{equation}
where $Z_{W}$ denotes the partition function of the VOA $W$. The
TQFT gravity duality then suggests a physical interpretation of these
quantities as partition functions of chiral Virasoro gravity, with
the former being a sum over all closed 3-manifolds and the latter
a sum over 3-manifolds with boundaries akin to the Siegel-Weil formula,
namely
\begin{align}
\sum_{W\in{\rm gen}(V)}\frac{1}{|\Aut(W)|} & \stackrel{?}{=}\lim_{g\rightarrow\infty}\sum_{\gamma\in\Gamma^{(g)}\backslash\mcg(\Sigma_{g})/\Gamma^{(g)}}\mu_{0}(M_{\gamma}){\bf Z}^{{\rm Vir}}(M_{\gamma}),\qquad\,\,\,\,\,\partial M_{\gamma}=\emptyset,\\
\nonumber \\\sum_{W\in{\rm gen}(V)}\frac{Z_{W}(\Omega_{\tilde{g}})}{|\Aut(W)|} & \stackrel{?}{=}\lim_{g\rightarrow\infty}\sum_{\gamma\in\Gamma^{(g,\tilde{g})}\backslash\mcg(\Sigma_{g})/\Gamma^{(g)}}\mu_{\tilde{g}}(M_{\gamma}){\bf Z}^{{\rm Vir}}(M_{\gamma};\Omega_{\tilde{g}}),\,\,\,\,\,\partial M_{\gamma}=\Sigma_{\tilde{g}},
\end{align}
where $\mu_{\tilde{g}}(M)$ is defined as in (\ref{eq: Goeritz measure}).

\section{Discussion\label{sec:Discussion}}

In this paper, assuming the completeness of topological boundary condition
states at large genus, we have shown that a 3d TQFT summed over all
topologies gives rise to an ensemble of boundary theories in which
each member is weighted by an appropriate symmetry factor corresponding
to its invertible symmetry relative to the bulk. When viewed in terms
of Lagrangian algebras, the ensemble average has a natural interpretation
as the uniform-up-to-isomorphism average of boundary theories that
can be constructed from the vertex algebra associated with the bulk
TQFT. This gave us a generalization of the Siegel-Weil formula now
applied to Lagrangian algebras. 

As a toy model for holography, the duality we presented can be viewed
as a confirmation of the principle of maximum ignorance \cite{deboer2024_PrincipleMaximum}
where semiclassical gravity is viewed as a maximally agnostic coarse
graining of some fine grained microscopic description. The naive application
of the principle of maximal ignorance would lead us to deduce that
we should assign equal weights to all CFTs; however, as we have seen,
if we define the CFTs algebraically then the correct maximally agnostic
average that takes isomorphisms into account should include the appropriate
automorphism/symmetry factors.

The coarse-grained interpretation of ensemble averaging in our context
can be backed up by the following argument. The bulk calculation that
we did involved no lines ending on the boundary. Information-wise,
this amounts to having only knowledge about the asymptotic left and
right vertex (chiral) algebra symmetry on the boundary and the identity
operator. Maximal ignorance would then tell us that we should do a
uniform (up to isomorphism) average on all theories that have this
vertex algebra symmetry and the identity operator, and indeed the
bulk gravity calculation involving the sum over all manifolds does
exactly that. Having access to more knowledge on the boundary through,
for example, specific local operators should lead us then to a more
fine-grained average over just the theories that include these local
operators. This can be reflected in a similar bulk calculation involving
bulk lines that end on the boundary, where now the bulk sum of TQFT
gravity needs to be done with the mapping class group of the punctured
boundary surface $\mcg(\Sigma_{g,n})$, which will project us onto
only theories where the boundary operators are local. In semiclassical
gravity, we typically do not have access to fine grained boundary
operators above the black hole threshold; we can at best know the
energy and the spin of a primary operator since these are the macroscopic
quantities of a BTZ black hole state. With only this knowledge, we
will still get a coarse ensemble average due to the universality of
the Cardy formula. However, if we somehow have access to some light
operators below the threshold (which are sparse in the spectrum),
we can fine grain our ensemble average since these operators are not
expected to be universal (except for the identity operator of course).
This seems to agree, at least in spirit, with the conclusion of Schlenker
and Witten in \cite{schlenker2022_NoEnsemble} that there is no ensemble-averaging
below the black hole threshold. In the full theory of quantum gravity,
we should have access to black hole microstates and so our ``ensemble''
should be as fine-grained as it can get, meaning we should probably
only have one microscopic theory. This would mean full knowledge about
which $\alpha$-state we are in, and as it is well known, in contrast
with the no-boundary HH state, $\alpha$-states lead to factorization.

We conclude this section with some open questions and possible future
directions.

\begin{itemize}
\item While we have considered the implications of the TQFT gravity toy
model for VTQFT, which seems to conform to the growing notion that
VTQFT is the maximal SymTFT of all topological line defects in a CFT
\cite{bao2024_QGSymQRG,Torres:2025jcb}, we need to actually try to
implement our arguments explicitly in VTQFT to verify if these implications
hold. It is not obvious how to do this, but an important step is to
first understand the implications of the measure (\ref{eq: Goeritz measure})
of the sum over manifolds. Particularly, it would be interesting to
understand if this measure resolves the tension between 3d gravity
and VTQFT on off-shell manifolds \cite{yan2025_Puzzles3D}.  Another
direction to potentially understand these issues is by seeking a regularization
of VTQFT using for example recent Turaev-Viro approaches \cite{bao2024_QGSymQRG,hartman2025_ConformalTuraevViro,hartman2025_TriangulatingQuantum}.
\item Another direction is to study this approach in higher dimensions.
Gravity is of course not topological beyond 3d and so the direct analog
of our toy model is not a representative toy model for gravity in
this case. Perhaps this could be part of the reason why pure gravity
is not expected to be dual to an ensemble in $d>3$. From the baby
universe Hilbert space perspective, this would mean that $\dim\H_{{\rm BU}}=1$
in this case, i.e. the Hartle-Hawking state is the unique state as
conjectured in \cite{McNamara:2020uza} for $d>3$. Nonetheless, one
could study examples of ensembles in higher dimensions and see if
the dual ensemble has the symmetry factor structure similar to the
one presented here. This is trivially true for Abelian TQFTs as for
example the ensemble of 4d Maxwell theories considered in \cite{Barbar:2025krh},
so it would be more interesting to consider non-Abelian examples based
on general higher fusion categories. A simple example would be to
consider a $\mathbb{Z}_{2}$ charge conjugation gauging of the 5d
Abelian BF theory considered in \cite{Barbar:2025krh}. The resulting
ensemble should include ${\rm U}(1)$ Maxwell theories with different
couplings and their $\mathbb{Z}_{2}$ orbifolds, the ${\rm O}(2)$
gauge theories. We might expect to find results similar to the $\mathbb{Z}_{2}$
orbifold of the compact boson considered in section \ref{subsec: Z2_orbifold}.
\item Finally, the holographic duality we considered gave us a Siegel-Weil
formula for Lagrangian condensations. It is tempting then to ask if
one can similarly derive a Siegel-Weil formula for an average over
condensable --- but not Lagrangian --- algebras that condense to
the same phase. This would be the analog of a genus of non-self-dual
VOAs, i.e. ~the average is not modular invariant but instead transforms
in some representation of the mapping class group, which is the representation
of the condensed phase. A prime example is the average over non-self-dual
Narain lattices considered in \cite{Ashwinkumar:2021kav}. It would
be interesting to investigate if this could be derived in the large
genus limit by considering a projection on a non-trivial representation
of the mapping class group and then performing genus reduction. 
\end{itemize}

\acknowledgments
I am especially grateful to Alfred Shapere and Anatoly Dymarsky for various fruitful discussions, collaboration on related projects and extensive comments on the draft. I also thank Mohamed Radwan, Nico Cooper and Sahand Seifnashri for general discussions and comments. I thank Noah Snyder for a useful discussion about symmetries of module categories. This work was partially supported by NSF under grant 2310426.

\appendix

\section{Brief review of modular tensor categories\label{sec:Brief-review-of MTC}}

A simple object of the MTC is one that has endomorphisms proportional
to the identity, i.e. ${\rm End}(a)\simeq\mathbb{C}{\bf id}_{a}$.
A semisimple category is one where any object is a direct sum of simple
objects. In semisimple tensor categories, we can take the tensor product
(fusion) of two simple objects and express it in terms of a direct
sum of simple objects
\begin{equation}
a\otimes b=\bigoplus_{c}N_{ab}^{c}c.
\end{equation}
The fusion coefficient $N_{ab}^{c}$ is the dimension of the vector
space of morphisms $V_{ab}^{c}\equiv\Hom(a\otimes b,c)$. The fusion
is associative, and in this case of MTCs it is also commutative due
to the existence of braiding structure (to be explained below).

There always exists a unique trivial anyon (vacuum) $0$ such that
$a\otimes0=a$. For each object $a$ there is a dual object $a^{\vee}$
such that 
\begin{equation}
a\otimes a^{\vee}=0\oplus\dots\,.
\end{equation}
This allows us to define dual morphisms $V_{c}^{ab}\equiv\Hom(c,a\otimes b)\simeq V_{ab}^{c}$
. We denote the morphisms graphically as \begin{equation}
\label{eq:morphisms}
    \raisebox{-0.8cm}{\begin{tikzpicture}[scale=1,x=0.75pt,y=0.75pt,yscale=-1,xscale=1]

\draw  [color={rgb, 255:red, 0; green, 0; blue, 0 }  ,draw opacity=1 ][fill={rgb, 255:red, 0; green, 0; blue, 0 }  ,fill opacity=1 ] (218.08,120) .. controls (218.08,118.94) and (218.94,118.08) .. (220,118.08) .. controls (221.06,118.08) and (221.92,118.94) .. (221.92,120) .. controls (221.92,121.06) and (221.06,121.92) .. (220,121.92) .. controls (218.94,121.92) and (218.08,121.06) .. (218.08,120) -- cycle ;
\draw    (190,160) -- (220,120) ;
\draw    (250,160) -- (220,120) ;
\draw    (220,120) -- (220,80) ;
\draw  [color={rgb, 255:red, 0; green, 0; blue, 0 }  ,draw opacity=1 ][fill={rgb, 255:red, 0; green, 0; blue, 0 }  ,fill opacity=1 ] (402.36,120.3) .. controls (402.37,121.36) and (401.52,122.23) .. (400.46,122.25) .. controls (399.39,122.26) and (398.52,121.41) .. (398.51,120.34) .. controls (398.5,119.28) and (399.35,118.41) .. (400.41,118.4) .. controls (401.48,118.39) and (402.35,119.24) .. (402.36,120.3) -- cycle ;
\draw    (430,80) -- (400.43,120.32) ;
\draw    (370,80.65) -- (400.43,120.32) ;
\draw    (400.43,120.32) -- (400.87,160.32) ;

\draw (178,151) node [anchor=north west][inner sep=0.75pt]   [align=left] {$\displaystyle a$};
\draw (252,149) node [anchor=north west][inner sep=0.75pt]   [align=left] {$\displaystyle b$};
\draw (222,75) node [anchor=north west][inner sep=0.75pt]   [align=left] {$\displaystyle c$};
\draw (201,111) node [anchor=north west][inner sep=0.75pt]  [font=\small] [align=left] {$\displaystyle \mu $};
\draw (359,77) node [anchor=north west][inner sep=0.75pt]   [align=left] {$\displaystyle a$};
\draw (431,75) node [anchor=north west][inner sep=0.75pt]   [align=left] {$\displaystyle b$};
\draw (402,149) node [anchor=north west][inner sep=0.75pt]   [align=left] {$\displaystyle c$};
\draw (411,112) node [anchor=north west][inner sep=0.75pt]  [font=\small] [align=left] {$\displaystyle \mu $};

\end{tikzpicture}}
\end{equation} We also have orthogonality and completeness relations as follows:\begin{equation}
\label{eq:orth}
    \raisebox{-1.4cm}{\begin{tikzpicture}[scale=0.8,x=0.75pt,y=0.75pt,yscale=-1,xscale=1]

\draw  [color={rgb, 255:red, 0; green, 0; blue, 0 }  ,draw opacity=1 ][fill={rgb, 255:red, 0; green, 0; blue, 0 }  ,fill opacity=1 ] (159.42,78.86) .. controls (159.42,77.79) and (160.28,76.93) .. (161.34,76.93) .. controls (162.4,76.93) and (163.27,77.79) .. (163.27,78.86) .. controls (163.27,79.92) and (162.4,80.78) .. (161.34,80.78) .. controls (160.28,80.78) and (159.42,79.92) .. (159.42,78.86) -- cycle ;
\draw    (131.34,119.51) -- (161.34,78.86) ;
\draw    (191.34,118.86) -- (161.34,78.86) ;
\draw    (161.34,78.86) -- (161.34,38.86) ;
\draw  [color={rgb, 255:red, 0; green, 0; blue, 0 }  ,draw opacity=1 ][fill={rgb, 255:red, 0; green, 0; blue, 0 }  ,fill opacity=1 ] (163.7,159.16) .. controls (163.71,160.22) and (162.86,161.09) .. (161.8,161.1) .. controls (160.73,161.11) and (159.86,160.26) .. (159.85,159.2) .. controls (159.84,158.14) and (160.69,157.27) .. (161.76,157.26) .. controls (162.82,157.24) and (163.69,158.1) .. (163.7,159.16) -- cycle ;
\draw    (191.34,118.86) -- (161.78,159.18) ;
\draw    (131.34,119.51) -- (161.78,159.18) ;
\draw    (161.78,159.18) -- (162.21,199.18) ;

\draw (118,116.36) node [anchor=north west][inner sep=0.75pt]   [align=left] {$\displaystyle a$};
\draw (193.02,114) node [anchor=north west][inner sep=0.75pt]   [align=left] {$\displaystyle b$};
\draw (164.77,22) node [anchor=north west][inner sep=0.75pt]   [align=left] {$\displaystyle c'$};
\draw (166.67,66) node [anchor=north west][inner sep=0.75pt]  [font=\small] [align=left] {$\displaystyle \mu '$};
\draw (167.54,151.89) node [anchor=north west][inner sep=0.75pt]  [font=\small] [align=left] {$\displaystyle \mu $};
\draw (167.1,186.67) node [anchor=north west][inner sep=0.75pt]   [align=left] {$\displaystyle c$};

\end{tikzpicture}} \, = \, \delta_{\mu \mu '} \delta_{cc'} \,\,\raisebox{-1.4cm}{\begin{tikzpicture}[scale=0.8,x=0.75pt,y=0.75pt,yscale=-1,xscale=1]

\draw    (310,38.67) -- (310.33,89.52) -- (310.15,150.24) -- (310,198.67) ;

\draw (314.67,184.33) node [anchor=north west][inner sep=0.75pt]   [align=left] {$\displaystyle c$};

\end{tikzpicture}} \qquad \qquad \qquad     \raisebox{-1.4cm}{\begin{tikzpicture}[scale=0.8,x=0.75pt,y=0.75pt,yscale=-1,xscale=1]

\draw    (429.33,30) -- (429.66,80.86) -- (429.48,141.57) -- (429.33,190) ;
\draw    (390,30) -- (390.33,80.86) -- (390.15,141.57) -- (390,190) ;

\draw (433,179) node [anchor=north west][inner sep=0.75pt]   [align=left] {$\displaystyle b$};
\draw (373,181) node [anchor=north west][inner sep=0.75pt]   [align=left] {$\displaystyle a$};

\end{tikzpicture}}= \,\enlargeop[1.6]{\sum}_{c} \enlargeop[1.6]{\sum}_{\mu} \raisebox{-1.4cm}{\begin{tikzpicture}[scale=0.8,x=0.75pt,y=0.75pt,yscale=-1,xscale=1]

\draw  [color={rgb, 255:red, 0; green, 0; blue, 0 }  ,draw opacity=1 ][fill={rgb, 255:red, 0; green, 0; blue, 0 }  ,fill opacity=1 ] (498.07,150) .. controls (498.07,148.94) and (498.93,148.08) .. (500,148.08) .. controls (501.06,148.08) and (501.92,148.94) .. (501.92,150) .. controls (501.92,151.06) and (501.06,151.92) .. (500,151.92) .. controls (498.93,151.92) and (498.07,151.06) .. (498.07,150) -- cycle ;
\draw    (470,190.65) -- (500,150) ;
\draw    (530,190) -- (500,150) ;
\draw    (500,150) -- (500.43,68.29) ;
\draw  [color={rgb, 255:red, 0; green, 0; blue, 0 }  ,draw opacity=1 ][fill={rgb, 255:red, 0; green, 0; blue, 0 }  ,fill opacity=1 ] (502.35,68.27) .. controls (502.37,69.33) and (501.51,70.2) .. (500.45,70.21) .. controls (499.39,70.22) and (498.52,69.37) .. (498.51,68.31) .. controls (498.5,67.25) and (499.35,66.38) .. (500.41,66.36) .. controls (501.47,66.35) and (502.34,67.2) .. (502.35,68.27) -- cycle ;
\draw    (530,27.97) -- (500.43,68.29) ;
\draw    (470,28.61) -- (500.43,68.29) ;

\draw (455,26.8) node [anchor=north west][inner sep=0.75pt]   [align=left] {$\displaystyle a$};
\draw (532,27.06) node [anchor=north west][inner sep=0.75pt]   [align=left] {$\displaystyle b$};
\draw (505.32,139.14) node [anchor=north west][inner sep=0.75pt]  [font=\small] [align=left] {$\displaystyle \mu $};
\draw (506.19,61) node [anchor=north west][inner sep=0.75pt]  [font=\small] [align=left] {$\displaystyle \mu $};
\draw (503.4,99) node [anchor=north west][inner sep=0.75pt]   [align=left] {$\displaystyle c$};
\draw (455,181) node [anchor=north west][inner sep=0.75pt]   [align=left] {$\displaystyle a$};
\draw (532,179) node [anchor=north west][inner sep=0.75pt]   [align=left] {$\displaystyle b$};

\end{tikzpicture}} 
\end{equation}

Using the isomorphism between $V_{ab}^{cd}$ and $V_{ac^{\vee}}^{b^{\vee}d}$,
we can show that
\begin{equation}
\sum_{e}N_{ab}^{e}N_{cd}^{e}=\sum_{f}N_{ac^{\vee}}^{f}N_{b^{\vee}d}^{f}.
\end{equation}
Thus if we define the matrices $({\bf N}_{a})_{bc}:=N_{ab}^{c}$,
we get $[{\bf N}_{a},{\bf N}_{b}]=0$ and hence they can be simultaneously
diagonalized. Since these matrices have nonnegative entries, one can
use an analog of the Frobenius-Perron theorem to show that there is
a unique vector ${\bf d}$ of maximal positive eigenvalues where we
have
\begin{equation}
{\bf N}_{a}{\bf d}=d_{a}{\bf d}.\label{eq: N_matrix_qd}
\end{equation}
The eigenvalues $d_{a}$ are called the quantum dimensions (also known
as Frobenius-Perron dimensions). These can be defined as the trace
of the identity morphism. Graphically, this is given by the unknot
of an anyon\begin{equation}
\label{eq: quantum_dimension}
    \raisebox{-0.8cm}{\begin{tikzpicture}[scale=0.8,x=0.75pt,y=0.75pt,yscale=-1,xscale=1]

\draw   (316.67,118.54) .. controls (316.67,99.88) and (331.79,84.76) .. (350.46,84.76) .. controls (369.12,84.76) and (384.24,99.88) .. (384.24,118.54) .. controls (384.24,137.21) and (369.12,152.33) .. (350.46,152.33) .. controls (331.79,152.33) and (316.67,137.21) .. (316.67,118.54) -- cycle ;

\draw (345.27,154.67) node [anchor=north west][inner sep=0.75pt]   [align=left] {$\displaystyle a$};

\end{tikzpicture}} = \rm{Tr}(\rm{\bold{id}}_{a})= d_a
\end{equation} From (\ref{eq: N_matrix_qd}) we can see that the quantum dimensions
are preserved under fusion
\begin{equation}
d_{a}d_{b}=\sum_{c}N_{ab}^{c}d_{c}.
\end{equation}
From $d_{a}$, we can define the total quantum dimension of the category
as\footnote{$\D^{2}$ is also known as the Frobenius-Perron dimension (FPdim)
of the category.}
\begin{equation}
\D=\sqrt{\sum_{a}d_{a}^{2}}.\label{eq: total_quantum_dimension}
\end{equation}

The associativity of the fusion rules is encoded in the fusion F matrix
(6j symbols) defined as\begin{equation}
\label{eq:F_symbols}
    \raisebox{-1.4cm}{\begin{tikzpicture}[scale=0.8,x=0.75pt,y=0.75pt,yscale=-1,xscale=1]

\draw  [color={rgb, 255:red, 0; green, 0; blue, 0 }  ,draw opacity=1 ][fill={rgb, 255:red, 0; green, 0; blue, 0 }  ,fill opacity=1 ] (259.08,140) .. controls (259.08,138.94) and (259.94,138.08) .. (261,138.08) .. controls (262.06,138.08) and (262.92,138.94) .. (262.92,140) .. controls (262.92,141.06) and (262.06,141.92) .. (261,141.92) .. controls (259.94,141.92) and (259.08,141.06) .. (259.08,140) -- cycle ;
\draw    (231,180) -- (261,140) ;
\draw    (291,180) -- (261,140) ;
\draw    (261,140) -- (231,100) ;
\draw  [color={rgb, 255:red, 0; green, 0; blue, 0 }  ,draw opacity=1 ][fill={rgb, 255:red, 0; green, 0; blue, 0 }  ,fill opacity=1 ] (229.08,100) .. controls (229.08,98.94) and (229.94,98.08) .. (231,98.08) .. controls (232.06,98.08) and (232.92,98.94) .. (232.92,100) .. controls (232.92,101.06) and (232.06,101.92) .. (231,101.92) .. controls (229.94,101.92) and (229.08,101.06) .. (229.08,100) -- cycle ;
\draw    (231,100) -- (201,60) ;
\draw    (161,180) -- (231,100) ;

\draw (153,179) node [anchor=north west][inner sep=0.75pt]   [align=left] {$\displaystyle a$};
\draw (223,179) node [anchor=north west][inner sep=0.75pt]   [align=left] {$\displaystyle b$};
\draw (292,179) node [anchor=north west][inner sep=0.75pt]   [align=left] {$\displaystyle c$};
\draw (266,121) node [anchor=north west][inner sep=0.75pt]  [font=\small] [align=left] {$\displaystyle \mu $};
\draw (192,39) node [anchor=north west][inner sep=0.75pt]   [align=left] {$\displaystyle d$};
\draw (232,81) node [anchor=north west][inner sep=0.75pt]  [font=\small] [align=left] {$\displaystyle \nu $};
\draw (232,119) node [anchor=north west][inner sep=0.75pt]   [align=left] {$\displaystyle e$};

\end{tikzpicture}}= \,\enlargeop[1.6]{\sum}_{f,\rho,\sigma}\,\, [F^{abc}_d]_{(e; \mu\nu)(f; \rho\sigma)} \,\,\raisebox{-1.4cm}{\begin{tikzpicture}[scale=0.8,x=0.75pt,y=0.75pt,yscale=-1,xscale=1]

\draw    (451,140) -- (481,100) ;
\draw    (551,180) -- (481,100) ;
\draw  [color={rgb, 255:red, 0; green, 0; blue, 0 }  ,draw opacity=1 ][fill={rgb, 255:red, 0; green, 0; blue, 0 }  ,fill opacity=1 ] (449.08,140) .. controls (449.08,138.94) and (449.94,138.08) .. (451,138.08) .. controls (452.06,138.08) and (452.92,138.94) .. (452.92,140) .. controls (452.92,141.06) and (452.06,141.92) .. (451,141.92) .. controls (449.94,141.92) and (449.08,141.06) .. (449.08,140) -- cycle ;
\draw    (421,180) -- (451,140) ;
\draw    (481,180) -- (451,140) ;
\draw  [color={rgb, 255:red, 0; green, 0; blue, 0 }  ,draw opacity=1 ][fill={rgb, 255:red, 0; green, 0; blue, 0 }  ,fill opacity=1 ] (479.08,100) .. controls (479.08,98.94) and (479.94,98.08) .. (481,98.08) .. controls (482.06,98.08) and (482.92,98.94) .. (482.92,100) .. controls (482.92,101.06) and (482.06,101.92) .. (481,101.92) .. controls (479.94,101.92) and (479.08,101.06) .. (479.08,100) -- cycle ;
\draw    (481,100) -- (511,60) ;

\draw (413,179) node [anchor=north west][inner sep=0.75pt]   [align=left] {$\displaystyle a$};
\draw (482,179) node [anchor=north west][inner sep=0.75pt]   [align=left] {$\displaystyle b$};
\draw (552,179) node [anchor=north west][inner sep=0.75pt]   [align=left] {$\displaystyle c$};
\draw (435,126) node [anchor=north west][inner sep=0.75pt]  [font=\small] [align=left] {$\displaystyle \rho $};
\draw (465,118) node [anchor=north west][inner sep=0.75pt]   [align=left] {$\displaystyle f$};
\draw (512,42) node [anchor=north west][inner sep=0.75pt]   [align=left] {$\displaystyle d$};
\draw (466,88) node [anchor=north west][inner sep=0.75pt]  [font=\small] [align=left] {$\displaystyle \sigma $};

\end{tikzpicture}}
\end{equation}

The braiding structure is given by the braiding matrix $R$, defined
as\begin{equation}
\label{eq:R_symbols}
    \raisebox{-1.0cm}{\begin{tikzpicture}[scale=0.8,x=0.75pt,y=0.75pt,yscale=-1,xscale=1]

\draw  [color={rgb, 255:red, 0; green, 0; blue, 0 }  ,draw opacity=1 ][fill={rgb, 255:red, 0; green, 0; blue, 0 }  ,fill opacity=1 ] (218.08,122) .. controls (218.08,120.94) and (218.94,120.08) .. (220,120.08) .. controls (221.06,120.08) and (221.92,120.94) .. (221.92,122) .. controls (221.92,123.06) and (221.06,123.92) .. (220,123.92) .. controls (218.94,123.92) and (218.08,123.06) .. (218.08,122) -- cycle ;
\draw    (220,122) -- (220,80) ;
\draw    (190,160) .. controls (219.6,154) and (243.92,129.41) .. (220,122) ;
\draw    (215.47,143.39) .. controls (213.18,139.25) and (204.5,125.12) .. (220,122) ;
\draw    (220.89,150.27) .. controls (222.22,152.82) and (242.43,160.57) .. (250,160) ;

\draw (182,163) node [anchor=north west][inner sep=0.75pt]   [align=left] {$\displaystyle a$};
\draw (251,162) node [anchor=north west][inner sep=0.75pt]   [align=left] {$\displaystyle b$};
\draw (216.86,61.43) node [anchor=north west][inner sep=0.75pt]   [align=left] {$\displaystyle c$};
\draw (197.29,97.86) node [anchor=north west][inner sep=0.75pt]  [font=\small] [align=left] {$\displaystyle \mu $};

\end{tikzpicture}}= \,\enlargeop[1.6]{\sum}_{\nu}\,\, [R^{c}_{ab}]_{\mu \nu} \,\,\raisebox{-1.0cm}{\begin{tikzpicture}[scale=0.8,x=0.75pt,y=0.75pt,yscale=-1,xscale=1]

\draw  [color={rgb, 255:red, 0; green, 0; blue, 0 }  ,draw opacity=1 ][fill={rgb, 255:red, 0; green, 0; blue, 0 }  ,fill opacity=1 ] (318.08,120) .. controls (318.08,118.94) and (318.94,118.08) .. (320,118.08) .. controls (321.06,118.08) and (321.92,118.94) .. (321.92,120) .. controls (321.92,121.06) and (321.06,121.92) .. (320,121.92) .. controls (318.94,121.92) and (318.08,121.06) .. (318.08,120) -- cycle ;
\draw    (290,160) -- (320,120) ;
\draw    (350,160) -- (320,120) ;
\draw    (320,120) -- (320,80) ;

\draw (287,160) node [anchor=north west][inner sep=0.75pt]   [align=left] {$\displaystyle a$};
\draw (350.14,158.14) node [anchor=north west][inner sep=0.75pt]   [align=left] {$\displaystyle b$};
\draw (315.43,60.29) node [anchor=north west][inner sep=0.75pt]   [align=left] {$\displaystyle c$};
\draw (301,111) node [anchor=north west][inner sep=0.75pt]  [font=\small] [align=left] {$\displaystyle \nu $};

\end{tikzpicture}}
\end{equation} The $F$ and $R$ matrices satisfy consistency conditions called
the pentagon and hexagon equations (Moore-Seiberg conditions) \cite{moore1988_PolynomialEquations}.
$F$ and $R$ depend on the choice of basis in the vector spaces
of morphisms $V_{ab}^{c}$; this is called a choice of ``gauge''.
The $R$ matrix allows us to define the following gauge invariant
quantities: the twist $\theta_{a}$ and the modular $S$ matrix.

\begin{equation}
\label{eq:twist}
    \raisebox{-1.2cm}{\begin{tikzpicture}[scale=1,x=0.75pt,y=0.75pt,yscale=-1,xscale=1]

\draw    (140,90) .. controls (140,92.5) and (138.8,127.6) .. (147.6,132.3) ;
\draw  [draw opacity=0] (150,130) .. controls (150,129.99) and (150,129.98) .. (150,129.96) .. controls (150.28,125.89) and (153.47,122.8) .. (157.11,123.05) .. controls (160.76,123.3) and (163.48,126.81) .. (163.2,130.87) .. controls (162.92,134.94) and (159.74,138.04) .. (156.09,137.79) .. controls (154.79,137.7) and (153.6,137.19) .. (152.62,136.39) -- (156.6,130.42) -- cycle ; \draw   (150,130) .. controls (150,129.99) and (150,129.98) .. (150,129.96) .. controls (150.28,125.89) and (153.47,122.8) .. (157.11,123.05) .. controls (160.76,123.3) and (163.48,126.81) .. (163.2,130.87) .. controls (162.92,134.94) and (159.74,138.04) .. (156.09,137.79) .. controls (154.79,137.7) and (153.6,137.19) .. (152.62,136.39) ;  
\draw    (150,130) -- (150,180) ;

\draw (142,182) node [anchor=north west][inner sep=0.75pt]   [align=left] {$\displaystyle a$};
\draw (142,182) node [anchor=north west][inner sep=0.75pt]   [align=left] {$\displaystyle a$};

\end{tikzpicture}} \,\, = \,\, \theta_{a} \,\,\raisebox{-1.2cm}{\begin{tikzpicture}[scale=1,x=0.75pt,y=0.75pt,yscale=-1,xscale=1]

\draw    (220,90) -- (220,180) ;

\draw (212,182) node [anchor=north west][inner sep=0.75pt]   [align=left] {$\displaystyle a$};

\end{tikzpicture}}
\end{equation}

\begin{equation}
\label{eq:S_matrix}
    S_{ab} = \frac{1}{\mathcal{D}} \,\,
    \raisebox{-0.8cm}{\begin{tikzpicture}[scale=0.5,x=0.75pt,y=0.75pt,yscale=-1,xscale=1]

\draw  [draw opacity=0] (344.6,125.37) .. controls (348.05,132.16) and (350,139.85) .. (350,148) .. controls (350,175.61) and (327.61,198) .. (300,198) .. controls (272.39,198) and (250,175.61) .. (250,148) .. controls (250,120.39) and (272.39,98) .. (300,98) .. controls (313.9,98) and (326.47,103.67) .. (335.53,112.82) -- (300,148) -- cycle ; \draw   (344.6,125.37) .. controls (348.05,132.16) and (350,139.85) .. (350,148) .. controls (350,175.61) and (327.61,198) .. (300,198) .. controls (272.39,198) and (250,175.61) .. (250,148) .. controls (250,120.39) and (272.39,98) .. (300,98) .. controls (313.9,98) and (326.47,103.67) .. (335.53,112.82) ;  
\draw  [draw opacity=0] (336.17,172.08) .. controls (332.24,164.93) and (330,156.73) .. (330,148) .. controls (330,120.39) and (352.39,98) .. (380,98) .. controls (407.61,98) and (430,120.39) .. (430,148) .. controls (430,175.61) and (407.61,198) .. (380,198) .. controls (366.52,198) and (354.29,192.67) .. (345.3,184) -- (380,148) -- cycle ; \draw   (336.17,172.08) .. controls (332.24,164.93) and (330,156.73) .. (330,148) .. controls (330,120.39) and (352.39,98) .. (380,98) .. controls (407.61,98) and (430,120.39) .. (430,148) .. controls (430,175.61) and (407.61,198) .. (380,198) .. controls (366.52,198) and (354.29,192.67) .. (345.3,184) ;  

\draw (294,205) node [anchor=north west][inner sep=0.75pt]  [font=\small] [align=left] {$\displaystyle a$};
\draw (374,204) node [anchor=north west][inner sep=0.75pt]  [font=\small] [align=left] {$\displaystyle b$};

\end{tikzpicture}}
\end{equation}where $\theta_{a}=\e^{2\pi\i s_{a}}$ and $s_{a}$ is the spin of
the anyon, which in the chiral case would be just the chiral dimension
$h_{a}$ while in the nonchiral case it is $h-\bar{h}$.

The fusion rules are related to the S matrix via the Verlinde formula
\begin{equation}
N_{ab}^{c}=\sum_{x}\frac{S_{ax}S_{bx}S_{cx}^{*}}{S_{0x}},
\end{equation}
so the S matrix diagonalizes the fusion matrices $\boldsymbol{N}_{a}$.
The quantum dimensions and the total quantum dimension are also related
to $S$ by $d_{a}=\frac{S_{0a}}{S_{00}}$ and $\D=\frac{1}{S_{00}}$.

These twists define the matrix $T_{ab}=\theta_{a}\delta_{ab}$ which,
together with $S$, gives us a representation of $SL(2,\mathbb{Z})$.
They satisfy
\begin{equation}
(ST)^{3}=\e^{2\pi\i\frac{c_{-}}{8}}C,\quad\quad S^{2}=C,\quad\quad C^{2}=\id,
\end{equation}
where $C$ is the charge conjugation matrix which maps an anyon $a$
to its dual $a^{\vee}$. The representation is projective in general,
but it becomes linear in the non-anomalous case $c_{-}=0\mod 8.$\footnote{Note that the representation of the $T$ matrix of the TQFT is related
to that of the CFT by a phase $\e^{2\pi\i\frac{c_{-}}{24}}$.}

The fusion data can sometimes be preserved under permuting some anyons.
We will call these anyon permutation symmetries. They preserve all
the gauge invariant quantities like $S$ and $T$ but can act on $F$
and $R$ by a gauge transformation. The anyon permutation symmetries
are 0-form symmetries of the TQFT. They do not correspond to the full
0-form symmetry of the TQFT though since one can have symmetries that
do not permute anyons. These are like inner versus outer automorphisms
of the underlying vertex algebra $\voa_{\T}$.

\section{Normalization of quantum dimensions of module categories\label{sec: FPdim of module categories}}

We will start with the formalities and then explain the physics picture.
We refer the reader to \cite{etingof2015_TensorCategories} for more
details about tensor and module categories. Let $\C$ be a unitary
semisimple tensor category, and let $\irr(\C)$ denote the set of
simple objects of $\C$. For any object $X$, let us denote its Frobenius-Perron
(FP) dimension as $\fpdim(X)$. The FP dimension of the category is
then defined as
\begin{equation}
\fpdim(\C)\equiv\sum_{X\in\irr(\C)}\fpdim(X)^{2}.
\end{equation}
This is related to the notation of the total quantum dimension defined
in (\ref{eq: total_quantum_dimension}) as $\D_{\C}^{2}=\fpdim(\C)$. 

Now let $A$ be a connected special symmetric Frobenius algebra object
in $\C$. The category of $A$-modules $\C_{A}$ is called a module
category \cite{etingof2015_TensorCategories}. Different algebras
can lead to the same module category, and in that case, these algebras
are called Morita equivalent. In other words, they are Morita representatives
of the same Morita class, where the module category depends on the
Morita class. For an algebra $A$, let us denote its Morita class
by $[A]$, i.e.~ $[A]\simeq[A']$ means $A$ and $A'$ are Morita
equivalent. Hence, in what follows we will denote the category of
$A$-modules by $\C_{[A]}$ since it depends only on the Morita class
of $A$.

Given a module category $\C_{[A]}$, Morita representatives of $[A]$
can be defined by what is known as the internal Hom \cite{ostrik2003module},
denoted by $\underline{\Hom}(M,M)$, for any object $M$ of the module
category $\C_{[A]}$. The object $\underline{\Hom}(M,M')$ is an object
in $\C$ defined by the following relation
\begin{equation}
\Hom_{\C_{[A]}}(X\otimes M,M')=\Hom_{\C}(X,\underline{\Hom}(M,M')),
\end{equation}
where $X$ is an object in $\C$, and $M$ and $M'$ are objects in
$\C_{[A]}$. Every Morita representative algebra can be obtained via
\cite{etingof2015_TensorCategories,fuchs2002_TFTConstruction}
\begin{equation}
A_{M}=\underline{\Hom}(M,M)\label{eq: A_M}
\end{equation}
for any object $M$ of $\C_{[A]}$.

We want to understand how to normalize the quantum/FP dimensions for
the simple objects of the module category $\C_{[A]}$, where a quantum
dimension here is a trace of the identity morphism of the simple object.
If $\C_{[A]}$ is tensor, then one can normalize the quantum dimensions
from the fusion ring of $\C_{[A]}$. However, the module category
$\C_{[A]}$ is not tensor in general.\footnote{It is always tensor, however, when $\C$ is braided and there exists
a Morita representative $A$ which is commutative.} In that case there is no preferred normalization for the quantum
dimensions of simple objects \cite{Schaumann_2013}. The normalization
we are interested in, which is used in eq. (\ref{eq: 2dTQFT part function generic}),
corresponds to setting the minimum quantum dimension to unity. Let
us denote the quantum dimension in this normalization by $\widetilde{\fpdim}$,
so if we denote an object of minimum dimension by $M_{\text{min}}$,
which is not necessarily unique, we have
\begin{equation}
\widetilde{\fpdim}(M_{\text{min}})=1.
\end{equation}
The question now is how to find the total quantum dimension $\tilde{\D}_{\C_{[A]}}$
of the module category $\C_{[A]}$ in this normalization in order
to prove eq. (\ref{eq: total quantum dim for gauged 2d}), where $\tilde{\D}_{\C_{[A]}}$
is defined by
\begin{equation}
\tilde{\D}_{\C_{[A]}}^{2}\equiv\sum_{M\in\irr(\C_{[A]})}\widetilde{\fpdim}(M)^{2},
\end{equation}
where $\irr(\C_{[A]})$ denotes the simple objects of $\C_{[A]}$.

We start with what is known as the canonical normalization of the
quantum dimensions, denoted in what follows by $\fpdim_{\text{can}}$,
which is defined such that

\begin{equation}
\sum_{M\in\irr(\C_{[A]})}\fpdim_{\text{can}}(M)^{2}={\rm FPdim}(\C).
\end{equation}
In this canonical normalization, one can show that (see \cite[Exercise 7.16.8]{etingof2015_TensorCategories})
\begin{equation}
\fpdim(\underline{\Hom}(M,M'))=\fpdim_{\text{can}}(M)\,\fpdim_{\text{can}}(M').
\end{equation}
Thus from (\ref{eq: A_M}), we get
\begin{equation}
{\rm \fpdim_{\text{can}}}(M_{\text{min}})^{2}=\fpdim(A_{\text{min}}),\qquad A_{\text{min}}=\underline{\Hom}(M_{\text{min}},M_{\text{min}}).\label{eq: A_min}
\end{equation}
This leads to our desired result

\begin{equation}
\sum_{M\in\irr(\C_{[A]})}\widetilde{{\rm FPdim}}(M)^{2}{\rm =}\frac{{\rm FPdim}(\C)}{{\rm FPdim}(A_{\text{min}})}.\label{eq: FPdim2}
\end{equation}
For the regular module category, which is the same as $\C$, $A_{\text{min}}$
is uniquely the trivial algebra (which is just the vacuum), so (\ref{eq: FPdim2})
recovers the definition of the quantum dimension of $\C$. For non-regular
module categories, $A_{\text{min}}$ must be a nontrivial algebra
and hence we always have ${\rm FPdim}(A_{\text{min}})>1$. Note that
from the definition in (\ref{eq: A_min}), $A_{\text{min}}$ is not
necessarily unique since $M_{\text{min}}$ may not be unique, but
its dimension is unique for a given module category so it is a Morita
invariant. Hence, (\ref{eq: FPdim2}) is a Morita invariant way of
specifying the relative Euler term in (\ref{eq: total quantum dim for gauged 2d}).

After choosing a representative $A_{\text{min}},$ if we use the dimensions
of $M$ as objects in $\C$, then we get \cite[Exercise 7.16.9(ii)]{etingof2015_TensorCategories}

\begin{equation}
\sum_{M\in\irr(\C_{[A]})}{\rm FPdim_{\C}}(M)^{2}={\rm FPdim}(A_{\text{min}}){\rm FPdim}(\C).\label{eq:FPdim1}
\end{equation}
In the physics picture, (\ref{eq:FPdim1}) can be derived by starting
from the $\C$-symmetric 2d TQFT on the sphere and gauging the algebra
$A_{\text{min}}$. The partition function of the gauged theory ${\cal Z}_{\C_{A_{\text{min}}}}$
can be written as the partition function of the original theory ${\cal Z}_{\C}$
with insertion of a mesh of $A_{\text{min}}$ lines. On the sphere,
since there are no non-contractible 1-cycles, the mesh can be reduced
(by removing any bubbles) to the diagram in (\ref{eq:algebra_normalization}),
which leads to
\begin{equation}
{\cal Z}_{\C_{A_{\text{min}}}}(S^{2})=\dim A_{\text{min}}\,{\cal Z}_{\C}(S^{2}),
\end{equation}
where the simple objects of the module category $\C_{A_{\text{min}}}$
(the category of boundary conditions of the gauged theory) have dimensions
as objects in $\C$. Using the normalization of (\ref{eq: FPdim2})
is equivalent to choosing an Euler counterterm $\left(\dim A_{\text{min}}\right)^{g-1}$
which is needed to normalize ${\cal Z}_{\C_{A_{\text{min}}}}(S^{2})$
to unity as per our 3d TQFT normalization choice.

\bibliographystyle{JHEP}
\bibliography{Biblio}

\end{document}